\documentclass{jfm}

\usepackage{multirow}%
\usepackage{booktabs}
\usepackage{graphicx}
\usepackage{xcolor}
\usepackage{caption}
\usepackage{subcaption}
\usepackage{newtxtext}
\usepackage{newtxmath}
\usepackage{natbib}
\usepackage{hyperref}
\usepackage{textcomp}%
\usepackage{manyfoot}%
\usepackage{algorithm}%
\usepackage{algorithmicx}%
\usepackage{algpseudocode}%
\usepackage{multirow}
\usepackage{listings}%
\hypersetup{
    colorlinks = true,
    urlcolor   = blue,
    citecolor  = black,
}

\newcommand{\RomanNumeralCaps}[1]
\linenumbers
\newcommand{\de}[2]{\frac{\partial #1}{\partial #2}}
\newcommand{\deq}[2]{\frac{\partial^{2} #1}{\partial #2^{2}}}
\newcommand{\dt}[2]{\frac{\mathrm{d} #1}{\mathrm{d} #2}}

\title{Reynolds number effects on turbulent flow in curved channels}

\author{Giulio Soldati\aff{1}
  \corresp{\email{giulio.soldati@uniroma1.it}},
  Paolo Orlandi\aff{1}
 \and Sergio Pirozzoli\aff{1}}

\affiliation{\aff{1}Sapienza Universit\`a di Roma, Dipartimento di Ingegneria Meccanica e Aerospaziale,
via Eudossiana 18, Roma, Italy}

\begin{document}
\maketitle

\begin{abstract}
In this work, we study the flow in curved channels, 
an archetypal configuration 
that allows insights into problems featuring turbulence bounded by curved walls. 
Besides its relevance to many engineering applications, 
it exhibits a rich physics due to the presence of turbulence superimposed 
to large-scale structures driven by centrifugal instabilities.
The resulting secondary motions, which depend on the channel geometry 
and Reynolds number, break symmetry between the convex and the concave surface. 
We investigate the effects of curvature by focusing on two cases 
of mildly and strongly curved channel, in which shear and inertia are 
supposed to control the general features of the flow, respectively. 
For each geometry, we examine systematically the effects of 
the Reynolds number: we run a campaign of direct numerical simulations 
covering flow regimes 
from laminar up to moderately high value of Reynolds number -- 
based on bulk velocity and channel height -- of 87000. 
Our analysis pivots around the friction coefficient, which is the macroscopic 
observable of the flow, and explores how the large-scale structures change 
their shape and role on turbulence for increasing Reynolds numbers. 
Special attention is paid to the longitudinal large-scale structures 
(resembling Dean vortices), and how their dynamics of splitting and merging 
is influenced by curvature. 
In addition, we also observe and characterise transverse large-scale structures 
populating the convex wall of strongly curved channels, which are originated by 
streamwise instabilities and contribute to the negative production of turbulence kinetic energy.
\end{abstract}


\section{Introduction}
\label{intro}
Turbulence bounded by curved walls is a key aspect of various engineering 
applications, such as highly cambered aerofoils, turbo-machinery blades 
and cooling channels. In these configurations, curvature alters significantly 
turbulence structures, impacting friction, heat transfer 
and flow stability. The spatially evolving and geometry-dependent nature of 
these flows poses major challenges in developing a unified and comprehensive 
description of curvature effects on turbulence, which can be achieved 
by considering the time-evolving curved channel flow. 
Direct numerical simulations (DNS) of plane channel flow led to a scientific 
breakthrough in the theory of wall-bounded turbulence through the pioneering 
work of~\cite{kim1987turbulence}, and subsequent efforts at higher Reynolds 
numbers were made by~\cite{lee2015direct} and by \cite{pirozzoli2021one} 
for the circular pipe flow. In a similar vein, we scrutinise the effect of 
the Reynolds number on flow in curved channels, with the aim of providing 
a benchmark dataset for Reynolds averaged Navier-Stokes (RANS) and 
large eddy simulations (LES) of turbulent flows over curved surfaces.  

Compared to flat walls, surface curvature produces additional strain 
rates an order of magnitude larger than what dimensional arguments would 
suggest~\citep{bradshaw1973effects}. 
The laminar solution of the equations of motion indicates that 
curvature effects on quantities such as the 
friction coefficient and boundary layer thickness, $\delta$, 
are of the order $\mathcal{O}(\kappa\delta)$, where $\kappa$ 
is the surface curvature~\citep{patel1997longitudinal}. 
However, experimental measurements in turbulent flow reveal much 
stronger effects, underscoring the direct impact of curvature 
on turbulence.
A crucial role is played by the onset of centrifugal instabilities 
that break the symmetry of the flow. Wall curvature generates a centrifugal 
force per unit mass, $\rho U^2/r$, where 
$U$ is the mean streamwise velocity at radius $r$ from the centre of curvature, 
counteracted by the radial pressure gradient, $\partial P/\partial r$. 
\citet{rayleigh1917dynamics} first investigated the stability of curved flow 
for an ideal fluid, showing that for flow over a concave surface the centrifugal 
force of the displaced fluid element is greater than the centripetal 
pressure gradient, leading to flow instability, whereas the reverse occurs 
for convex walls. This idea was later supported by experimental 
evidence~\citep{wattendorf1935study, eskinazi1956investigation, 
ellis1974turbulent}. 
\citet{dean1928fluid} showed analytically that fully developed flow between 
two concentric cylinders becomes unstable when the Reynolds number exceeds 
a critical value, which decreases as the channel curvature increases. 
The centrifugal instability leads to the so-called Dean vortices, 
pairs of counter-rotating longitudinal roll cells. Similar structures, 
the Taylor-G{\"o}rtler vortices, form when the flow over a concave surface 
exceeds a critical Reynolds number~\citep{gortler1954three}.
Several experiments on boundary layers over concave walls~\citep{so1975experiment, 
smits1979effect, hoffmann1985effect} and on curved duct 
flow~\citep{brewster1959stability, ellis1974turbulent, hunt1979effects} 
revealed the presence of regular spanwise variations in mean velocity and 
friction coefficient, which were considered evidence for the existence of 
longitudinal large-scale structures. Due to these stable structures, 
an upwash (downwash) motion is generated between a pair of counter-rotating 
vortices, resulting in a corresponding peak (trough) in boundary layer thickness 
and a trough (peak) in the friction coefficient. 
Other experiments~\citep{eskinazi1956investigation, ramaprian1978structure, 
kobayashi1989two}, however, did not detect any variation along the span.
\citet{barlow1988structure} noted the presence of large-scale eddies, 
which did not cause spanwise variations due to their unsteadiness. 
 
The dynamics of longitudinal vortices may be unsteady due to events 
of splitting and merging, depending on curvature and Reynolds number. 
Through experiments in a curved duct, 
\cite{ligrani1988flow} and \cite{ligrani1994splitting} observed the splitting 
of vortex pairs, during which new vortex pairs seem to emerge from the 
concave wall between existing pairs. 
A similar splitting mechanism was noted 
by~\cite{alfredsson1989instabilities} in channels with rotation and 
by~\cite{matsson1990curvature} in a channel with both curvature and rotation. 
Flow visualisations with reflective flakes allowed the authors to recognised
the disappearance of vortex pairs from the merging of adjacent bright streaks. 
\cite{guo1991splitting}, using linear stability theory and spectral methods, 
found that vortex splitting and merging in rotating or curved channels 
are caused by Eckhaus instability, namely a secondary instability of steady 
periodic flows (such as the Dean vortex flow) 
with respect to spanwise perturbations~\citep{eckhaus1965studies}.  
In addition, a secondary instability with respect to streamwise perturbations
affects the curved channel flow. By simulated roll cells 
in curved channels, \cite{finlay1988instability} observed a 
`wavy Dean vortex flow' in the form of traveling waves superposed 
on the secondary flow (the Dean vortices). The authors suggested that 
traveling waves are originated by a shear-layer instability 
induced by the Dean vortices. 
Consistently, the stability analysis carried out by~\cite{yu1991secondary} 
for the G{\"o}rtler flow identified the inflectional profile of the mean 
velocity as the mechanism driving the secondary instability. 
Similar results were obtained through experiments of  
curved channel flow by~\cite{matsson1990curvature}, who 
found a secondary instability of traveling-wave type at a Reynolds number 
about three times higher than the critical one for the primary 
(centrifugal) instability. 
The secondary instability was further characterised 
by~\cite{matsson1992experiments}, who described it as a 
`wave train riding on the primary instability', 
traveling in the downstream direction with about $80\%$ 
of the bulk velocity and with a streamwise wavelength slightly larger 
than the spanwise wavelength of the primary disturbance. 

The distinct three-dimensional effects associated with concave curvature 
are absent in the case of convex curvature, whose main effect is to reduce 
turbulence intensity~\citep{gillis1983turbulent}. \citet{so1973experiment} 
found experimentally that the turbulent stress decreases near the wall 
and vanishes about midway between the convex surface and the edge of the 
velocity gradient layer. 
The strong reduction in turbulent shear stress due to convex curvature  
leads to a shift of the zero-crossing point closer to the convex wall. 
This asymmetry is associated with a displacement between the locations 
where turbulent and viscous shear stresses vanish, creating a small 
region with negative production of turbulence kinetic energy. 
The phenomenon of negative production in curved channel flow was 
examined by~\cite{eskinazi1969energy}, who related it to a local 
`energy reversal' mechanism, whereby energy is transferred from 
turbulent fluctuations to the mean flow.
In the attempt to interpret the effects of curvature on the velocity profile, 
\cite{patel1969measurements} observed an analogy between the effects 
of concave (convex) curvature and favorable (adverse) pressure gradient 
on boundary layers. That author concluded that curvature affects directly 
the inner layer, contrary to previous studies suggesting that  
the velocity profile is curvature-independent in the viscous sublayer. 
The analogy between the effects of curvature and streamwise pressure gradient, 
however, lacks general validity: a favorable pressure gradient can lead 
to flow relaminarisation, which may occur, rather, on the convex wall of 
the curved channel~\citep{brethouwer2022turbulent}.

Despite the large amount of laboratory data available in the literature, 
the effects of curvature on turbulence have neither been fully quantified nor 
understood. Due to the narrow range of both Reynolds numbers and curvatures 
spanned by the experiments conducted so far, it is difficult to draw general 
conclusions. Additionally, it is often unclear whether the flow is fully developed, 
inflow conditions are free of disturbances and three-dimensional 
effects due to side walls are negligible. DNS 
is an invaluable tool that can help to understand the flow physics more thoroughly.
The first major contribution was made by~\cite{moser1987effects}, who performed 
a DNS of fully developed turbulent flow at bulk Reynolds number 
$\Rey_b=u_b\delta/\nu=5200$ (where $u_{b}$ is the bulk velocity, 
$\delta$ the channel height and $\nu$ the kinematic viscosity) 
in a curved channel with curvature radius $r_c/\delta=39.5$ at the channel 
centreline. They concluded that most turbulence quantities, 
except for the turbulent shear stress, are equivalent on the convex and concave
sides of the channel when scaled in local wall units, inferring the existence 
of flow similarity. However,~\cite{nagata2004spatio}, who carried out 
several DNS of mildly ($r_c/\delta=39.5$) to moderately ($r_c/\delta=2.5$) 
curved channel flow at $\Rey_{b}\approx4600$, did not observe near-wall similarity. 
More recently,~\cite{brethouwer2022turbulent} studied by DNS 
the fully developed turbulent flow in channels with mild ($r_c/\delta=30$) 
to moderate ($r_c/\delta=3$) curvature at $\Rey_{b}=40000$. 
Velocity fluctuations in local scaling turned out to collapse 
into a single curve in the viscous wall region, while in the outer layer 
were found to depend heavily on curvature. 

\begin{figure}
\centering
\includegraphics[width=.85\textwidth]{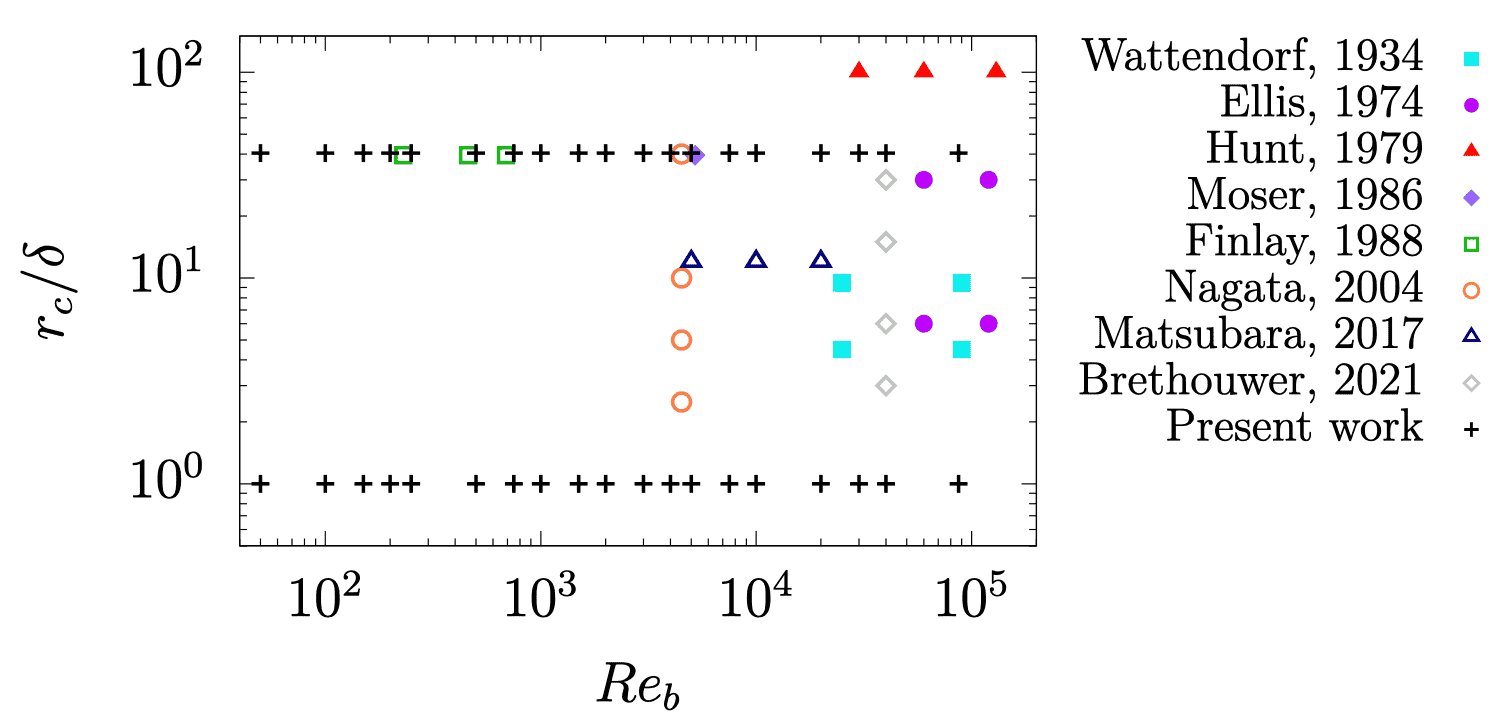}
\caption{Overview of previous experimental and computational studies
of curved channel flow in terms of bulk Reynolds number ($\Rey_b$), 
and relative curvature radius ($r_c/\delta$). 
Crosses indicate the flow cases computed in the present work.}
\label{fig:preview}
\end{figure}
 
Given this background, with this work we aim to extend our understanding of turbulence 
in curved channels by widening the range of Reynolds numbers and examining cases with 
extreme curvature, thus filling existing gaps in the current literature.
This is well illustrated in figure~\ref{fig:preview}, which reports 
an overview of the controlling flow parameters, namely bulk Reynolds number ($\Rey_b$)
and relative curvature radius ($r_c/\delta$), from
previous experimental and computational studies of curved channel 
flow, along with our DNS.
We investigate the effects of channel curvature by focusing 
on two extreme cases, namely a mildly ($r_c/\delta=40.5$) 
and strongly ($r_c/\delta=1$) curved channel. 
According to the definition proposed by~\cite{hunt1979effects}, these two 
cases correspond to `shear-dominated' and `inertia-dominated' flows, 
respectively. For both cases, we examine the effect of the Reynolds number, 
which has never been studied systematically for this flow configuration, 
by carrying out a DNS campaign covering a wide range of flow regimes. 
Special attention is paid to friction, specifically to how
the changes in flow organisation and turbulence structures induced 
by curvature affect its behavior. This analysis reveals important insights 
into the flow transition, which varies according to the type and magnitude 
of curvature.
In addition, we focus on characterising the secondary motions induced by 
large-scale structures that develop in curved channel flow. 
 
The paper is structured as follows: in \S\ref{sec:method} we describe the 
numerical methodology used for the analysis; in \S\ref{sec:results} we 
present the main results of the DNS campaign: 
specifically, the friction coefficient and flow transition 
are studied in \S\ref{sec:fric}; 
the flow organisation is explored through instantaneous velocity fields 
in \S\ref{sec:visua} and time-averaged velocity spectra in \S\ref{sec:spectra}; 
longitudinal large-scale structures are addressed in \S\ref{sec:long}, 
with a focus on splitting and merging events in \S\ref{sec:split}, 
and on their role on velocity fluctuations in \S\ref{sec:fluc};  
transverse large-scale structures are discussed in \S\ref{sec:transv}, 
and their role on shear stress and pressure fluctuations 
at the convex wall is analysed in \S\ref{sec:wall}; the structure of the 
turbulent shear stress is investigated in \S\ref{sec:rey} through a quadrant 
analysis, and a region with negative production is observed and examined 
in \S\ref{sec:reversal}; 
finally, in \S\ref{sec:conclusions} 
we conclude with a discussion of the results.

\section{Methodology}\label{sec:method}

\begin{figure}
\centering
(a)\includegraphics[width=.53\textwidth]{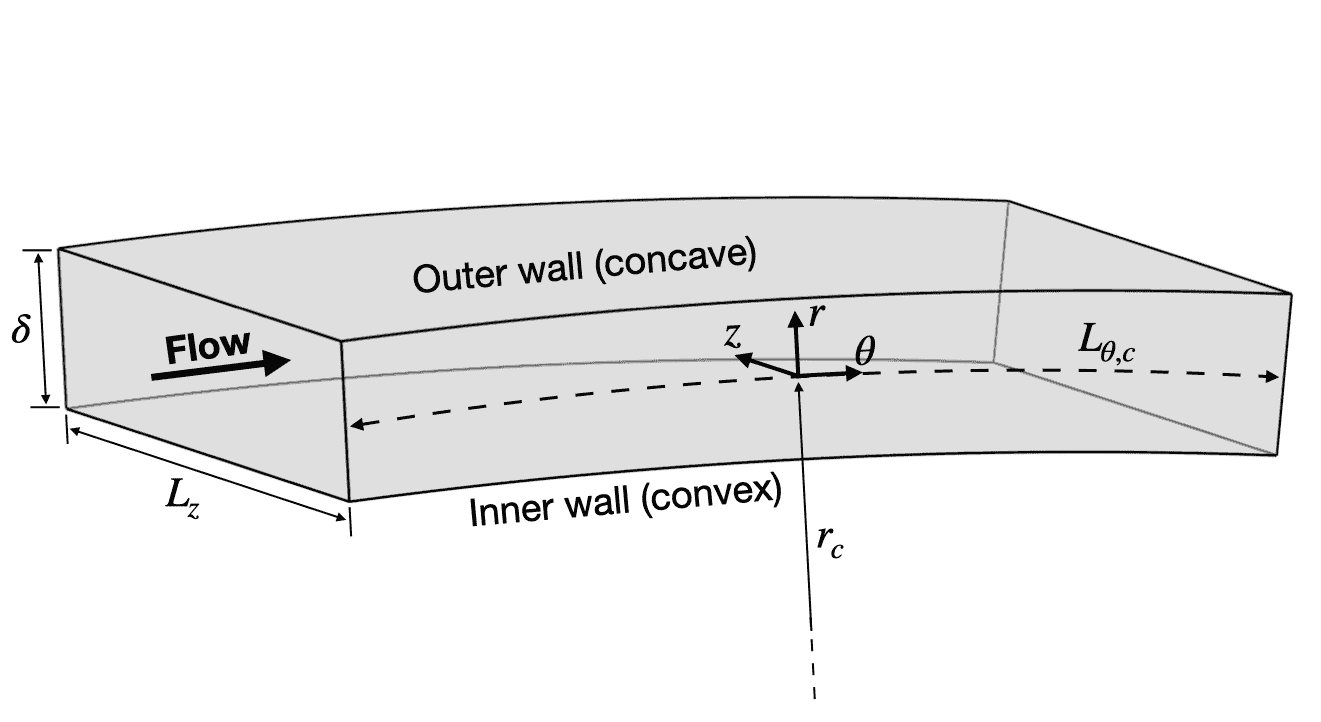}
(b)\includegraphics[width=.40\textwidth]{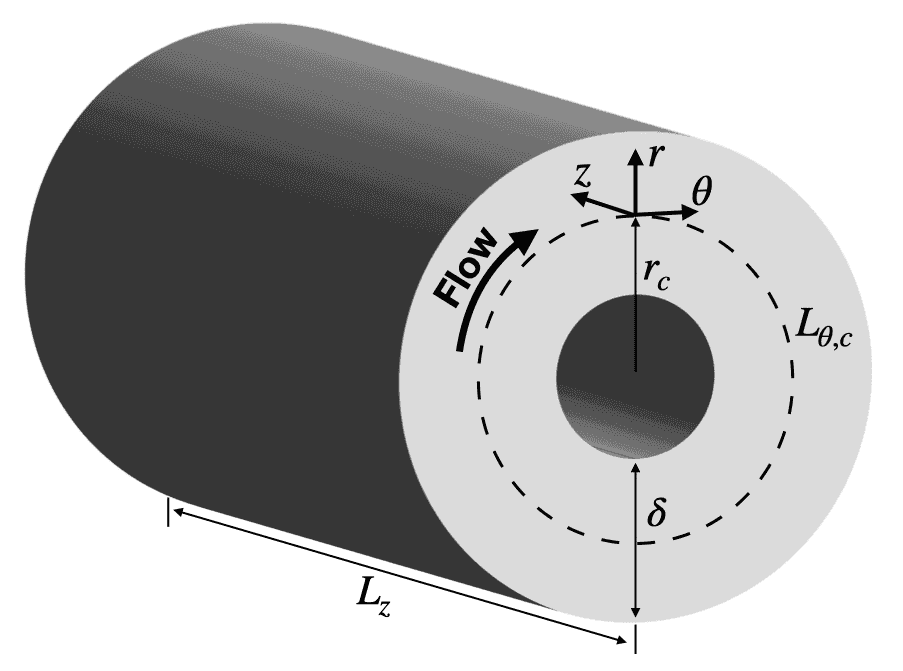}
\caption{Computational setup for flow in mildly (a) and strongly (b) curved channels.} 
\label{fig:domain}
\end{figure}

The turbulent flow in curved channels is simulated in a computational domain 
bounded by sectors of concentric cylinders, as shown in figure~\ref{fig:domain}.
The velocity components along the streamwise ($\theta$), radial ($r$) 
and spanwise ($z$) directions are denoted by $u$, $v$ and $w$, respectively.
The flow is driven by a mean-pressure gradient ($\partial P/\partial\theta$),   
which is imposed as a volumetric forcing to maintain constant mass flow 
rate in time. Numerical simulations are carried out assuming periodicity 
conditions in the streamwise and spanwise directions, so that the flow is 
fully developed. 

The in-house code used for DNS, which solves the 
incompressible Navier-Stokes equations in cylindrical coordinates, 
stems from a previous solver developed by~\cite{verzicco1996finite} and 
used for DNS of pipe flow by~\cite{orlandi1997direct}.
The switch from the pipe setup to the curved channel was attained 
implementing two main modifications to the code. The first is 
the addition of an inner cylinder (the inner wall of the channel) 
concentric to the outer cylinder. The second is the change in direction 
of the mean-pressure gradient, which is imposed along the azimuthal 
direction ($\theta$), namely the streamwise direction of the curved channel. 
In this respect, we note that the pressure term in the streamwise momentum equation is 
$(\partial P/\partial\theta)/r$, with constant $\partial P/\partial\theta$, 
hence the volumetric forcing varies along the radial direction. 
The spatial discretisation is based on second-order finite-difference schemes, 
which are implemented in the classical marker-and-cell 
framework~\citep{harlow1965numerical}. The pressure is located at the cell 
centres, whereas the velocity components at the cell faces, thus removing 
odd-even decoupling phenomena and guaranteeing discrete conservation 
of the total kinetic energy in the inviscid limit~\citep{pirozzoli2023searching}. 
The governing equations are advanced in time by means of a hybrid third-order 
low-storage Runge-Kutta algorithm, whereby the diffusive terms are handled 
implicitly and convective terms explicitly. Further details regarding the 
numerical methods implemented in the code can be found e.g. 
in~\cite{pirozzoli2021one}. The code was adapted to run on clusters 
of graphic accelerators (GPUs), using a combination of CUDA Fortran 
and OpenACC directives, and relying on the CUDA libraries for efficient 
execution of fast Fourier transforms.
\begin{table}
\begin{center}
\begin{tabular}{ccccc}
\multicolumn{5}{c}{Mild curvature (R40): $r_c/\delta=40.5$, $L_{\theta}/\delta\times L_z/\delta=2\pi\times4$}\\
\toprule
$\Rey_{b}$ & $\Rey_{\tau,i}$ & $\Rey_{\tau,o}$ & $N_\theta\times N_r\times N_z$ & $r_o\Delta\theta^+\times\Delta r^+\times\Delta z^+$ \\ 	       
\midrule
$25$    &$6    $&$6  $ &$129\times65 \times129$&$0.6\times0.01\times0.4$                      \\ 
$50$    &$9    $&$9  $ &$129\times65 \times129$&$0.9\times0.01\times0.5$                      \\ 
$100$   &$12   $&$12 $ &$129\times65 \times129$&$1.2\times0.01\times0.8$                      \\ 
$150$   &$15   $&$15 $ &$129\times65 \times129$&$1.5\times0.01\times0.9$                      \\ 
$200$   &$17   $&$17 $ &$129\times65 \times129$&$1.7\times0.01\times1.0$                      \\ 
$250$   &$19   $&$19 $ &$129\times65 \times129$&$1.9\times0.01\times1.2$                      \\ 
$500$   &$28   $&$31 $ &$129\times65 \times129$&$3.1\times0.01\times2.0$                      \\ 
$750$   &$35   $&$40 $ &$129\times65 \times129$&$4.0\times0.02\times2.5$                      \\ 
$1000$  &$41   $&$50 $ &$129\times65 \times129$&$5.0\times0.01\times3.1$                      \\ 
$1500$  &$51   $&$65 $ &$161\times97 \times161$&$5.2\times0.01\times3.2$                      \\ 
$2000$  &$61   $&$78 $ &$193\times97 \times193$&$5.2\times0.01\times3.3$                      \\ 
$2500$  &$69   $&$92 $ &$225\times97 \times225$&$5.3\times0.02\times3.3$ \\ 
$3000$  &$82   $&$112$ &$257\times129\times257$&$5.6\times0.01\times3.5$                      \\ 
$4000$  &$103  $&$143$ &$321\times129\times321$&$5.8\times0.01\times3.6$                      \\ 
$5000$  &$139  $&$170$ &$321\times145\times353$&$6.8\times0.01\times3.8$ \\
$7500$  &$211  $&$241$ &$449\times145\times449$&$6.9\times0.02\times4.3$ \\
$10000$ &$270  $&$305$ &$513\times161\times513$&$7.7\times0.02\times4.8$ \\
$20000$ &$506  $&$583$ &$769\times205\times1025$&$9.8\times0.03\times4.6 $\\
$30000$ &$720  $&$839$ &$1153\times261\times1537$&$9.4\times0.03\times4.4 $\\
$40000$ &$922 $&$1085$ &$1409\times303\times1793$&$9.9\times0.03\times4.8$ \\
$87000$ &$1824 $&$2177$&$3073\times489\times3585$&$9.1\times0.03\times4.8$ \\
\bottomrule
\multicolumn{5}{c}{$\quad$}\\
\multicolumn{5}{c}{Strong curvature (R1): $r_c/\delta=1.0$, $L_{\theta}/\delta\times L_z/\delta=2\pi\times8$}\\
\toprule
$\Rey_{b}$ & $\Rey_{\tau,i}$ & $\Rey_{\tau,o}$ & $N_\theta\times N_r\times N_z$ & $r_o\Delta\theta^+\times\Delta r^+\times\Delta z^+$ \\ 	       
\midrule
$25$    &$8   $&$5   $&$129\times65\times129$&$0.8\times0.01\times0.7$ \\
$50$    &$11  $&$8   $&$129\times65\times129$&$1.1\times0.01\times1.0$ \\
$75$    &$14  $&$10  $&$129\times65\times129$&$1.5\times0.01\times1.3$ \\
$100$   &$16  $&$12  $&$129\times65\times129$&$1.8\times0.01\times1.5$ \\
$150$   &$20  $&$15  $&$129\times65\times129$&$2.2\times0.01\times1.9$ \\
$200$   &$23  $&$18  $&$129\times65\times129$&$2.6\times0.01\times2.2$ \\
$250$   &$26  $&$21  $&$129\times65\times129$&$3.0\times0.01\times2.6$ \\
$500$   &$39  $&$35  $&$129\times65\times129$&$5.1\times0.01\times4.4$ \\
$750$   &$49  $&$47  $&$161\times65\times193$&$5.5\times0.02\times3.9$ \\
$1000$  &$58  $&$58  $&$193\times97\times257$&$5.7\times0.01\times3.6$ \\
$1500$  &$73  $&$79  $&$225\times97 \times289$&$6.7\times0.01\times4.4$ \\
$2000$  &$86  $&$98  $&$225\times129\times385$&$8.2\times0.01\times4.0$ \\
$2500$  &$98  $&$115 $&$289\times129\times449$&$7.6\times0.01\times4.1$ \\
$3000$  &$110 $&$132 $&$385\times129\times513$&$6.5\times0.01\times4.1$ \\
$4000$  &$132 $&$165 $&$449\times145\times513$&$6.9\times0.01\times5.1$ \\
$5000$  &$152$ &$193$ &$449\times145\times769$&$8.1\times0.02\times4.0$ \\ 
$7500$  &$199$ &$263$ &$577\times145\times961$&$8.6\times0.02\times4.4$ \\ 
$10000$&$244$ &$327$  &$641\times161\times1153$&$9.6\times0.04\times4.5$ \\
$20000$&$403$ &$563$  &$1153\times205\times2049$&$9.2\times0.03\times4.4$\\
$30000$&$551$ &$784$  &$1665\times261\times2817$&$8.8\times0.02\times4.4$\\
$40000$&$697$ &$995$  &$2049\times303\times3585$&$9.2\times0.02\times4.4$\\
$87000$&$1362$&$1952$ &$4097\times489\times7169$&$9.0\times0.02\times4.4$\\
\bottomrule
\end{tabular}
\caption{Flow parameters: 
bulk Reynolds number, friction Reynolds number at the inner and outer wall,
number of grid points and grid spancing in inner units 
in the streamwise, radial and spanwise directions, respectively.
The title line reports curvature radius and domain extension 
in the streamwise (along the centreline) and spanwise directions.}
\label{tab:1}
\end{center}
\end{table}

The flow field is controlled by two parameters, namely the bulk Reynolds 
number, $\Rey_b$, and the radius of curvature at the centreline, $r_c/\delta$. 
To investigate the effect of both parameters, we split the simulation 
campaign in two main groups: 1) shear-dominated flow with mild curvature 
($r_c/\delta=40.5$), and 2) inertia-dominated with strong curvature ($r_c/\delta=1$).
Throughout the paper, we will refer to the first group as R40 and to the 
second group as R1. Within each group, we vary the Reynolds number from 
$\Rey_b=25$ up to $87000$. The domain extends in the radial direction 
from $r=r_i$ (inner wall), to $r=r_o$ (outer wall), 
where $r_o-r_i=\delta$ and $r_c=(r_i+r_o)/2$. 
To compare cases with different values of the curvature 
radius, we consider the $y$-coordinate, aligned with the radial direction 
and origin shifted at the inner wall, i.e. $y=r-r_{i}$. 
In all cases, 
the computational domain has a streamwise length $L_{\theta}/\delta=2\pi$ 
along the centreline, meaning that for the R1 cases the domain resolves a  
full cylinder circumference (see figure~\ref{fig:domain}). 
The spanwise width is set to $L_z/\delta=4$ 
for the R40 cases, and it is doubled for R1 cases, in order 
to minimise the influence of spanwise periodicity on the large-scale structures.  
The effects of the domain sizes on the smulations results are assessed 
in appendix~\ref{app:domain}. 
Domain sizes, number of grid points and grid spacings are listed 
in table~\ref{tab:1}, along with the resulting friction 
Reynolds number at the inner ($\Rey_{\tau,i}=u_{\tau,i}\delta/2\nu$) 
and outer wall ($\Rey_{\tau,o}=u_{\tau,o}\delta/2\nu$). Distinction 
between inner and outer wall is necessary as the friction velocity changes 
depending on the wall at which it is evaluated.
A global friction velocity can also be defined based on 
the mean-pressure gradient, as shown in previous 
works~\citep[e.g.][]{moser1987effects, brethouwer2022turbulent}. 
Inner, outer and global friction velocities are defined respectively as 
\begin{equation}
u_{\tau,i}=\sqrt{\nu\frac{\partial U}{\partial r}\bigg|_{r_i}},\quad
u_{\tau,o}=\sqrt{\nu\frac{\partial U}{\partial r}\bigg|_{r_o}},\quad 
u_{\tau,g}=\sqrt{\frac{u_{\tau,i}^2 r_i^2+u_{\tau,o}^2 r_o^2}{2r_c^2}}, 
\label{eq:utau}
\end{equation}
where $u_{\tau,g}$ is derived in appendix~\ref{app:utaug}. 
 
Throughout the paper, we will use two types of normalisation: local wall scaling, 
based on $u_{\tau,i}$ and $u_{\tau,o}$ (denoted with the `plus' superscript), 
and global wall scaling, based on $u_{\tau,g}$ (denoted with the `star' superscript). 
Brackets denote the averaging operator (where the subscripts denote the variables 
over which the averaging is done), the overline is reserved for temporal averages,  
capital letters denote flow properties averaged in the homogeneous spatial 
directions and in time, and lower-case letters denote instantaneous values. 
Hence, $u=U+u'$, where $u'$ is the fluctuation from the mean.
The grid spacings in local wall scaling ($r_o\Delta\theta^+$ and $\Delta r^+_{o}$) 
listed in table~\ref{tab:1} are evaluated at the outer wall, which is the
most critical as the wall shear stress is higher and the grid spacing in the streamwise 
direction is larger.
All simulations are run for sufficient time to reach a statistically steady state, 
and further for at least $600\delta/u_b$ to ensure convergence of statistics, 
which is verified in appendix~\ref{app:stats}. 

\section{Results}\label{sec:results}

\subsection{Friction coefficient and flow transition}\label{sec:fric}

\begin{figure}
\centering
(a)\includegraphics[width=.47\textwidth]{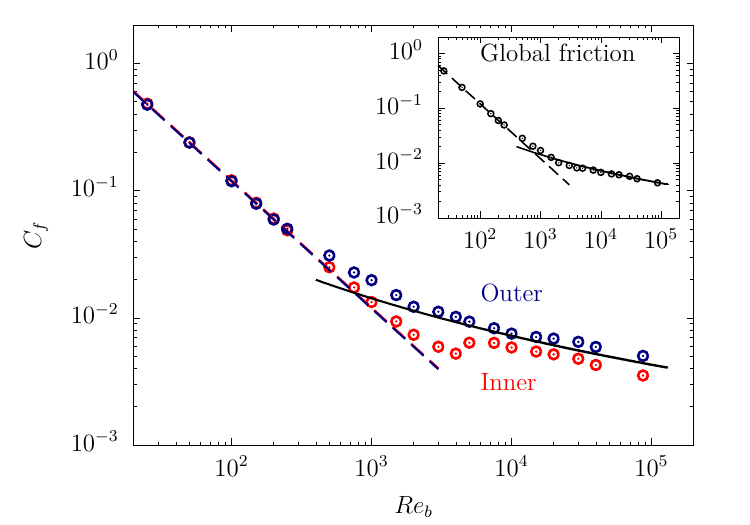}
(b)\includegraphics[width=.47\textwidth]{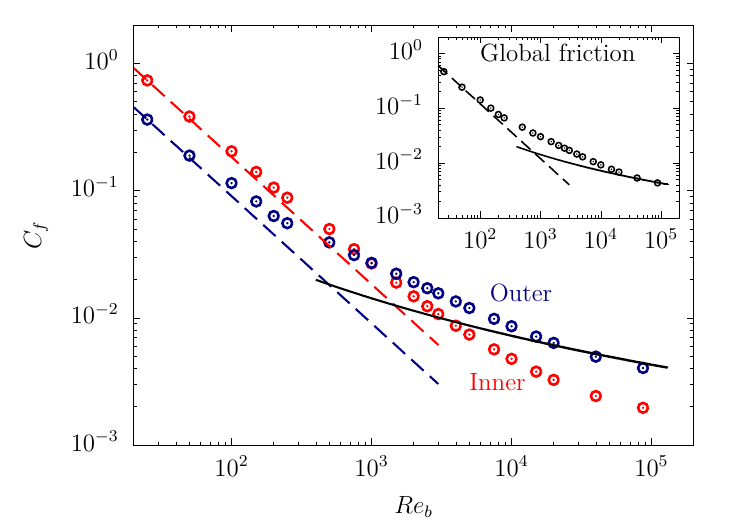}
\caption{Friction coefficient as a function of the bulk Reynolds number 
for the R40 flow cases (a) and R1 flow cases (b). 
Red circles denote the local friction coefficient at the inner wall 
($C_{f,i}$), blue circles at the outer wall ($C_{f,o}$), 
black circles denote in the insets the global friction coefficient ($C_{f,g}$);
dashed lines denote the analytical friction law for laminar flow 
(red for the convex wall, blue for the concave wall, black 
for the plane channel), black solid lines the 
logarithmic friction relation for plane channel flow~\eqref{eq:fric}. 
}
\label{fig:cf}
\end{figure}

The friction coefficient is a key parameter for characterising 
the transition from laminar to turbulent flow in curved channels.
According to the various definitions of friction velocity given in~\eqref{eq:utau}, 
we define local friction coefficients at the inner and 
at the outer wall, as well as a global friction coefficient, namely
\begin{equation}
C_{f,i}=2\left(\frac{u_{\tau,i}}{u_b}\right)^2,\quad 
C_{f,o}=2\left(\frac{u_{\tau,o}}{u_b}\right)^2,\quad
C_{f,g}=2\left(\frac{u_{\tau,g}}{u_b}\right)^2. 
\end{equation}
The global friction, based on $u_{\tau,g}$, 
is related to the mean-pressure gradient~\eqref{eq:utau_g},  
hence it is useful for comparing 
the power required to drive the flow in a curved channel 
versus a plane channel. 
In the inset of figure~\ref{fig:cf} we show the global friction 
coefficient (circles) as a function of the bulk Reynolds number, 
which can be compared with the friction coefficient of a plane channel 
in the laminar regime, 
$C_f=12/\Rey_b$ (dashed line), and in the turbulent regime 
via the logarithmic friction relation, 
\begin{equation}
\sqrt{\frac{2}{C_f}}=\frac{1}{k}\left[\ln\left(\frac{\Rey_b}{2}
	\sqrt{\frac{C_f}{2}}\right)-1\right]+A
\label{eq:fric}
\end{equation}
(solid line), obtained by~\cite{zanoun2009refined} from the logarithmic
law of the wall, where the log-law constants are $k=0.41$ and $A=5.17$.  
For the R40 flow cases (a) the global friction coefficient 
is nearly equivalent to that of a plane channel at the same 
Reynolds number both in the laminar and turbulent regime,  
whereas for the R1 flow cases (b) it is slightly higher, 
especially in the transitional regime ($500\le\Rey_b\le3000$).  
This indicates that frictional resistance in curved channel flows is only 
moderately higher than in plane channel flow, consistent with previous 
experimental~\citep{wattendorf1935study} and 
computational~\citep{brethouwer2022turbulent} results. 
 
\begin{figure}
(a)\includegraphics[width=.47\textwidth]{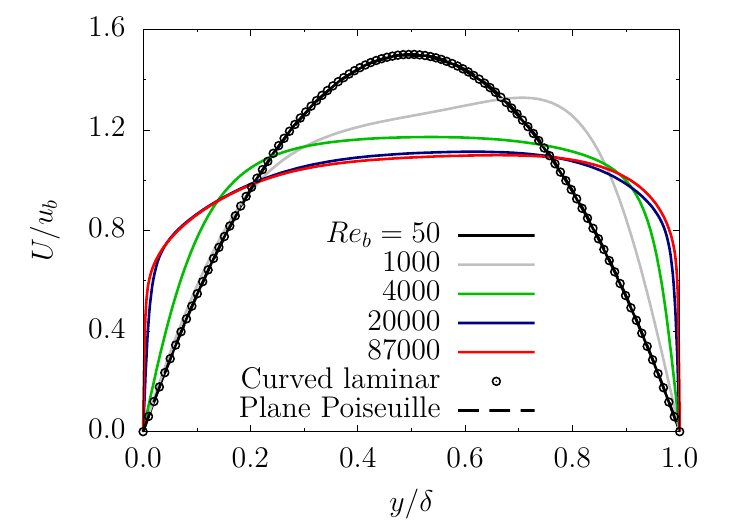}
(b)\includegraphics[width=.47\textwidth]{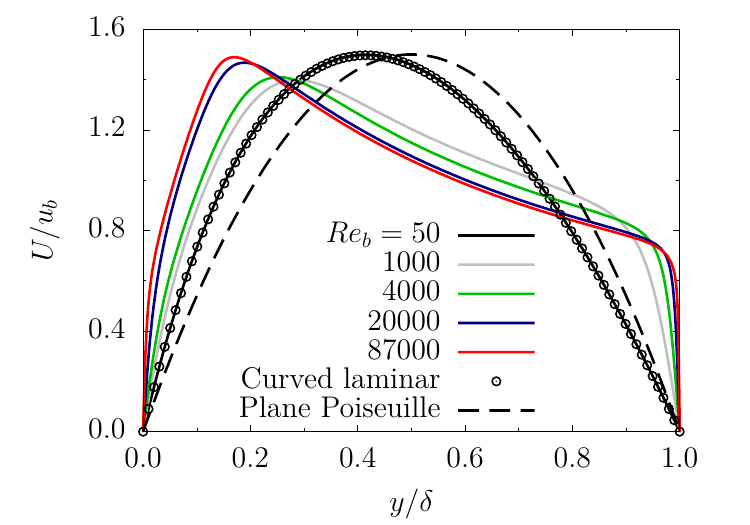}
\caption{Mean streamwise velocity ($U/u_b$) 
at various Reynolds numbers for the R40 flow cases (a) and R1 flow cases (b). 
Black dashed lines refer to the Poiseuille profile, black circles 
to the analytical profile of laminar curved channel flow~\eqref{eq:ulam}.}
\label{fig:umean}
\end{figure}

In figure~\ref{fig:cf} we show the local friction coefficient 
at the two walls as a function of the bulk Reynolds number. 
In the turbulent regime, friction at the outer wall is always 
higher than at the inner wall, due to increased turbulence intensity. 
For the R40 flow cases (a), the friction coefficient of a plane channel 
at the same Reynolds number (black solid line) is about halfway 
between the values at the two walls of the curved channel.  
For the R1 flow cases (b), the outer-wall friction matches 
that of an equivalent plane channel, 
and the inner-wall friction is significantly lower.
To comment more clearly on the friction trend, especially  
in the laminar and transitional regime, 
it is useful to consider the mean velocity profiles
at representative Reynolds numbers, as reported in figure~\ref{fig:umean}.
At sufficiently low Reynolds number, 
the flow in a curved channel is laminar and the velocity profile 
can be derived analytically as  
\begin{equation}
U(r)=\alpha r-\frac{\beta}{r}-r\ln r,
\label{eq:ulam}
\end{equation}
where 
\begin{equation}
\alpha=\frac{r_o^2\ln r_o - r_i^2\ln r_i}{r_o^2-r_i^2},\quad
\beta=\frac{r_i^2 r_o^2 \ln(r_o/r_i)}{r_o^2-r_i^2}.
\end{equation}
%
For the R40 flow cases, the laminar velocity profile is nearly identical 
to the parabolic profile of plane channel flow, 
meaning that the shear stress distribution is almost symmetrical. 
As the curvature increases, the pressure gradient in the radial 
direction becomes stronger
and the location of maximum velocity shifts towards the inner wall.
As a result, the shear stress at the inner wall is greater than at the outer.
In the laminar regime, the friction coefficient at the two walls 
can be determined analytically 
by differentiating equation~\eqref{eq:ulam}, which yields
\begin{equation}
\dt{U(r)}{r}=\alpha+\frac{\beta}{r^2}-\ln r-1.
\label{eq:dudr}
\end{equation}
For the R40 flow cases, we find $C_{f,i} \approx 12.03/\Rey_b$, hence nearly identical 
to plane Poiseuille flow, and $C_{f,o} \approx 11.83/\Rey_b$.
In contrast, for the R1 flow cases the friction coefficient at the inner wall, 
$C_{f,i} \approx 18.34/\Rey_b$, is about twice that at the outer wall, 
$C_{f,o} \approx 9.04/\Rey_b$. 
%
\begin{figure}
(a)\includegraphics[width=.47\textwidth]{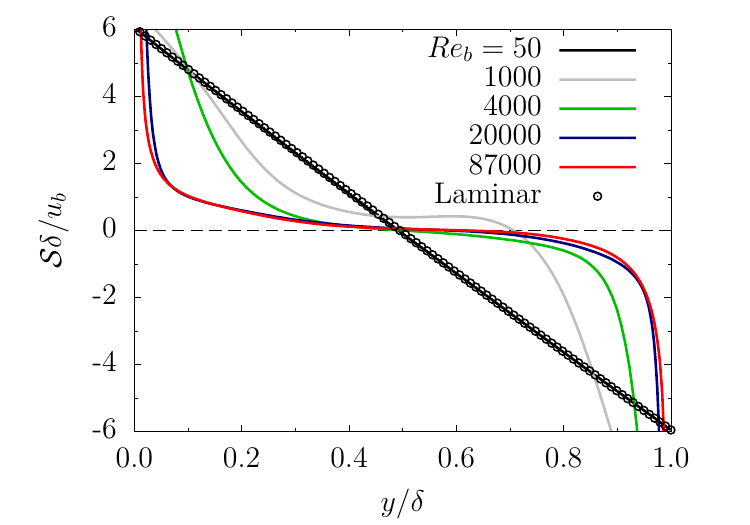}
(b)\includegraphics[width=.47\textwidth]{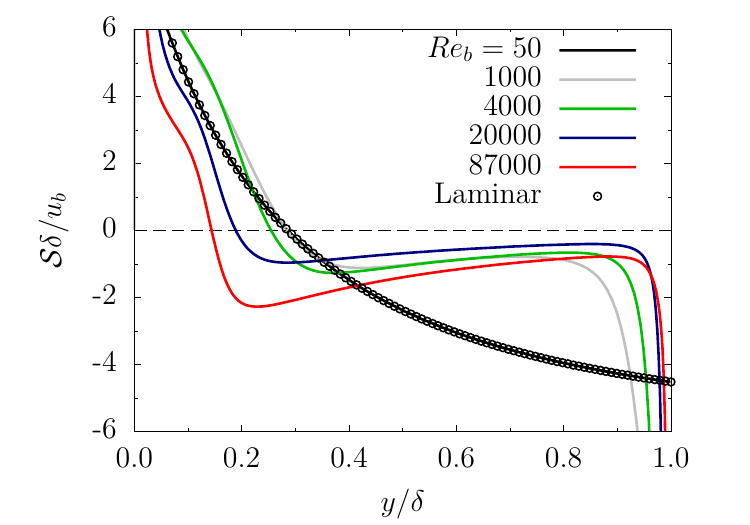}
\caption{Mean shear rate ($\mathcal{S}\delta/u_b$) 
at various Reynolds numbers 
for the R40 flow cases~(a) and R1 flow cases~(b).  
Circles denote the analytical profile for the laminar case.}
\label{fig:shear}
\end{figure}
Flow transition occurs when the Reynolds number reaches the critical value, 
which we define as the point where the friction coefficient deviates from 
the laminar trend by at least $1\%$. For the R40 flow cases  
the critical Reynolds number is $\Rey_b\approx250$, which is in good agreement 
with the value of $228.5$ found by~\cite{finlay1988instability}, whereas 
for the R1 flow cases transition starts earlier at $\Rey_b\approx50$. 
The transitional regime is greatly affected by curvature, 
depending both on its radius and on its type,
with convex curvature (inner wall) tending to stabilise the flow, 
and concave curvature (outer wall) having the opposite effect.
This is particularly evident in the R40 flow cases: at $\Rey_b=1000$, the flow 
near the inner wall is laminar, as visible from the mean velocity 
profile in figure~\ref{fig:umean}(a). In contrast, near the outer wall  
secondary motions are generated, which result in stronger momentum exchange. 
The peak velocity then shifts towards the outer wall, resulting in higher friction. 
At $\Rey_b=4000$, the flow near the outer wall is fully turbulent 
(as will be shown later through flow visualisations), whereas
the velocity profile near the inner wall is still close to the laminar state. 
Hence, mild convex curvature tends to delay transition compared to 
the plane channel, in which configuration the flow at $\Rey_b\approx2600$ 
is close to fully turbulent~\citep{yimprasert2021flow}. 
By increasing the Reynolds number to $5000$, the flow near the inner wall 
undergoes abrupt transition to a fully turbulent state, as indicated 
by the jump in the 
friction trend indicated by red circles 
in figure~\ref{fig:cf}(a). 
This type of transition from laminar to turbulent regime 
is typical of canonical wall flows \citep[see e.g.][]{patel1969some}, 
yet it is profoundly different from that at the outer wall, 
where transition is facilitated by centrifugal instabilities 
and occurs more smoothly. 

For the R1 flow cases, friction in the laminar regime is higher at the inner wall.
By increasing the Reynolds number, the velocity profiles tend to flatten 
at the outer wall, where friction increases. Turbulence is inhibited near 
the inner wall, and friction deviates less markedly from the laminar trend. 
An inversion occurs past $\Rey_b \approx 1000$, 
whereby friction at the outer wall becomes higher that at the inner wall.
Flow transition is facilitated by strong channel curvature, 
and it occurs smoothly at both walls. At the outer wall, 
the different transition from canonical wall flows is due to centrifugal 
(primary) instabilities. 
As we will show, in the R1 flow cases secondary motions associated with
centrifugal instabilities do not reach the inner wall,
and transition is affected by streamwise (secondary) instabilities, 
which lead to the formation of large-scale cross-stream structures. 
 
In figure~\ref{fig:shear} we show the mean shear rate, 
$\mathcal{S}=\mathrm{d} U/\mathrm{d} r-U/r$. Similar to the case of the plane 
channel, the location where the mean shear vanishes 
defines a bound for the regions of influence of the two walls.
For the R40 flow cases (a) the shear rate profile is nearly symmetrical, hence
this location coincides roughly with the channel centreline, 
with exception of the case $\Rey_b=1000$, for which it occurs
at $y/\delta\approx0.71$. 
In contrast, for the R1 flow cases (b) the shear rate profile is strongly asymmetrical, 
and the point where $\mathcal{S}=0$ shifts towards the inner wall, 
between $y/\delta\approx0.22$ and $y/\delta\approx0.14$, 
moving closer to the inner wall for higher Reynolds numbers. 
Hence, the region of influence of the inner wall
is much narrower than for the outer wall.
The shear rate profile also exhibits a local maximum 
near the outer wall and a local minimum near the inner wall, 
corresponding to inflectional points of the mean velocity profiles.
The local minimum near the inner wall is especially pronounced, 
showcasing the presence of a shear layer 
that can support inflectional instabilities. 

%
%
%
\subsection{Flow visualisations}\label{sec:visua}

\begin{figure}
\centering
(a)\includegraphics[width=.47\textwidth]{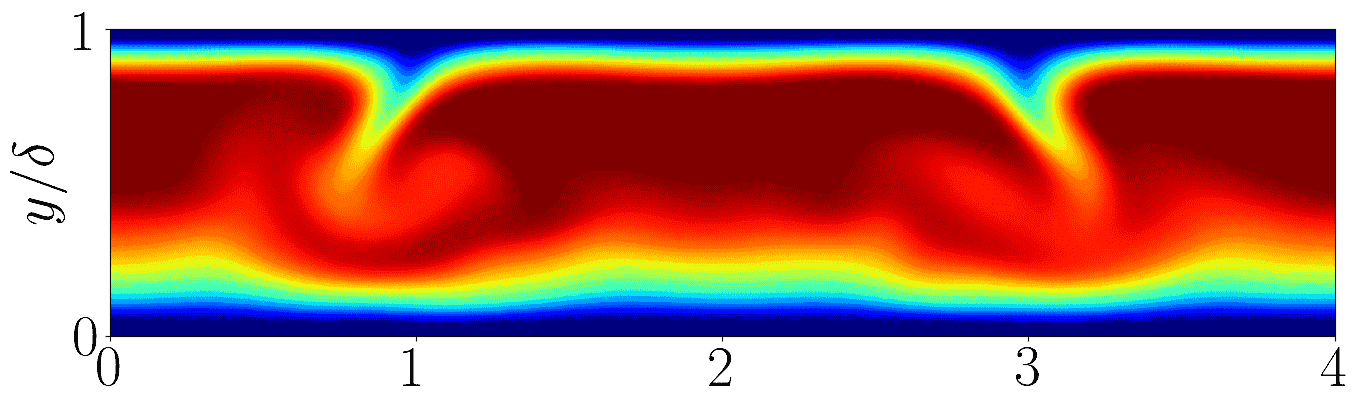}
(b)\includegraphics[width=.47\textwidth]{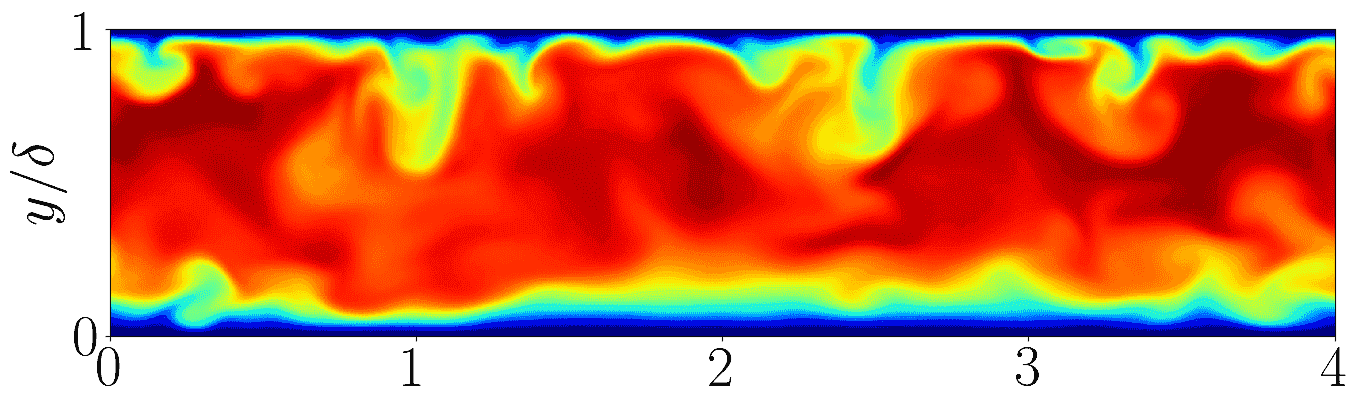}
(c)\includegraphics[width=.47\textwidth]{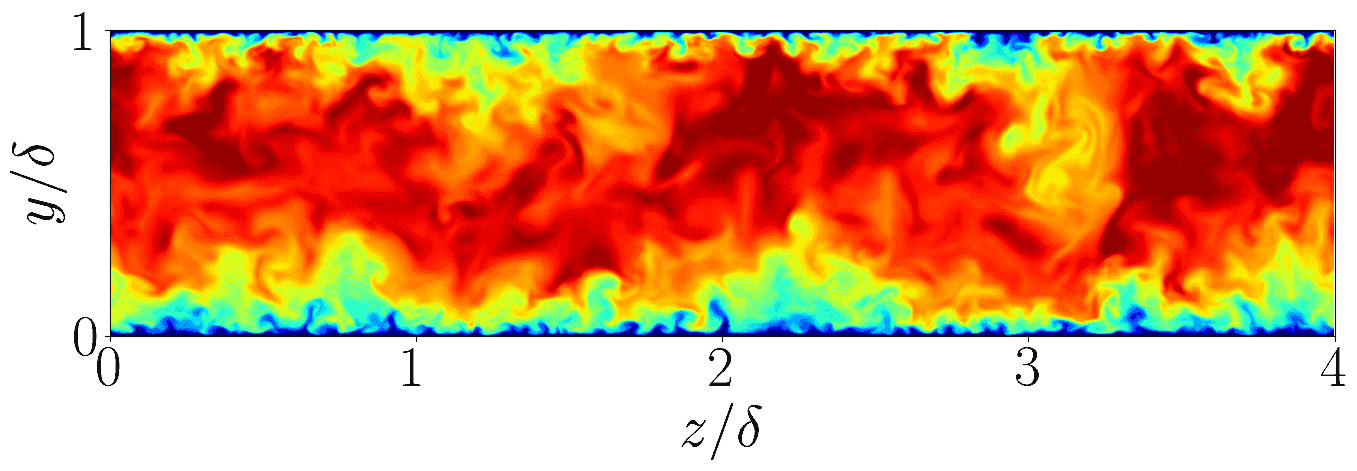}
(d)\includegraphics[width=.47\textwidth]{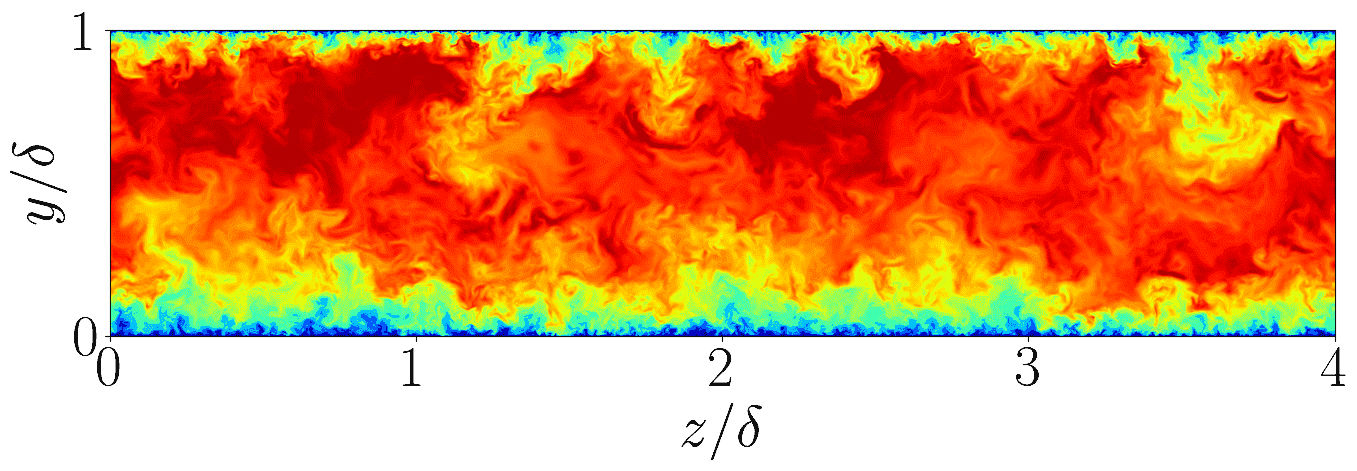}
\caption{Instantaneous streamwise velocity fields 
in a cross-stream plane for the R40 flow cases 
at $\Rey_b=1000$~(a), $4000$~(b), $20000$~(c), $87000$~(d).} 
\label{fig:utheta40}
\end{figure}

\begin{figure}
(a)\includegraphics[width=.48\textwidth]{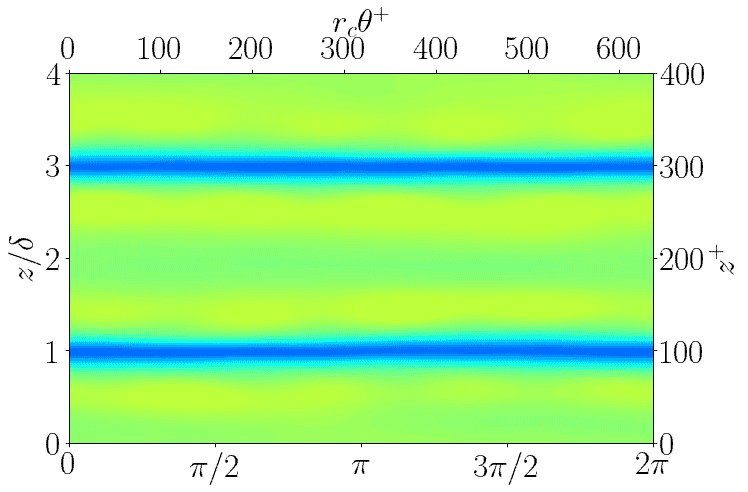}
(b)\includegraphics[width=.48\textwidth]{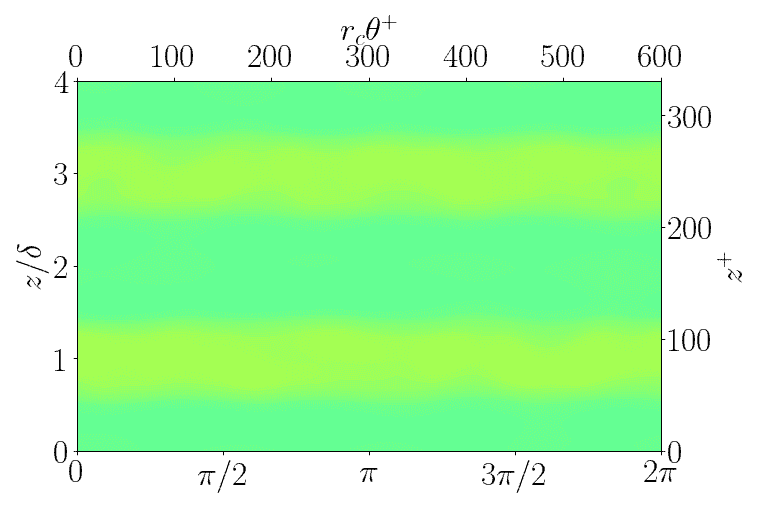}
(c)\includegraphics[width=.48\textwidth]{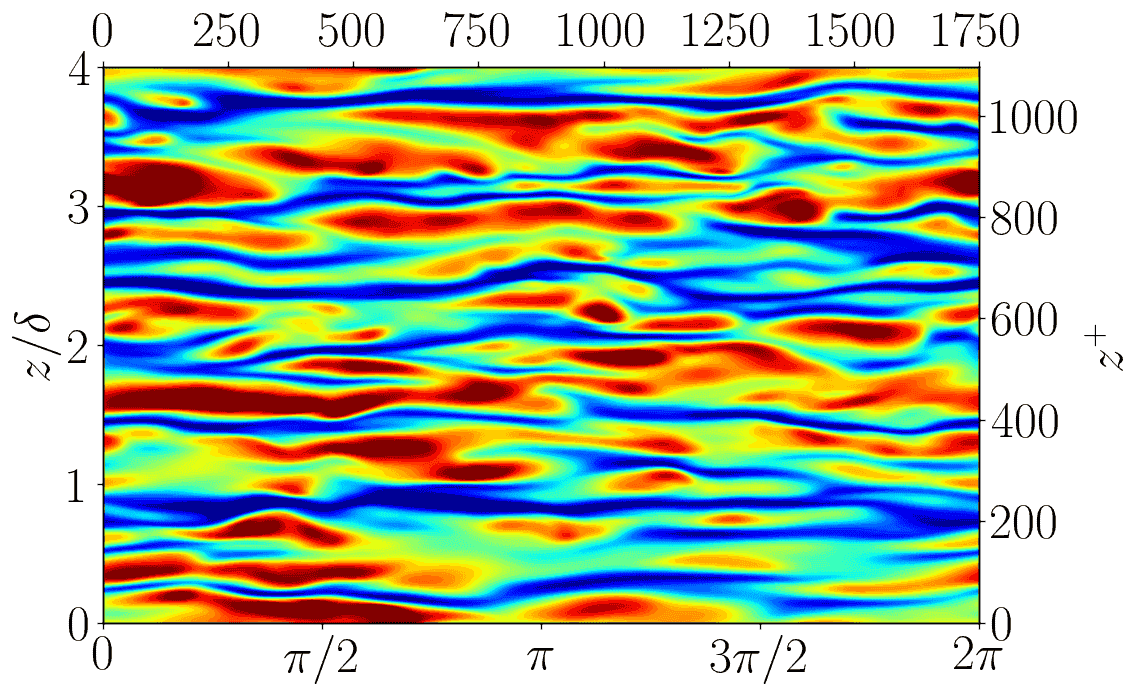}
(d)\includegraphics[width=.48\textwidth]{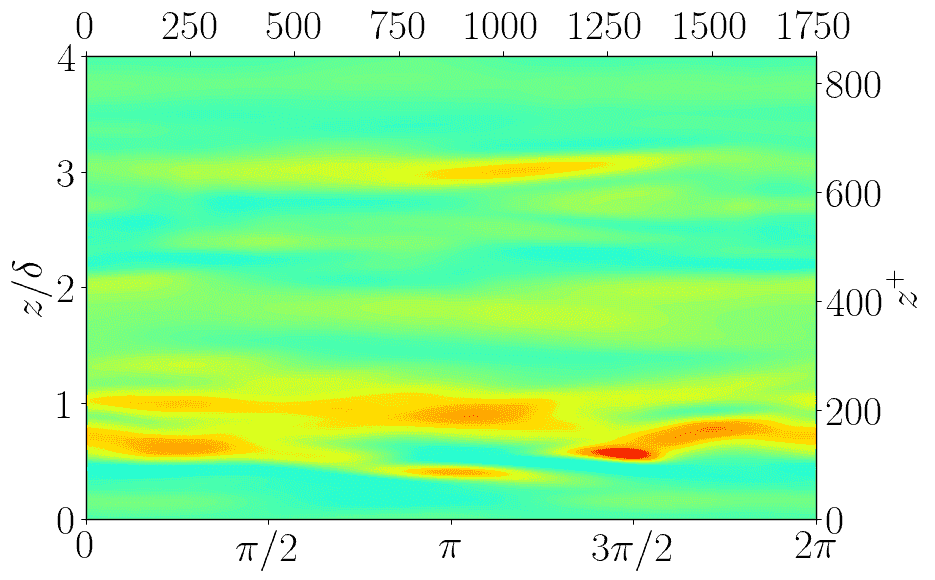}
(e)\includegraphics[width=.48\textwidth]{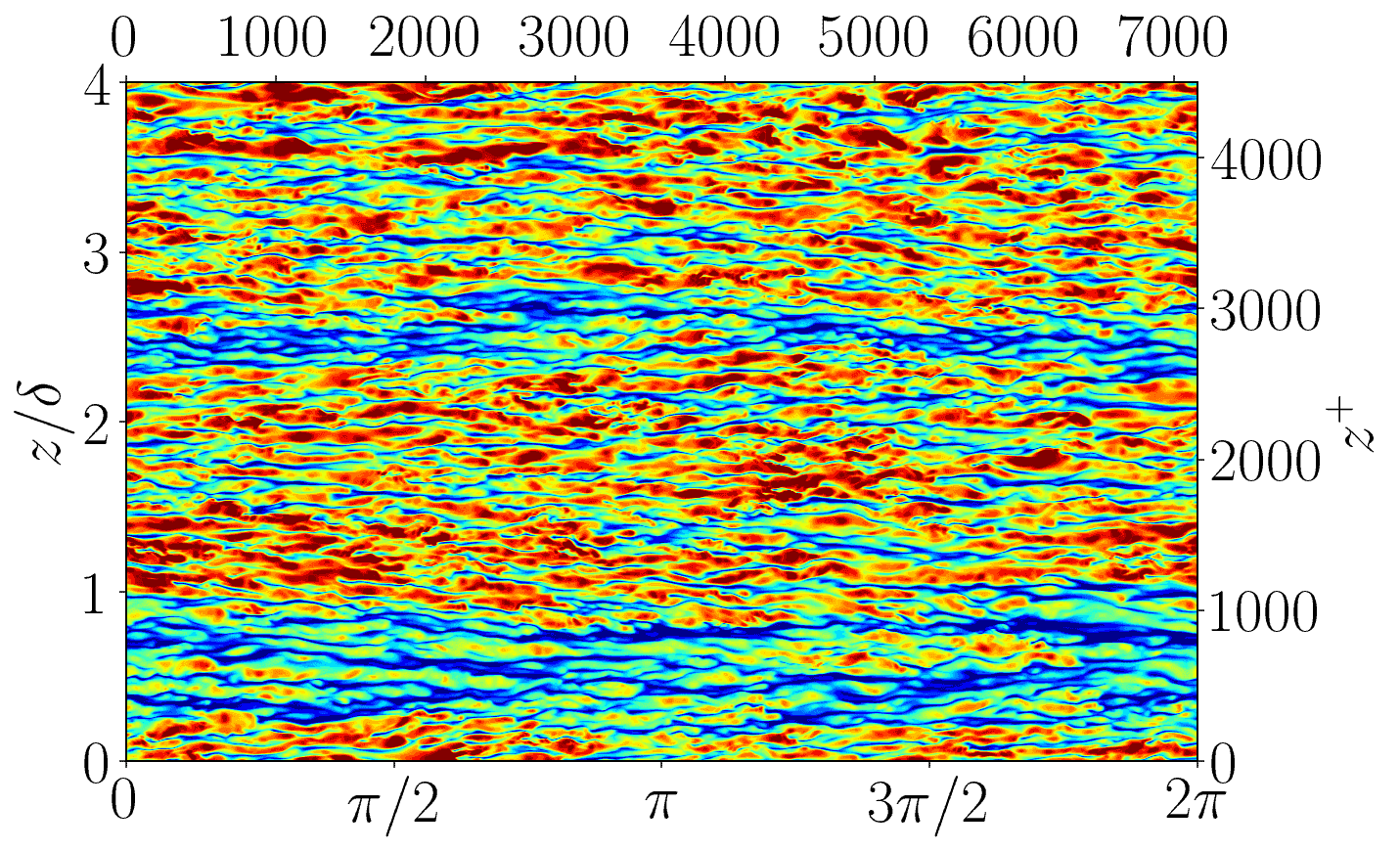}\
(f)\includegraphics[width=.48\textwidth]{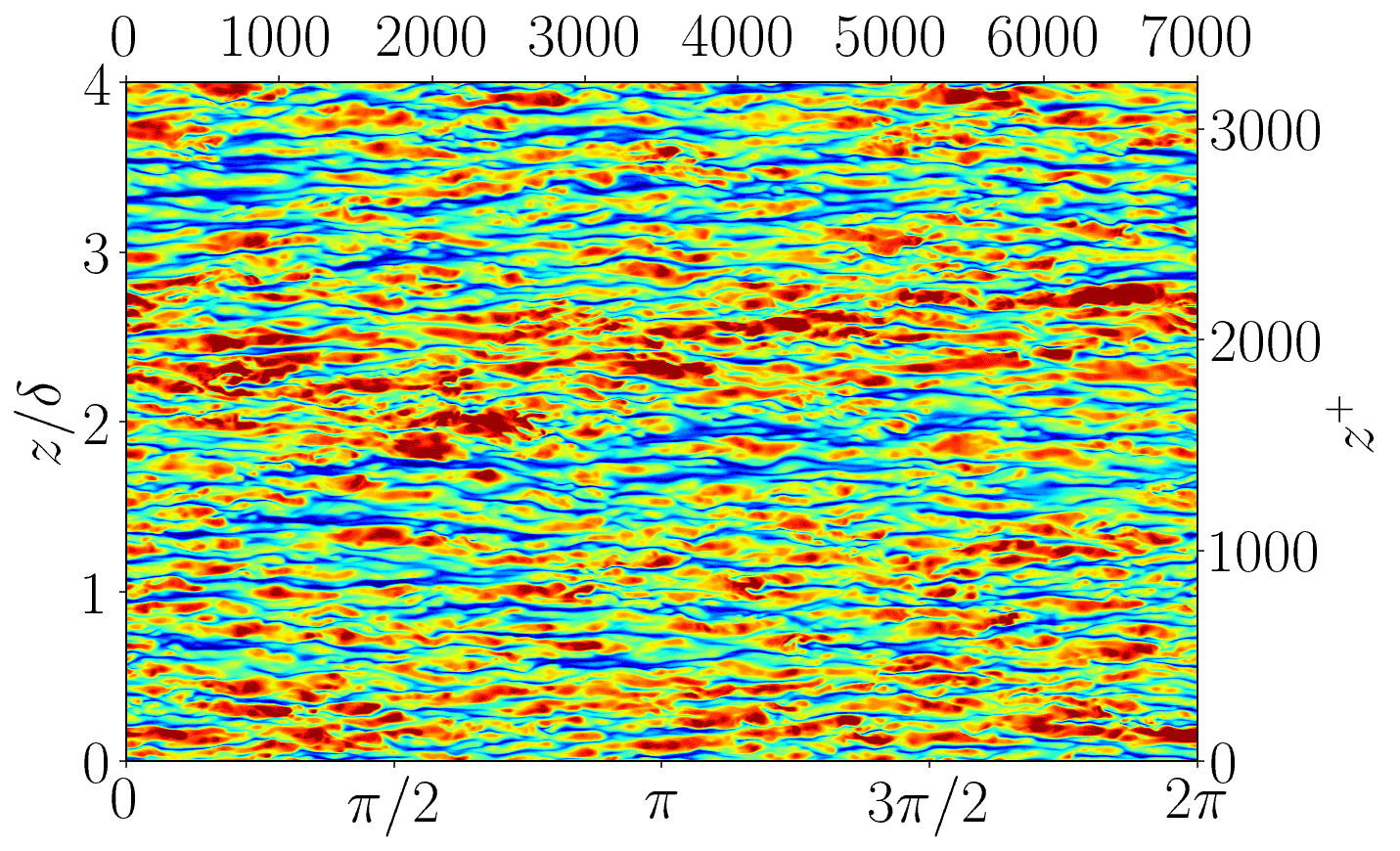}
(g)\includegraphics[width=.48\textwidth]{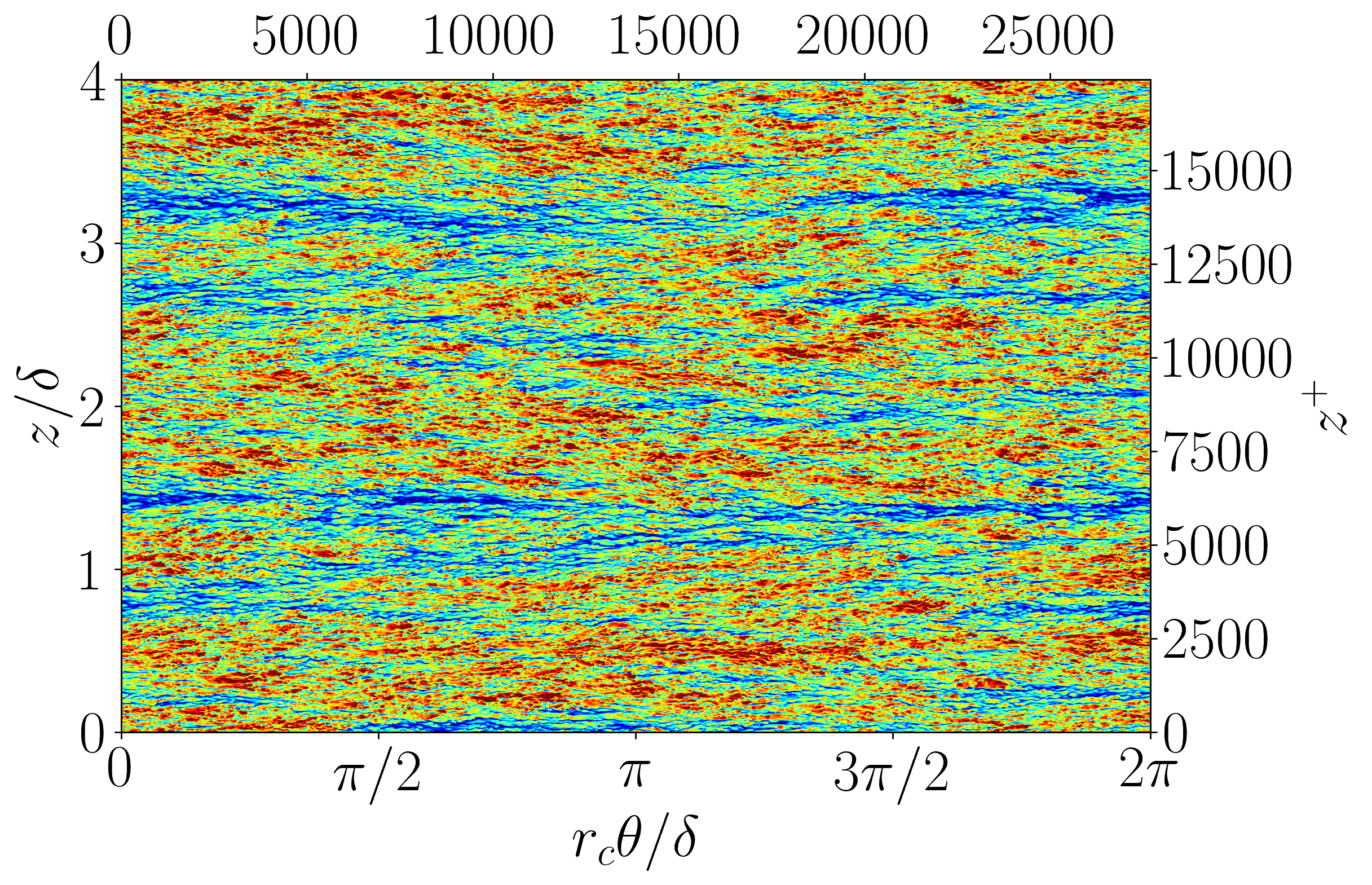}
(h)\includegraphics[width=.48\textwidth]{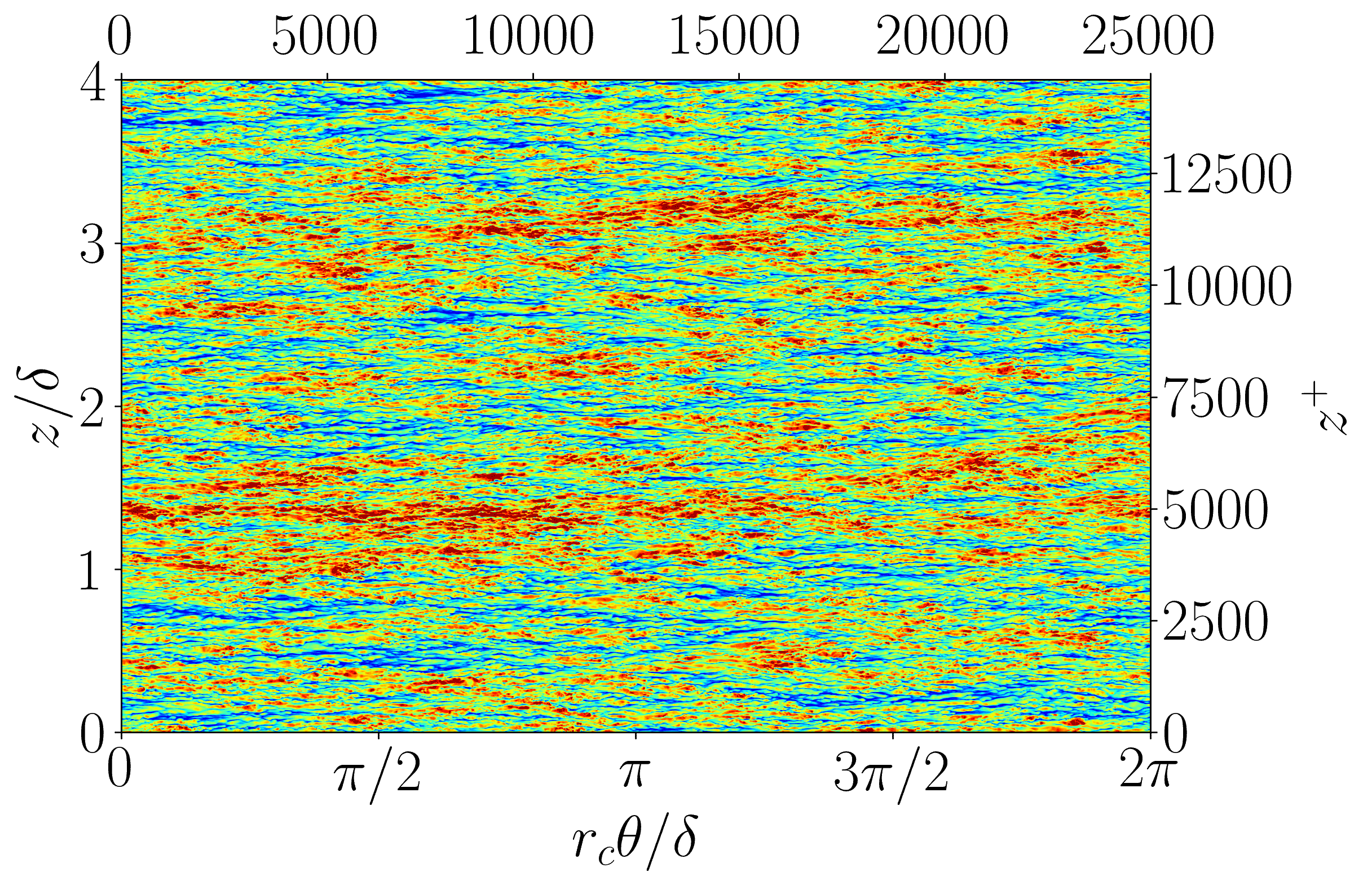}
\caption{Instantaneous fields of streamwise velocity fluctuations 
for the R40 flow cases in wall-parallel planes near the outer (left panels) and 
inner wall (right panels) at $y^+\approx12$.
From top to bottom, the panels correspond to 
$\Rey_b=1000$~(a, b), $4000$~(c, d), $20000$~(e, f), $87000$~(g, h). 
Streamwise and spanwise coordinates are shown in both outer units 
($r_c\theta/\delta$, $z/\delta$), and local wall units ($r_c\theta^+$, $z^+$).
Mean flow goes from left to right.}
\label{fig:ufluc40}
\end{figure}

\begin{figure}
\centering
(a)\includegraphics[width=.47\textwidth]{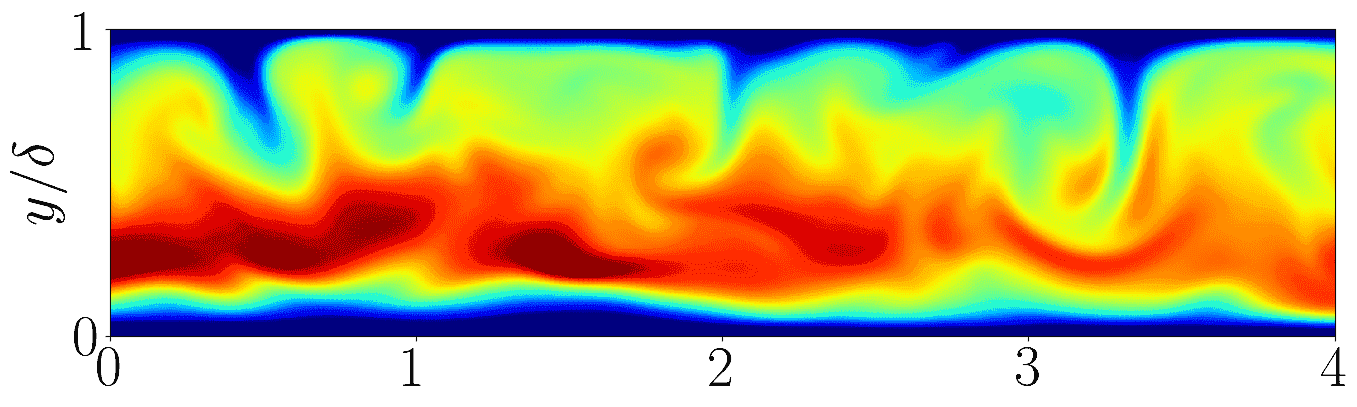}
(b)\includegraphics[width=.47\textwidth]{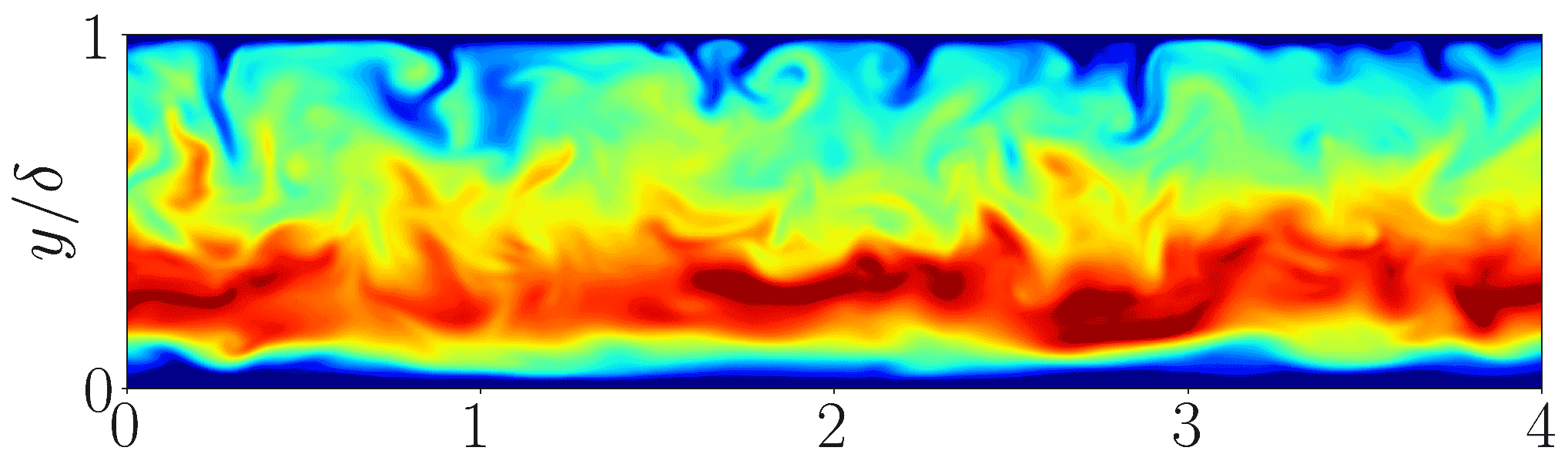}
(c)\includegraphics[width=.47\textwidth]{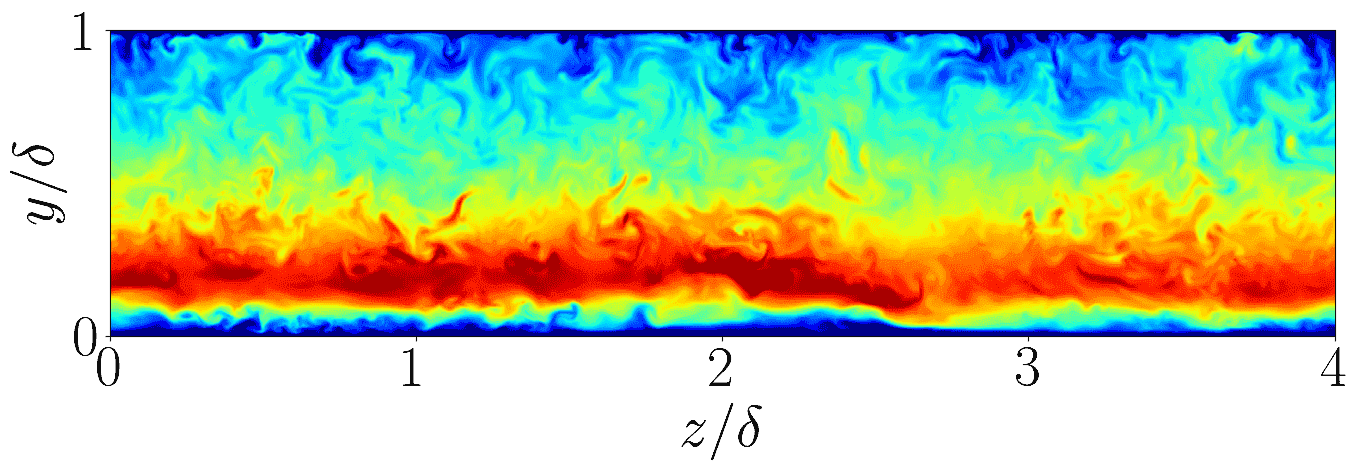}
(d)\includegraphics[width=.47\textwidth]{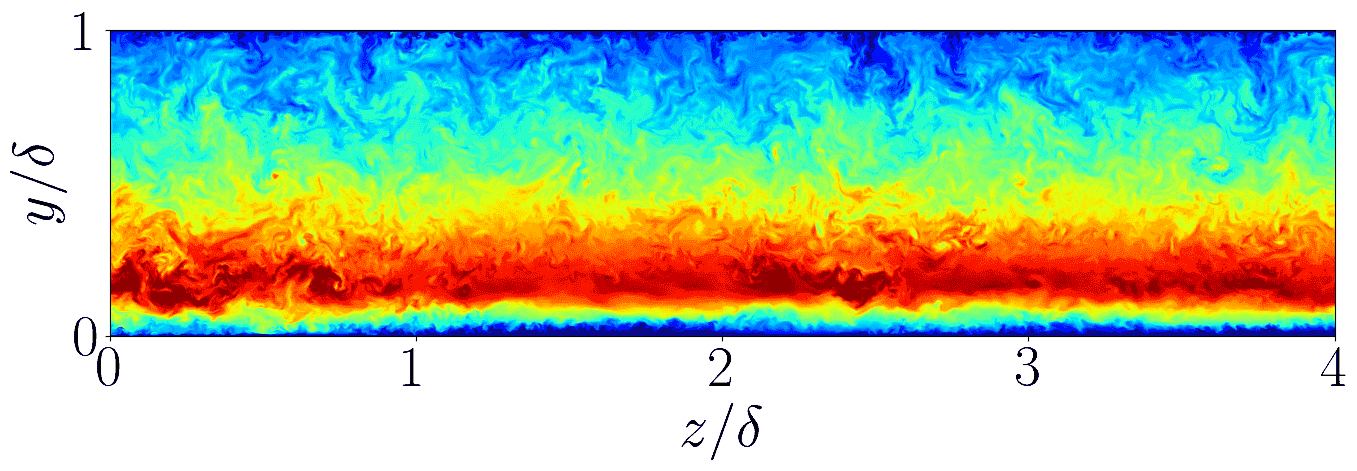}
\caption{Instantaneous streamwise velocity fields 
in a cross-stream plane for the R1 flow cases 
at $\Rey_b=1000$~(a), $4000$~(b), $20000$~(c), $87000$~(d). 
Only half of the domain is shown.} 
\label{fig:utheta1}
\end{figure}

\begin{figure}
(a)\includegraphics[width=.48\textwidth]{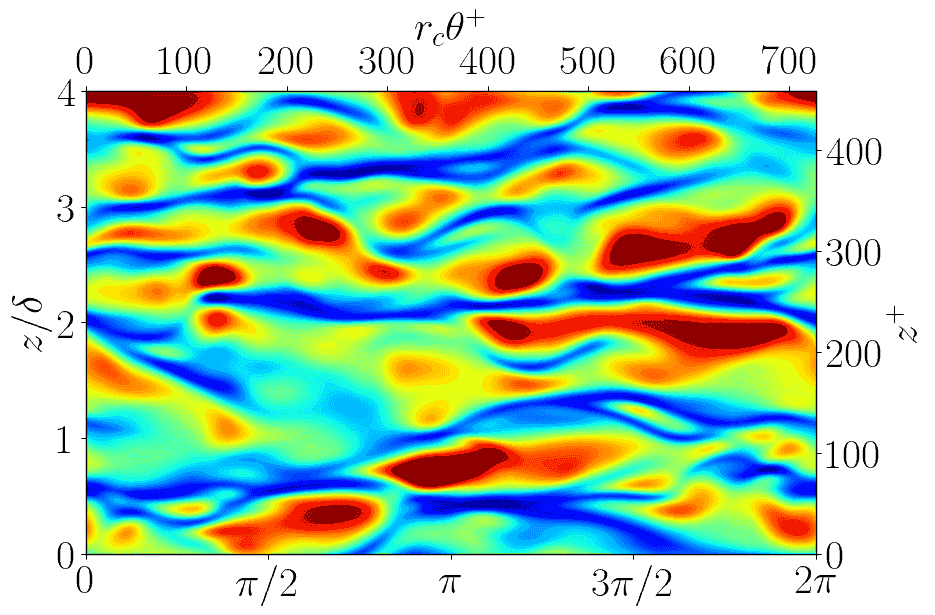}
(b)\includegraphics[width=.48\textwidth]{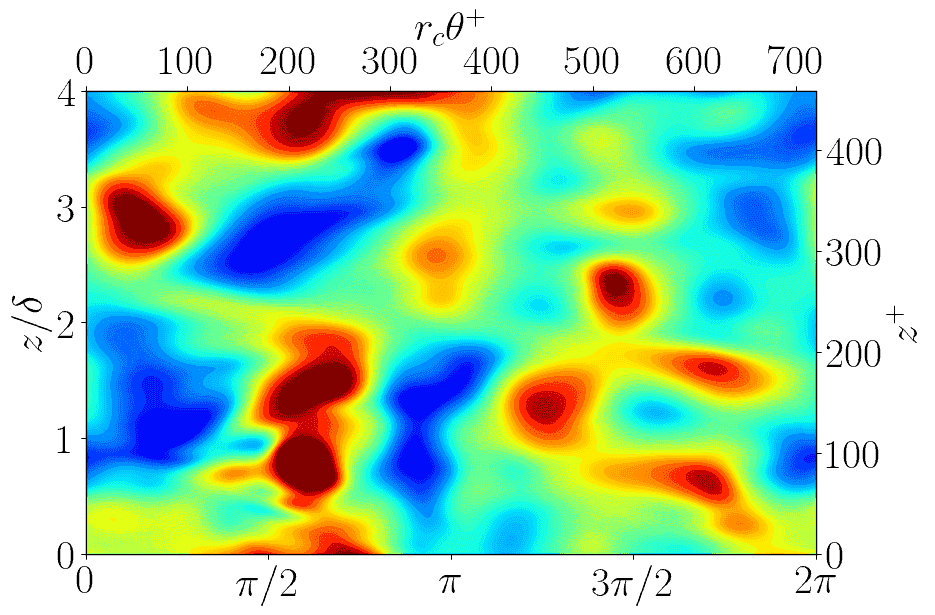}
(c)\includegraphics[width=.48\textwidth]{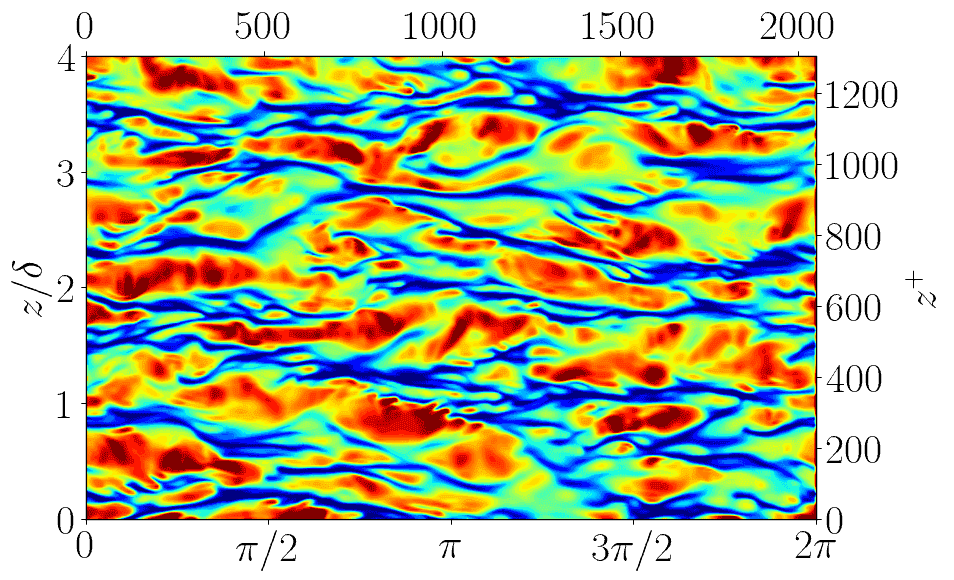}
(d)\includegraphics[width=.48\textwidth]{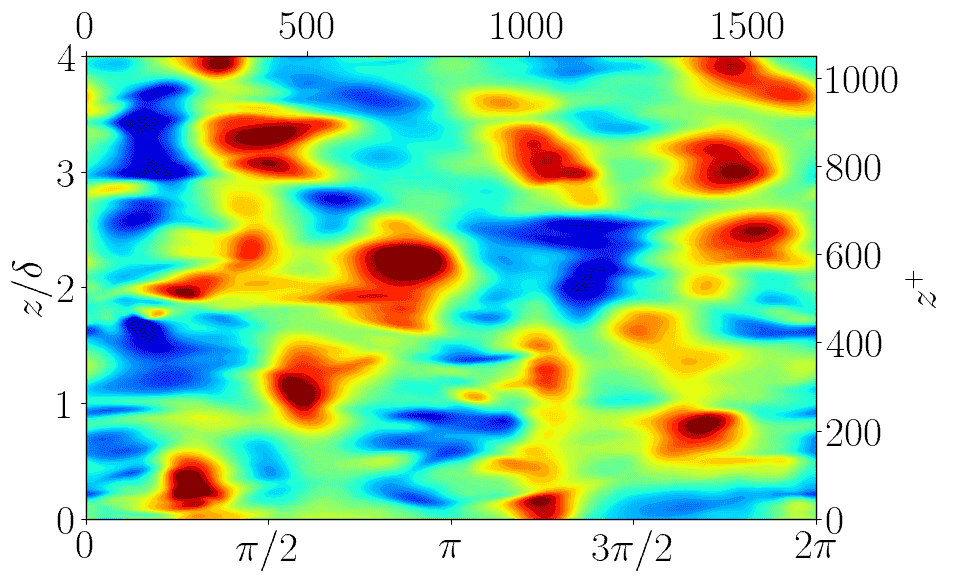}
(e)\includegraphics[width=.48\textwidth]{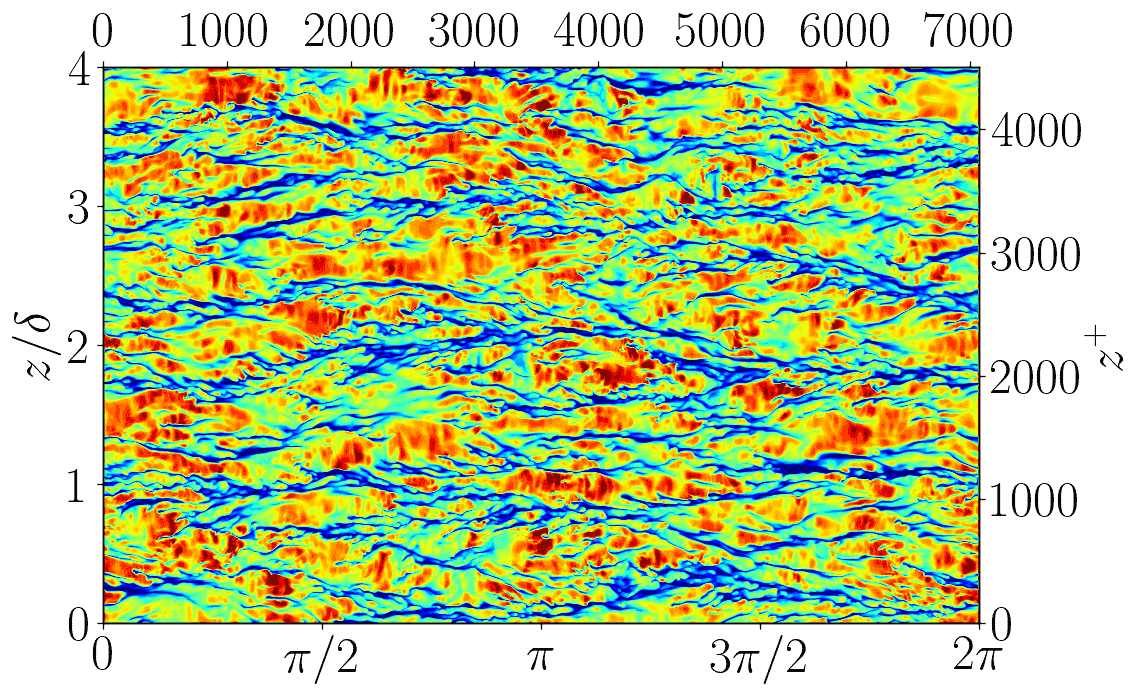}
(f)\includegraphics[width=.48\textwidth]{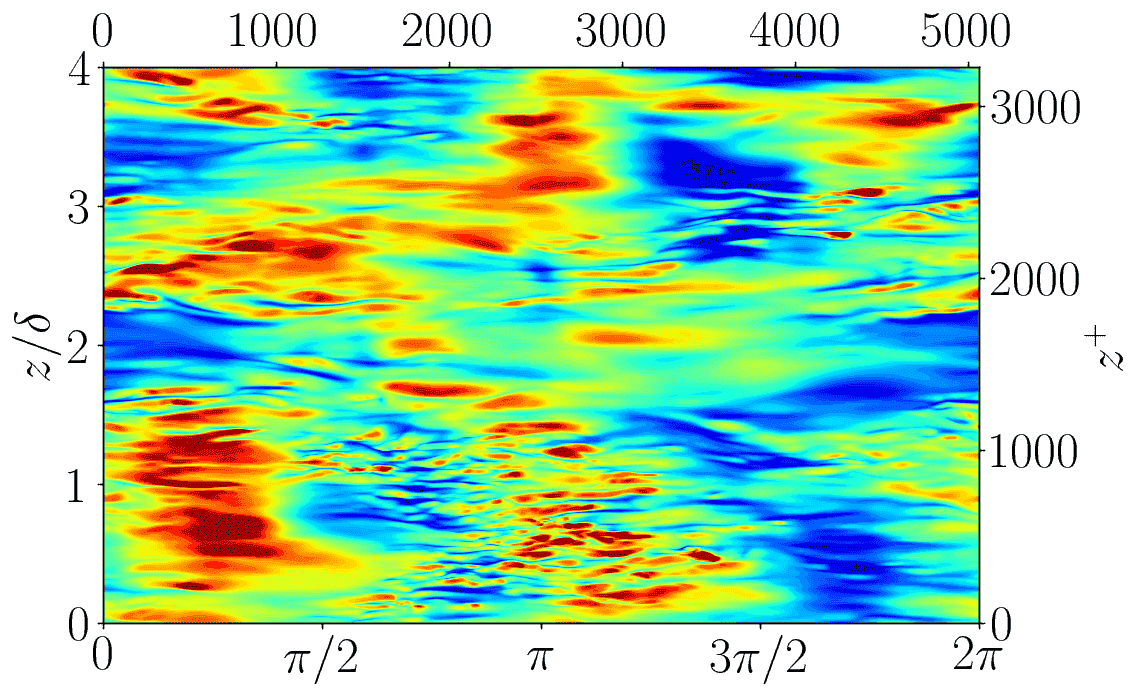}
(g)\includegraphics[width=.48\textwidth]{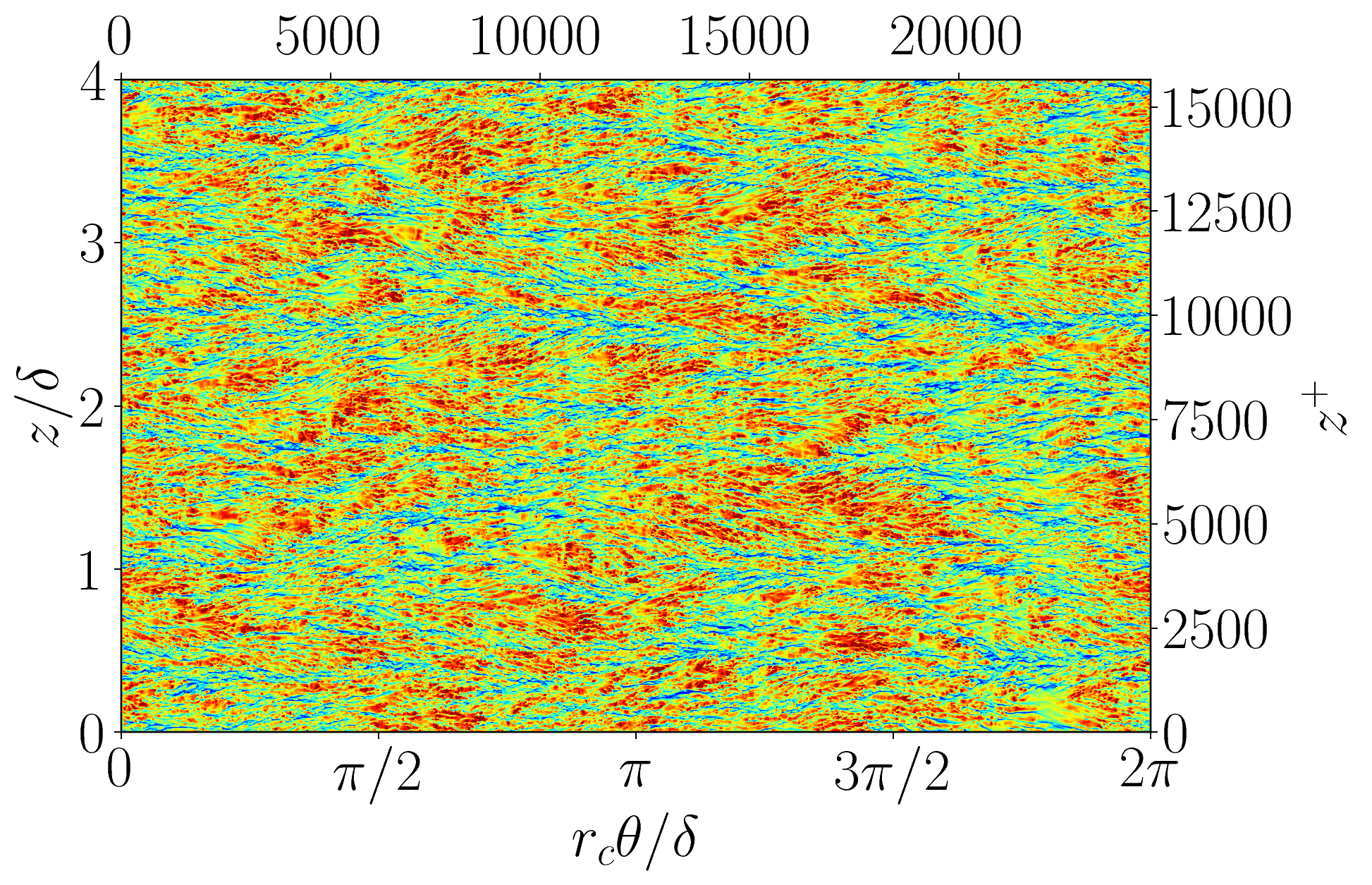}
(h)\includegraphics[width=.48\textwidth]{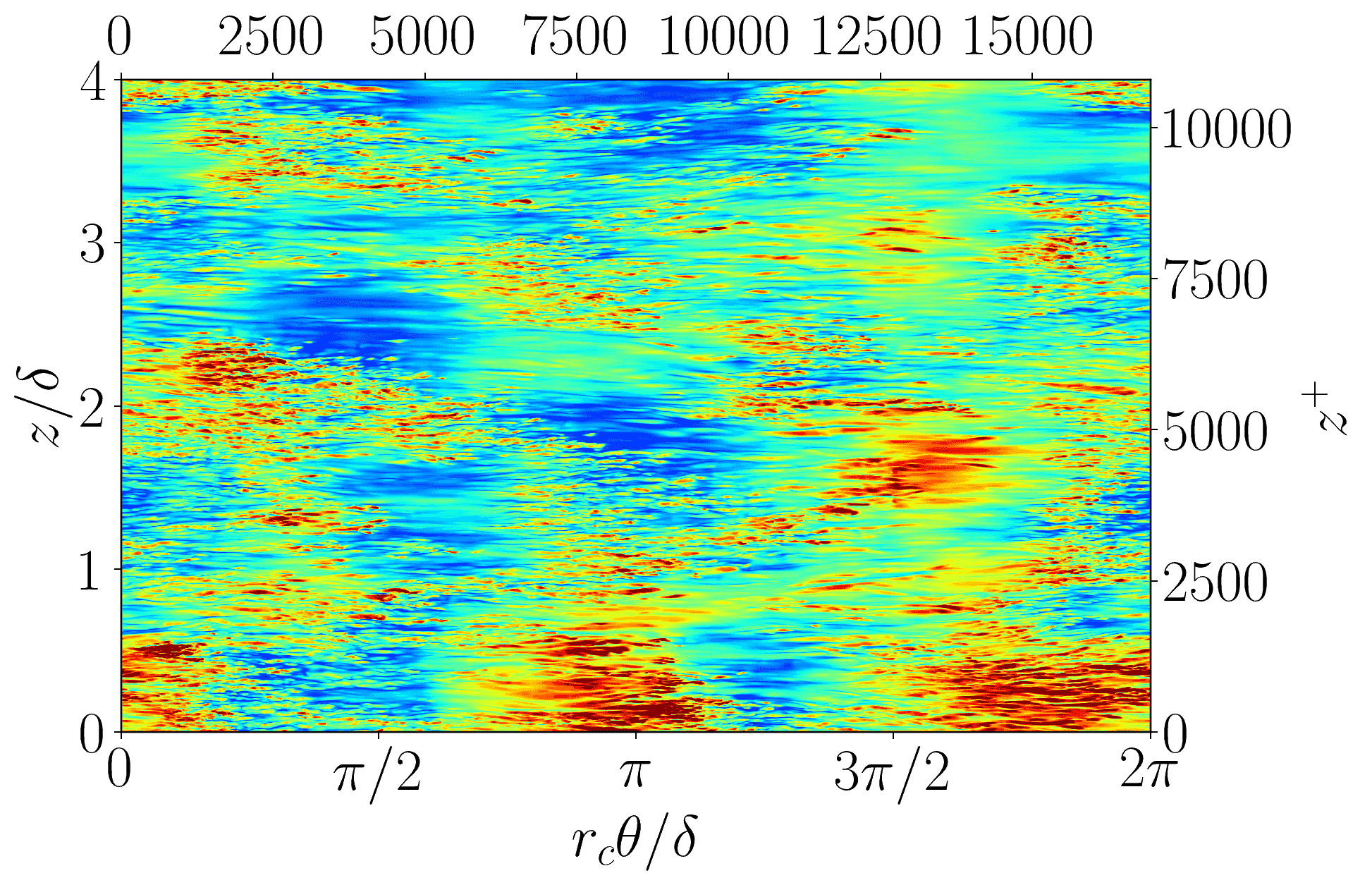}
\caption{Instantaneous fields of streamwise velocity fluctuations  
for the R1 flow cases in wall-parallel planes near the outer (left panels) 
and inner wall (right panels) at $y^+\approx 12$.
From top to bottom, the panels correspond to 
$\Rey_b=1000$~(a, b), $4000$~(c, d), $20000$~(e, f), $87000$~(g, h). 
Streamwise and spanwise coordinates are shown in both outer units 
($r_c\theta/\delta$, $z/\delta$), and local wall units ($r_c\theta^+$, $z^+$).
Only half of the domain is shown.}
\label{fig:ufluc1}
\end{figure}

The previous discussion on flow transition can be visualised
through the instantaneous fields of streamwise velocity in 
cross-stream and wall-parallel planes, shown in figures~\ref{fig:utheta40} 
and~\ref{fig:ufluc40} for the R40 flow cases and in 
figures~\ref{fig:utheta1} and~\ref{fig:ufluc1} for R1 flow cases. 

%
%
As for the R40 flow cases, the cross-stream and longitudinal planes 
at $\Rey_b=1000$, shown in figures~\ref{fig:utheta40}(a) 
and~\ref{fig:ufluc40}(a), highlight the onset of two symmetric 
large-scale ejections carrying low-speed fluid from the outer wall 
towards the channel core. These secondary motions are generated 
by two pairs of counter-rotating roll cells, which resemble closely 
Dean vortices. The inner-wall flow region is almost unaffected by these vortices, 
as one can see from figure~\ref{fig:ufluc40}(b). 
Figures~\ref{fig:utheta40}(b) and~\ref{fig:ufluc40}(c), depicting the 
cross-stream and longitudinal planes at $\Rey_b=4000$, show that fine-scale 
ejections and organised structures of momentum streaks characterise the 
outer wall, from which one can infer that the flow is fully turbulent. 
Traces of the longitudinal vortices appear as longitudinal bands of high-speed 
fluid near the inner wall, as visible in figure~\ref{fig:ufluc40}(d).  
Being centrifugally stable, the inner-wall flow region is still transitional, 
as no clear signs of turbulent activity appear. 
The cross-stream planes at $\Rey_b=20000$ and $87000$ (panels (c) and (d) 
of figure~\ref{fig:utheta40}) show clear signs of streaky patterns at both walls. 
Overlaying the small scales of turbulence, longitudinal large-scale structures 
are found near the outer wall, whose footprint appears in the wall-parallel 
planes of figures~\ref{fig:ufluc40}(e) and~\ref{fig:ufluc40}(g) 
as wide regions of alternating high and low speed fluid. 
These large-scale regions fill the entire 
streamwise extent of the domain, and their spanwise width is comparable with 
the channel height, hence two pairs of roll cells are present. 
The influence of the latter is found to extend to the inner wall 
at high Reynolds numbers. In fact, low-speed regions at the outer wall, 
corresponding to large-scale ejections, are found at the same spanwise 
location as high-speed regions at the inner wall, corresponding to large-scale 
sweeps (panels (f) and (h) of figure~\ref{fig:ufluc40}). 

%
Cross-stream visualisations (figure \ref{fig:utheta1}) show clearly how 
the region with higher momentum shifts towards the inner wall 
for the R1 flow cases, as expected from the 
mean velocity profiles (figure~\ref{fig:umean}). 
As well as for the R40 flow cases, signatures of longitudinal roll cells 
are visible as large-scale ejections from the outer wall, 
although those are more numerous and chaotic. 
As displayed in the wall-parallel plane of figure~\ref{fig:ufluc1}(a), 
streaky structures appear to be quite distinct at the outer wall 
at $\Rey_b=1000$, confirming that increasing 
curvature promotes the transition to turbulence.
Panels (c), (e), (g) of figure~\ref{fig:ufluc1} show that 
momentum streaks are less regular at the outer wall and form ripples along 
the streamwise direction. 
Furthermore, streaks are wider for the R1 flow cases than for R40, 
whereas the longitudinal large-scale structures are smaller and more chaotic. 
This makes the distinction between fine-scale turbulence and large-scale 
structures less clear-cut. In fact, longitudinal roll cells do not clearly 
show up in the wall-parallel planes of fluctuating streamwise velocity, 
however this does not convey that they are absent,
as secondary motions due to longitudinal vortices mainly affect
the wall-normal and spanwise velocity components.
As visible from the wall-parallel planes shown in panels (b), (d), (f) 
of figure~\ref{fig:ufluc1}, the inner-wall flow region is devoid of turbulent 
structures, or nearly so for the highest Reynolds number (h), 
whereas it is characterised by the presence of large-scale structures 
stretched in the spanwise direction. The latter are visible in the form 
of alternating regions of high and low speed fluid spanning the transverse 
direction with spanwise width comparable with $\delta$. 
The streamwise extent of each pair of transverse 
structures is about $\pi \delta$, meaning that the computational 
domain can fit two pairs, with exception of the case $\Rey_b=4000$ 
at which the streamwise wavelength seems slightly shorter.
  
\subsection{Velocity spectra}\label{sec:spectra}

\begin{figure}
\centering
(a)\includegraphics[width=.30\textwidth]{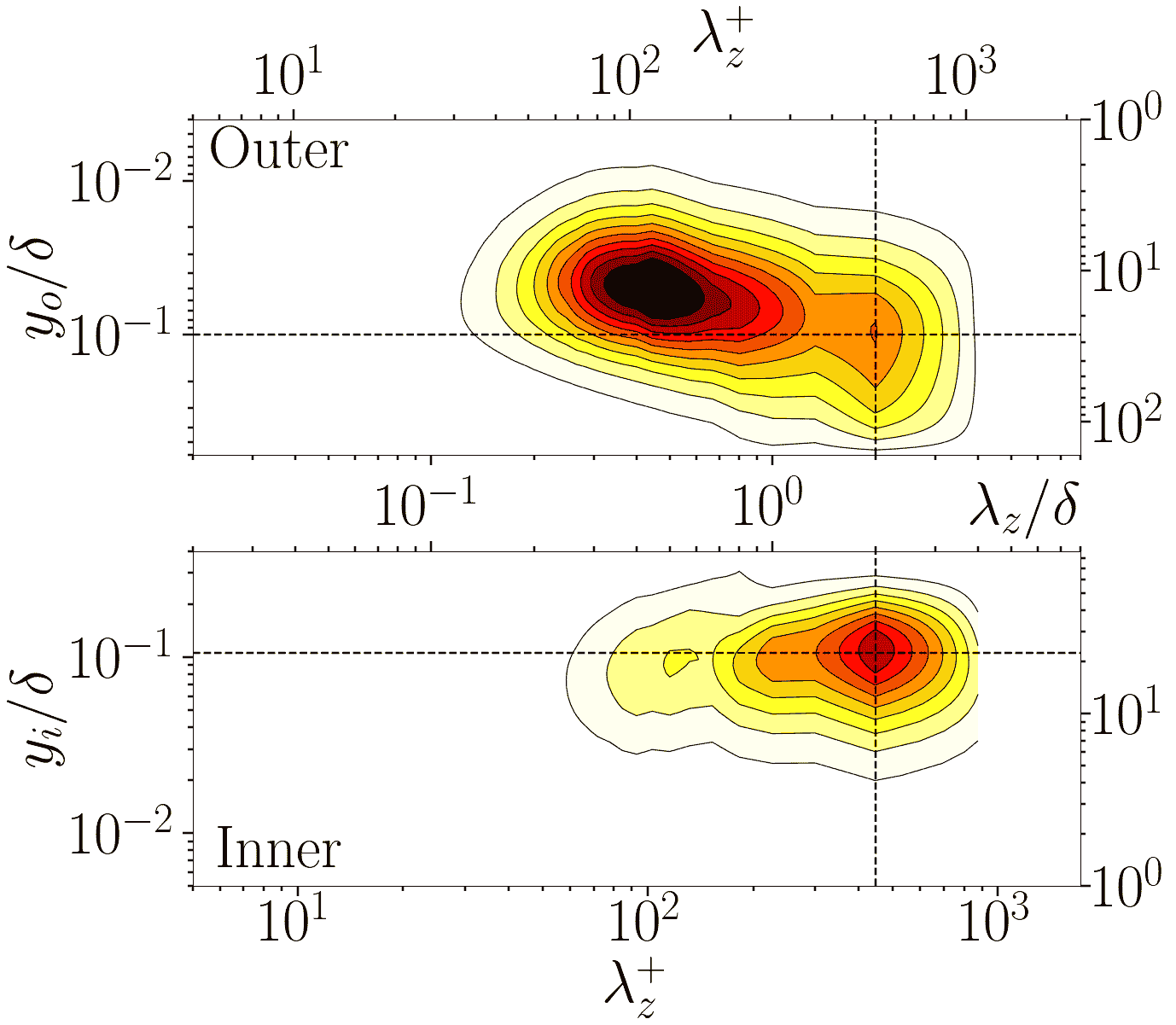} 
(b)\includegraphics[width=.29\textwidth]{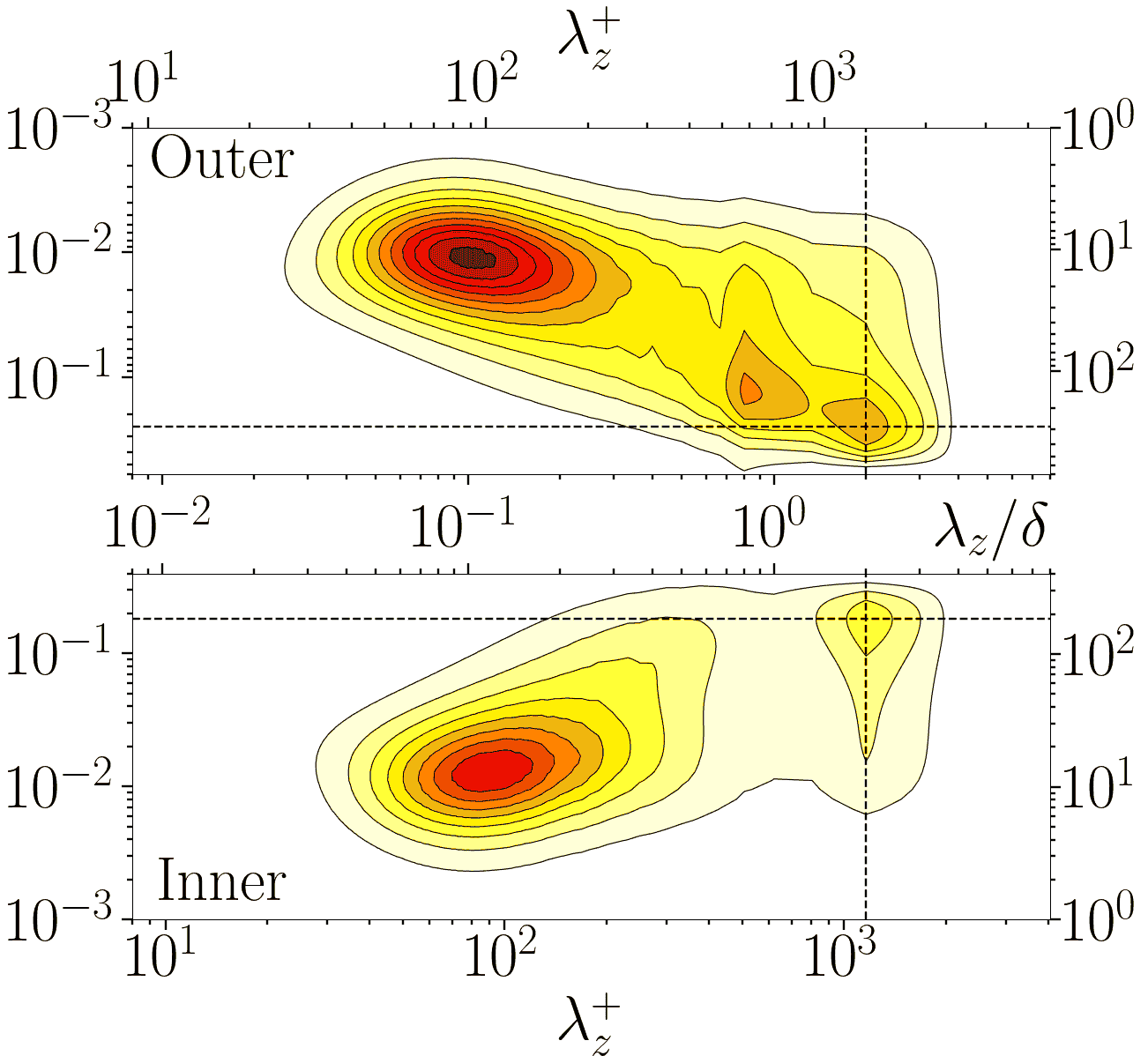}
(c)\includegraphics[width=.30\textwidth]{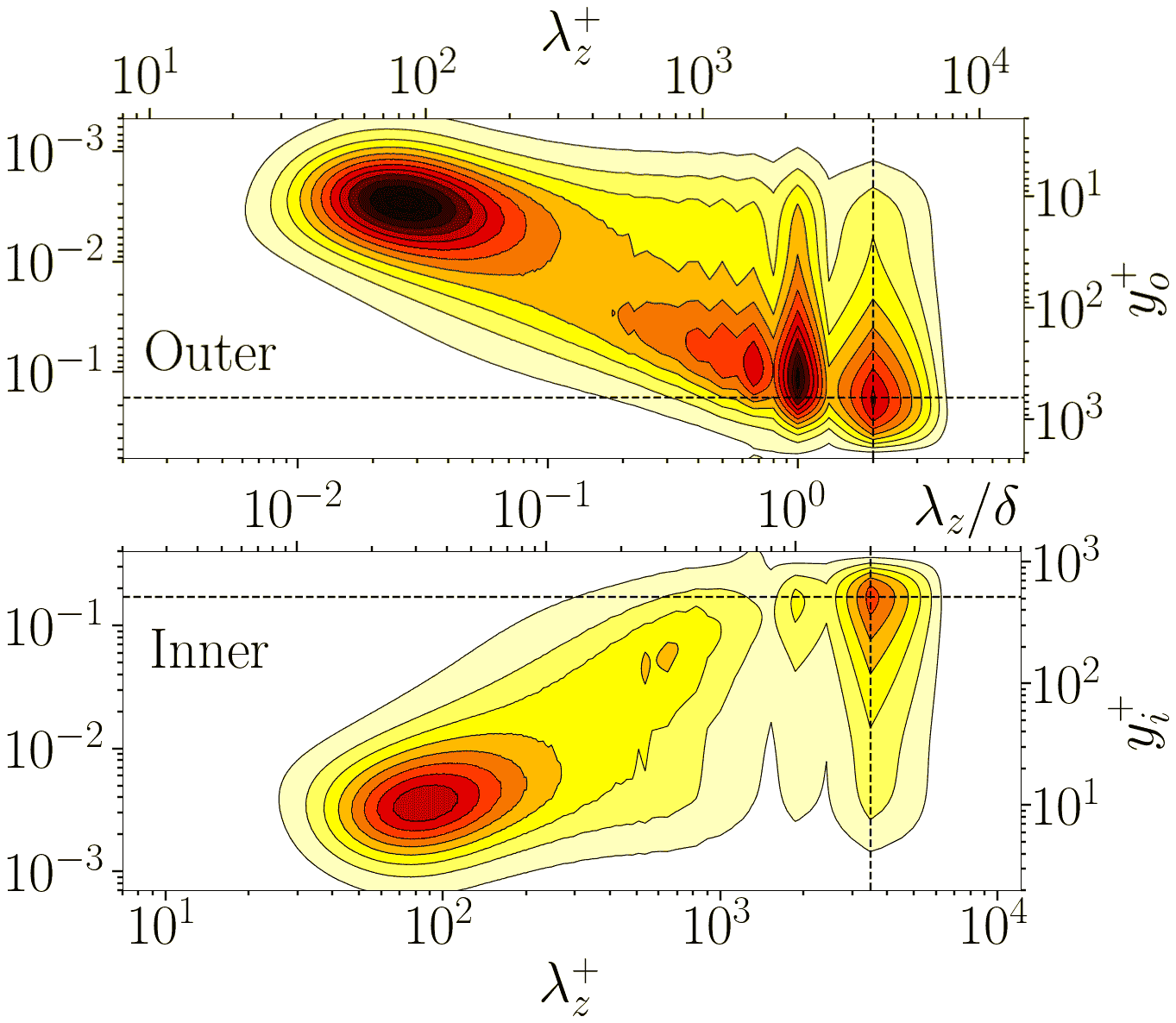}\\
(d)\includegraphics[width=.30\textwidth]{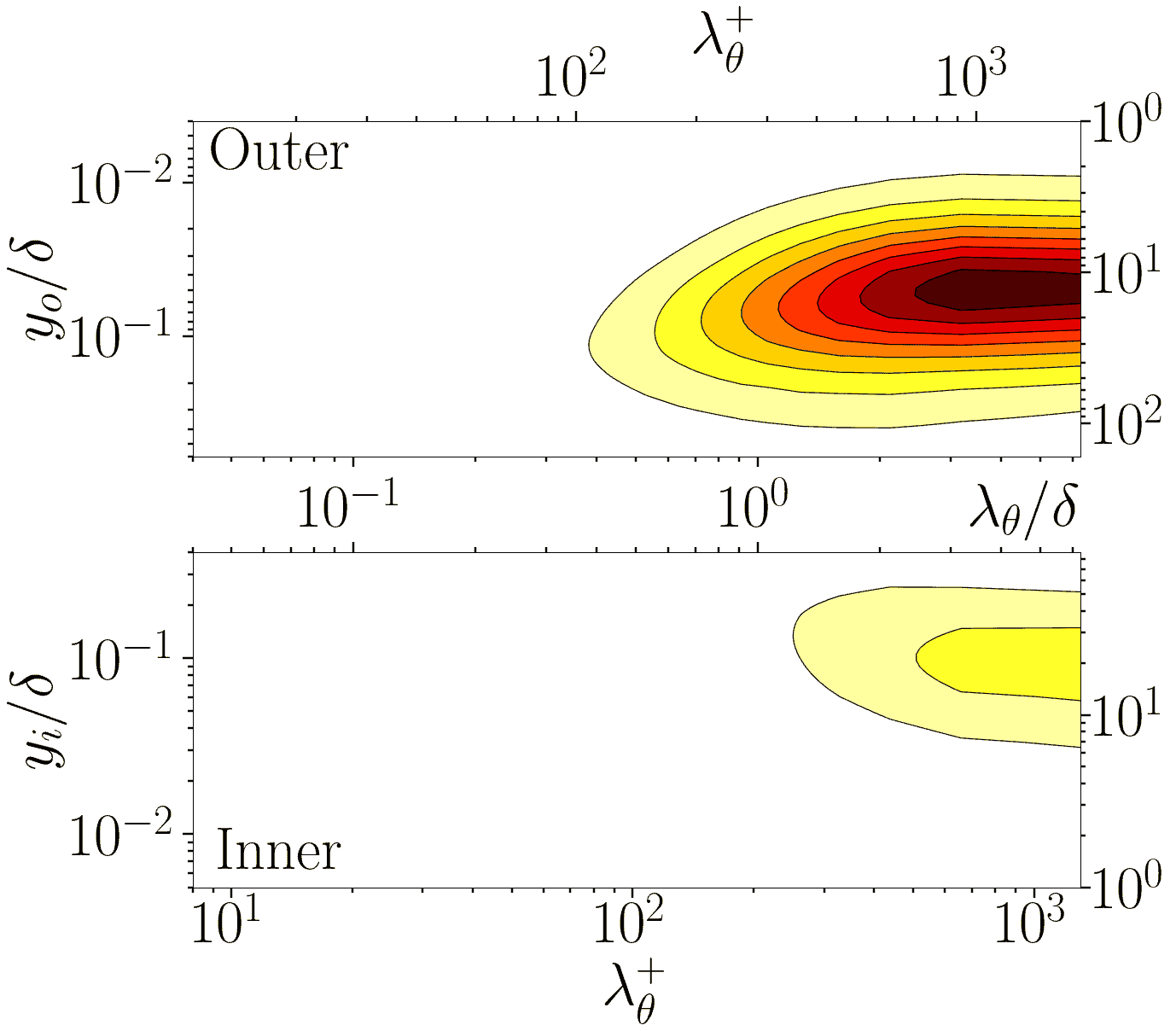} 
(e)\includegraphics[width=.29\textwidth]{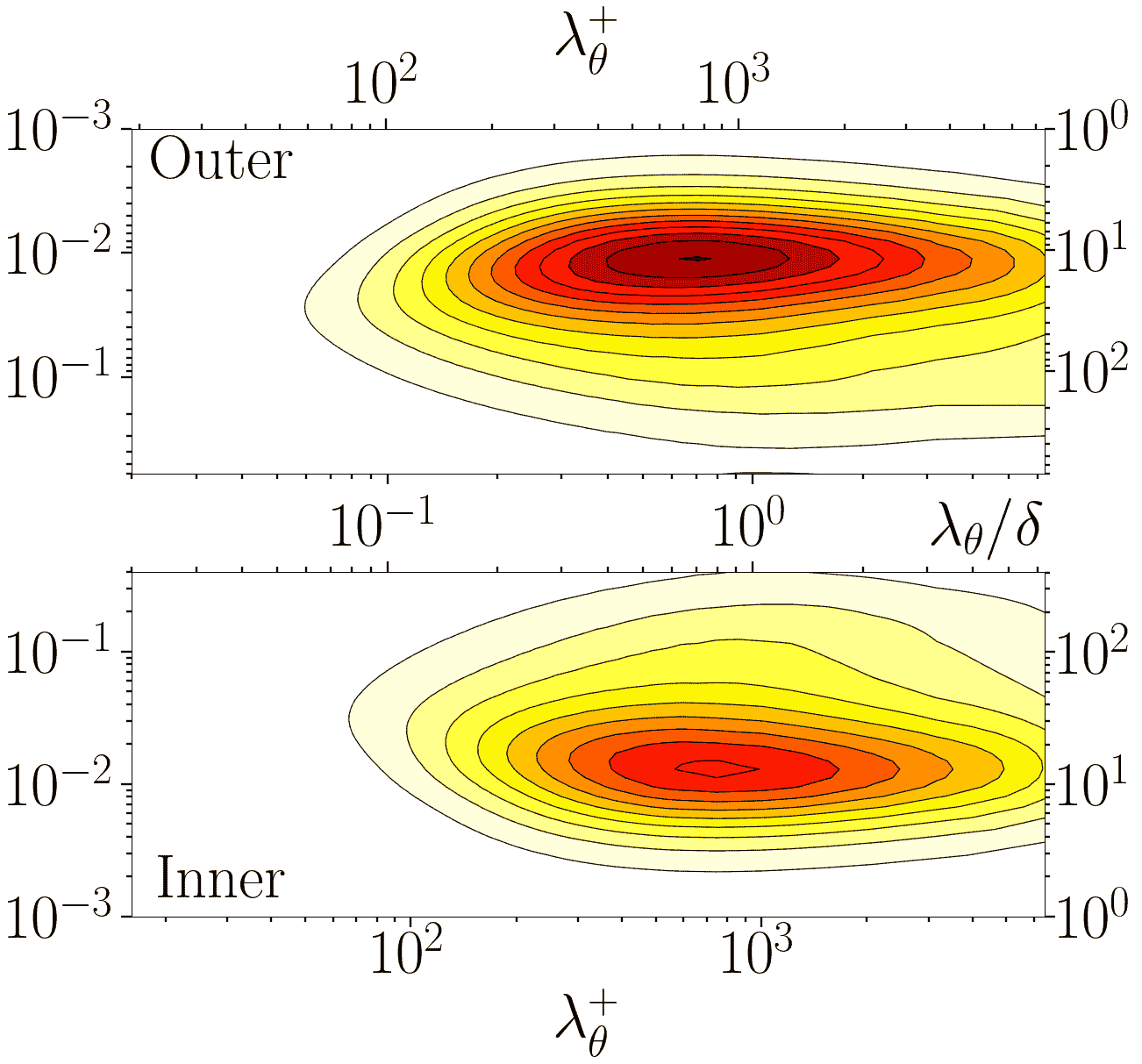} 
(f)\includegraphics[width=.30\textwidth]{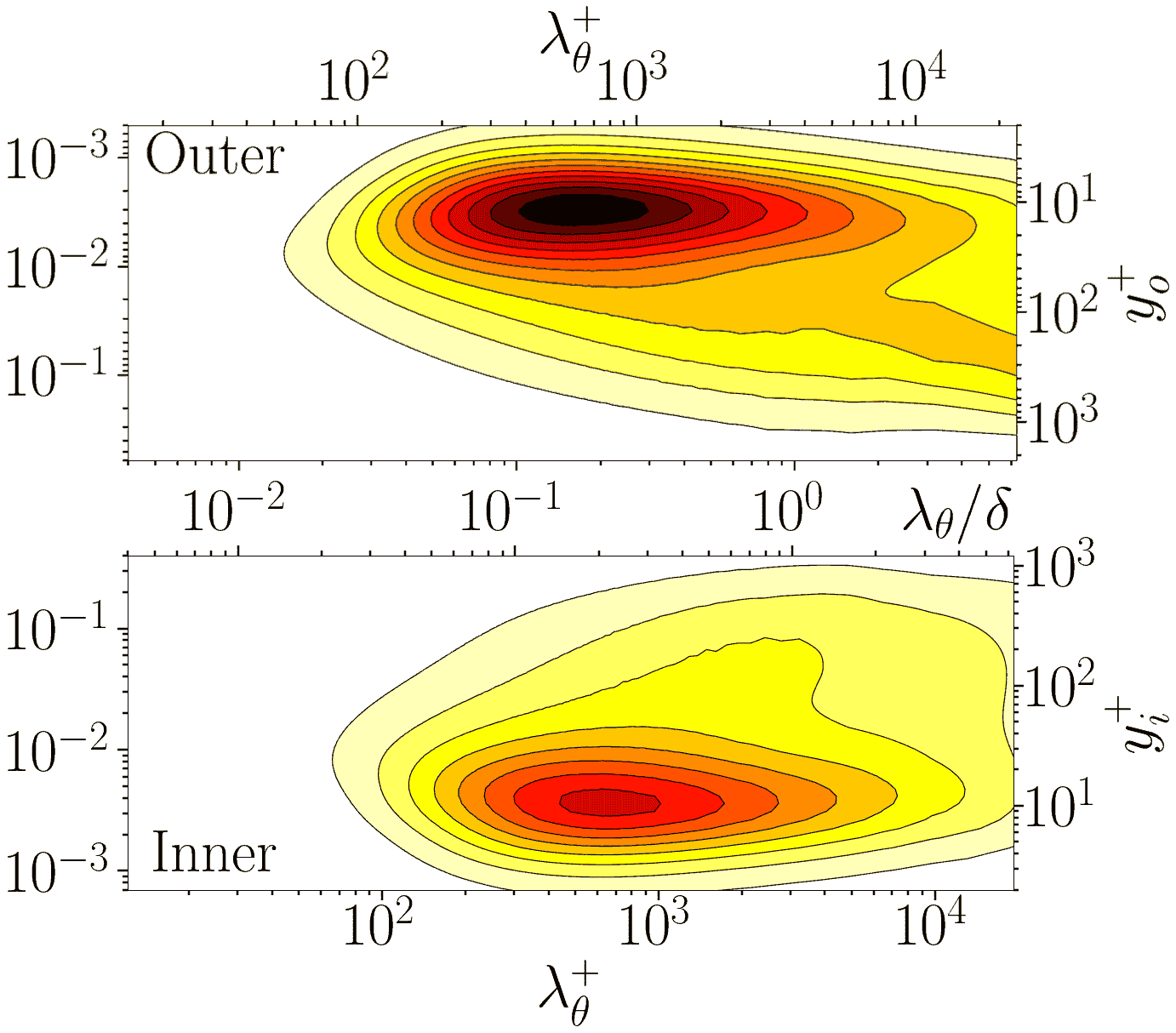} 
\caption{Pre-multiplied spectra of streamwise fluctuating velocity as a 
function of spanwise wavelength ($k_z^* E^*_{uu}$, upper panels) and of  
streamwise wavelength ($k_\theta^* E^*_{uu}$, lower panels) as the wall 
distance varies for the R40 flow cases.  
From left to right, the panels correspond to 
$\Rey_b=4000$~(a, d), $20000$~(b, e), $87000$~(c, f). 
Wall distance from the inner and outer wall, 
spanwise and streamwise wavelengths are reported in outer units 
($y_i/\delta$, $y_o/\delta$, $\lambda_z/\delta$, $\lambda_\theta/\delta$), 
and in local wall units ($y_i^+$, $y_o^+$, $\lambda_z^+$, $\lambda_\theta^+$), 
respectively. Dashed lines mark wavelength and wall distance 
of the energy peak associated with longitudinal large-scale structures.} 
\label{fig:usp40}
\end{figure}
 
\begin{figure}
\centering
(a)\includegraphics[width=.30\textwidth]{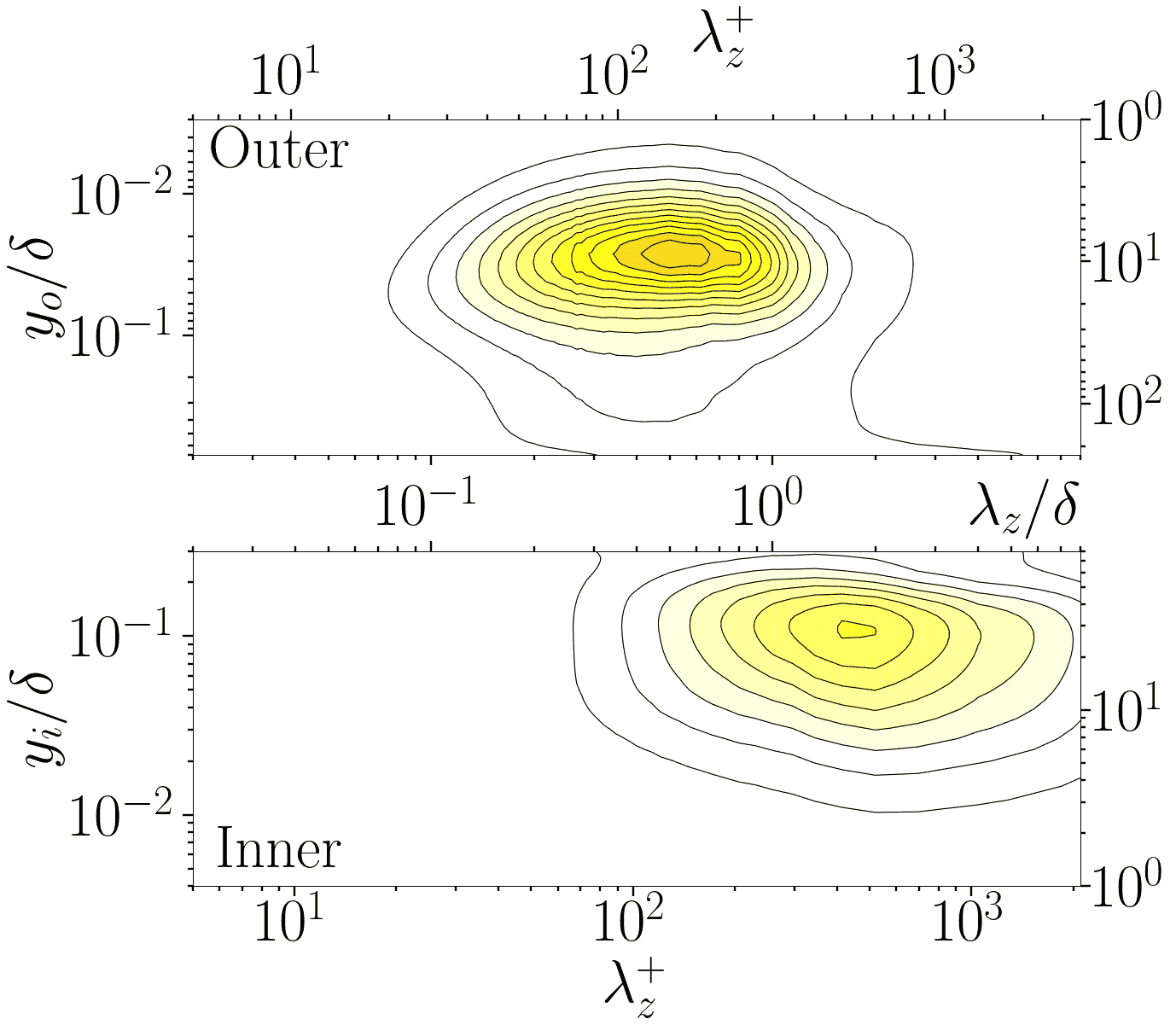}
(b)\includegraphics[width=.29\textwidth]{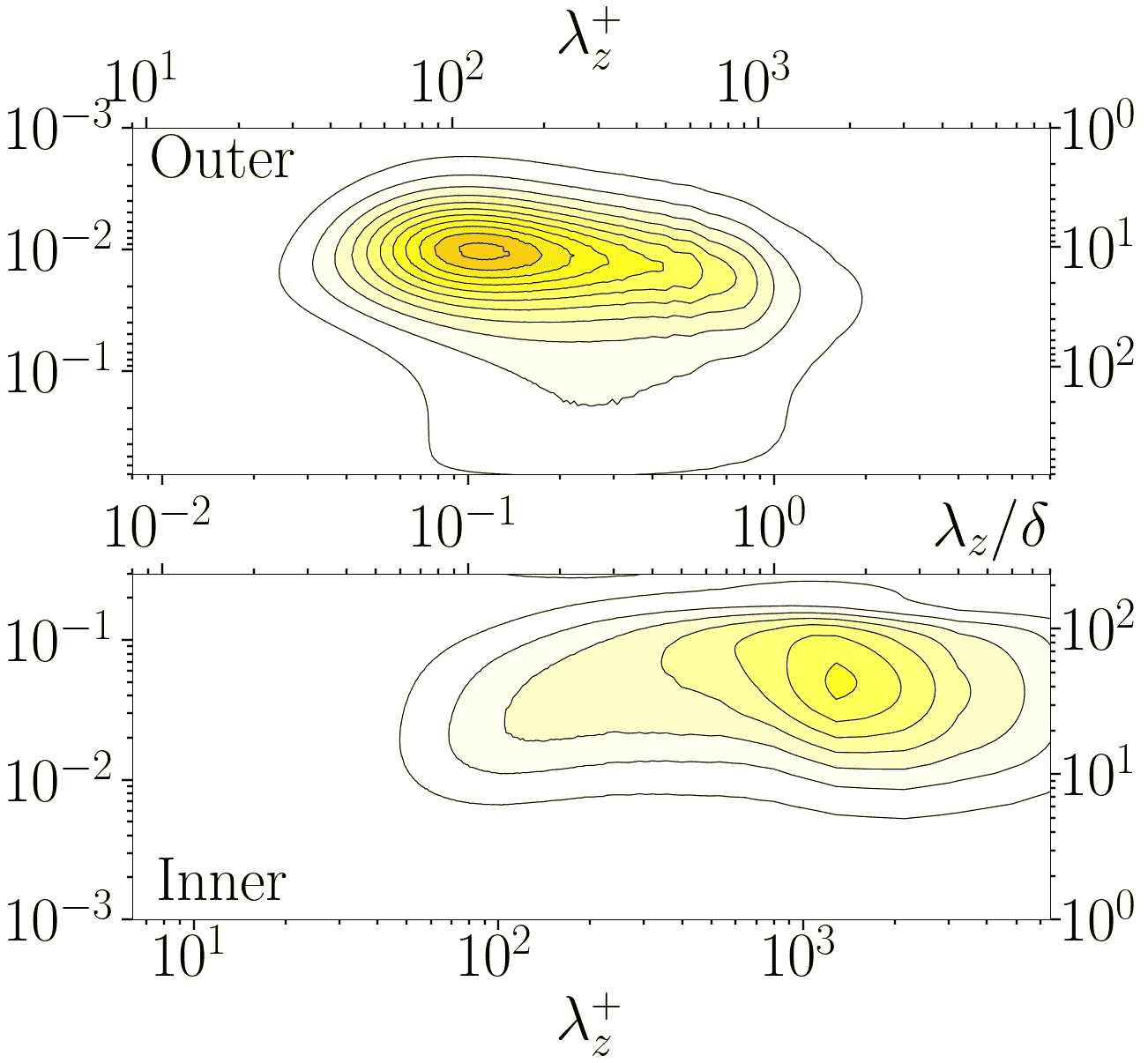}
(c)\includegraphics[width=.30\textwidth]{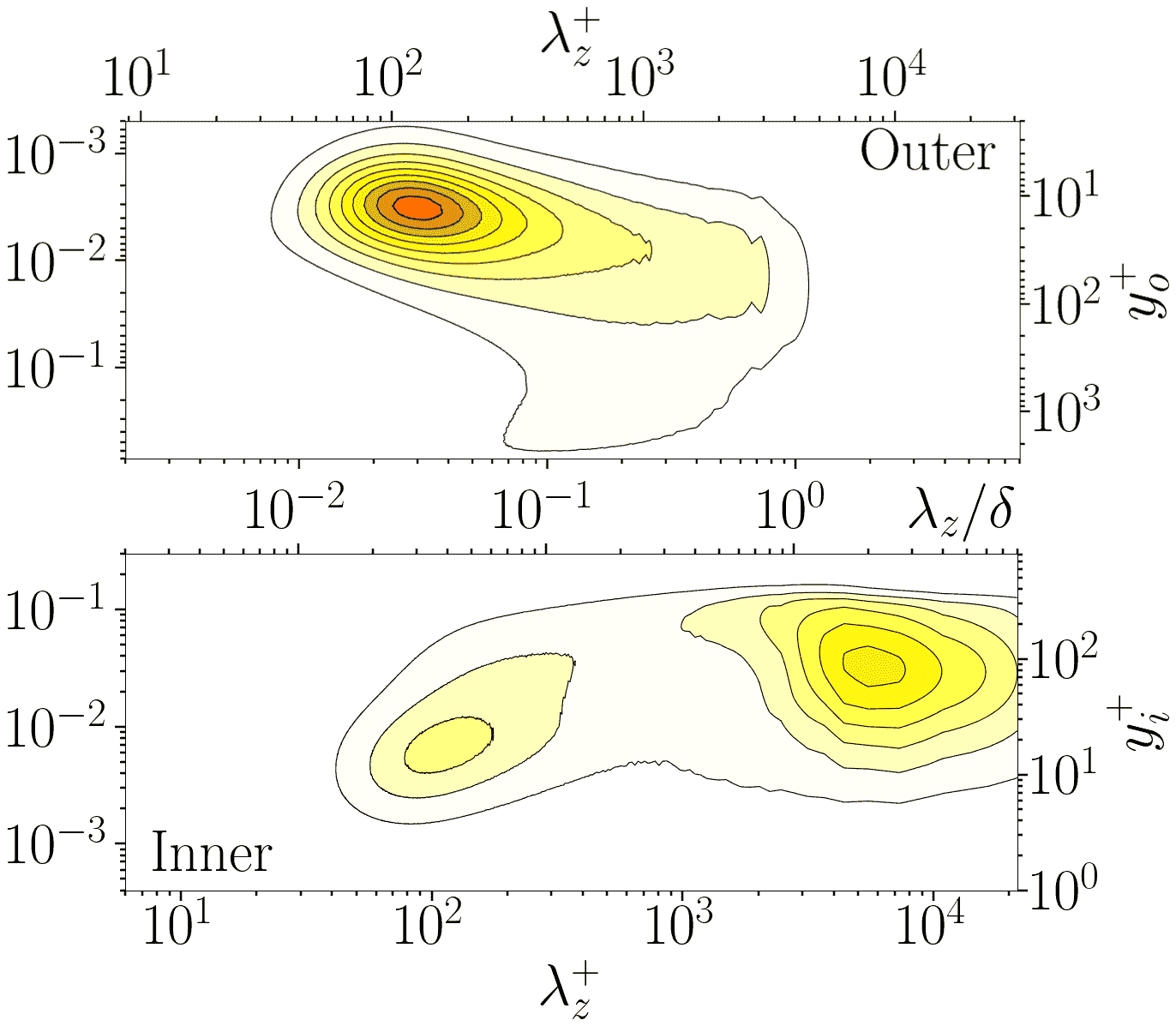}\\
(d)\includegraphics[width=.30\textwidth]{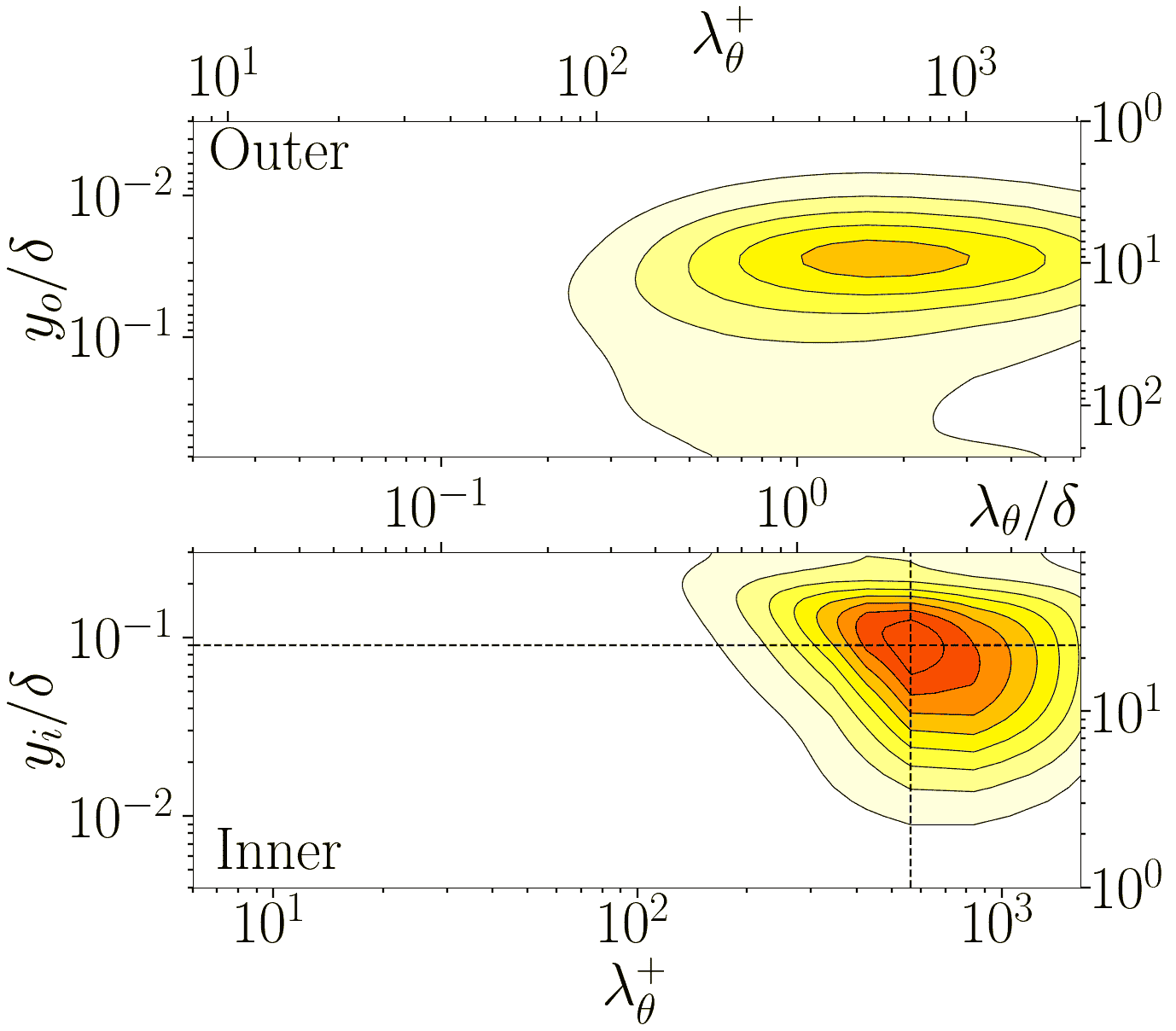}
(e)\includegraphics[width=.29\textwidth]{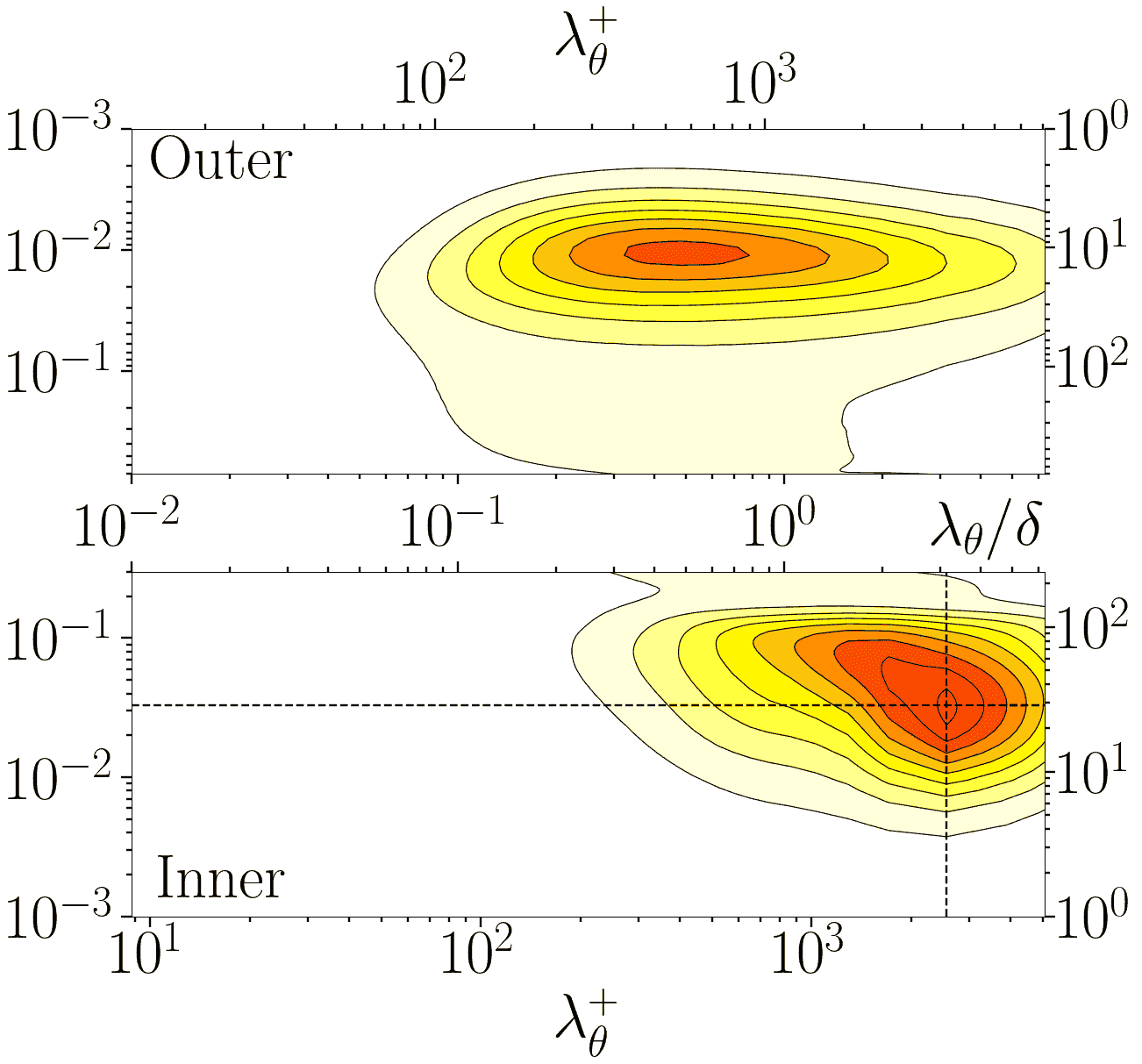}
(f)\includegraphics[width=.30\textwidth]{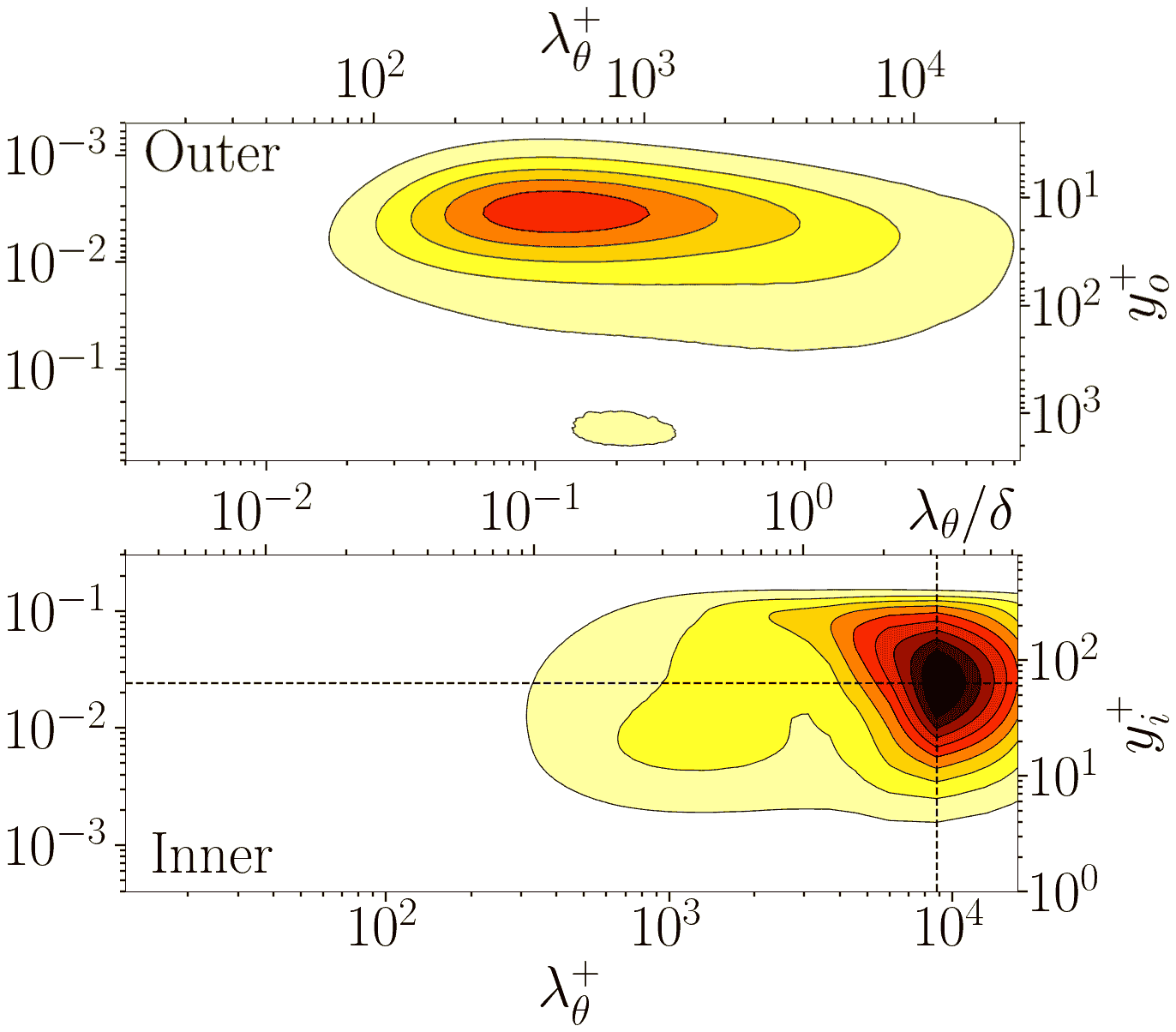}
\caption{Pre-multiplied spectra of streamwise fluctuating velocity as a 
function of spanwise wavelength ($k_z^* E^*_{uu}$, upper panels) and of  
streamwise wavelength ($k_\theta^* E^*_{uu}$, lower panels) 
as the wall distance varies for the R1 flow cases.  
From left to right, the panels correspond to 
$\Rey_b=4000$~(a, d), $20000$~(b, e), $87000$~(c, f). 
Wall distance from the inner and outer wall, 
spanwise and streamwise wavelengths are reported in outer units 
($y_i/\delta$, $y_o/\delta$, $\lambda_z/\delta$, $\lambda_\theta/\delta$), 
and in local wall units ($y_i^+$, $y_o^+$, $\lambda_z^+$, $\lambda_\theta^+$), 
respectively. Dashed lines mark wavelength and wall distance 
of the energy peak associated with transverse large-scale structures.} 
\label{fig:usp1}
\end{figure}

A clearer understanding of the energetic relevance of the various 
scales of motions can be achieved from inspecting the spectra 
of the velocity fluctuations. Spectra of both streamwise and wall-normal 
velocity fluctuations are presented, the former to detect large-scale 
structures typical of plane channel flows~\citep{lee2015direct}, 
whereas the latter are instrumental to reveal the presence of 
longitudinal roll cells~\citep{dai2016effects}. 
Figures~\ref{fig:usp40} and~\ref{fig:usp1} display the pre-multiplied 
spectra of streamwise velocity fluctuations in the spanwise 
($k_z^* E^*_{uu}$) and streamwise ($k_\theta^* E^*_{uu}$) directions 
at various wall-normal locations for R40 and R1 flow cases, respectively. 
For clarity, we refer to the distance from the inner wall  
as $y_i/\delta$ (where $y_i/\delta=y/\delta$) 
and from the outer wall as $y_o/\delta$ (where $y_o/\delta=1-y/\delta$), 
both reported on the left vertical axis. 
The distance from the inner and outer wall is also reported on the 
right vertical axis in local wall units, $y^+_i$ and $y^+_o$, respectively. 
 
In the R40 flow cases a primary energy peak appears in the spanwise 
spectra (upper panels) near both walls 
at $y_i^+\approx y_o^+\approx12$, $\lambda_z^+\approx100$, 
which is the typical signature of the regeneration cycle of momentum 
streaks~\citep{waleffe_95}. An exception is the case at $\Rey_b=4000$: 
the inner-wall primary peak is nearly absent, 
confirming the inferences from flow visualisations.
From the streamwise spectra (lower panels) we note that the 
streaks-related peak is at 
$\lambda_\theta^+\approx800$, meaning that the mean wavelength of 
momentum streaks is slightly shorter than the typical value of plane channel 
flows, namely $\lambda_x^+\approx 1000$~\citep{monty2009comparison}. 
A secondary peak appears in the spanwise spectra farther from walls, at approximately 
$y_i/\delta\approx y_o/\delta\approx0.1$ at $\Rey_b=4000$~(a)
and $y_i/\delta\approx y_o/\delta\approx0.2$ at $\Rey_b=20000$~(b) and $87000$~(c). 
As seen from the intersection of the dashed lines, 
the energy of the peaks is concentrated at $\lambda_z/\delta\approx2$. 
If the domain contains $n$ pairs of roll cells, we expect to find a peak in 
the spanwise spectra at $\lambda_z=L_z/n$, where $L_z=4\delta$ for the R40 flow cases. 
Flow visualisations indeed revealed the presence $n=2$ roll cells pairs,   
hence we interpret these peaks as the footprint of longitudinal 
large-scale structures. A third peak at $\lambda_z/\delta\approx1$ appears 
faintly at $\Rey_b=20000$, and sharply emerges at $\Rey_b=87000$. 
This peak corresponds to $n=4$ pairs of longitudinal cells, 
indicating that roll cells with various sizes could be present, 
which we will investigate below. 
The streamwise spectra reveal that the scales of motion with the 
maximum streamwise wavelength, $\lambda_\theta=L_\theta=2\pi$, 
are quite energetic, suggesting that the longitudinal vortices 
span the whole domain along the streamwise direction. 

As for the R1 flow cases (figure~\ref{fig:usp1}) 
the primary peak in the spanwise spectra due to momentum streaks 
still occurs near the outer wall at $y_o^+\approx12$ (upper panels). 
As noted from flow visualisations, streaky structures have a 
greater spanwise wavelength than the R40 flow cases, 
which is about $\lambda_z^+\approx120$. 
Their streamwise wavelength is instead shorter, about 
$\lambda_\theta^+\approx500$, as visible from the 
streamwise spectra (lower panels). 
Near the inner wall the picture is different: the primary near-wall peak 
in the spanwise spectra is absent at $\Rey_b=4000$~(d), starts to form 
at $\Rey_b=20000$~(e), and only becomes visible at $\Rey_b=87000$~(f). 
This trend, which is also found in the streamwise spectra, 
suggests that the inhibiting effect of convex curvature on turbulence 
is experienced more strongly at low Reynolds number. 
%
In contrast to the R40 flow cases, never does a secondary peak 
emerge in the spanwise spectra near the outer wall. 
This implies that the longitudinal roll cells, if present, 
have a negligible effect on the streamwise velocity fluctuations.
A secondary peak is instead present near the inner wall 
at $\lambda_z/\delta\approx2$,
which moves closer to the wall (in $\delta$ units) as $\Rey_b$ increases.
This peak cannot be related to longitudinal roll cells for two reasons: 
1) they are most intense at the outer wall, yet no secondary peak is 
detected there, and 2) the spanwise wavelength of longitudinal vortices 
must be shorter, as one can infer from figure~\ref{fig:utheta1} 
and as we will show more clearly below. 
%
To elaborate further, we consider the streamwise energy spectra, 
which highlight energetic modes near the inner wall 
at long wavelengths: 
$\lambda_\theta/\delta\approx2\pi/3$ at $\Rey_b=4000$~(d) and 
$\lambda_\theta/\delta\approx\pi$ at $20000$~(e) and $87000$~(f), 
as marked by the dashed lines. 
These energy peaks cannot be related neither to turbulence structures,
which are absent or strongly inhibited near the inner wall, 
neither to longitudinal roll cells, as we have just seen through the 
spanwise spectra that they do not affect streamwise velocity fluctuations. 
However, flow visualisations in wall-parallel planes near the inner wall 
(figure~\ref{fig:ufluc1}, right panels) revealed the presence of 
transverse large-scale structures, organised in pairs with a streamwise 
length of about $\pi\delta$. 
The wavelength of these peaks corresponds to $n=L_\theta/\lambda_\theta=3$ 
pairs of eddies at $\Rey_b=4000$ and 
$n=L_\theta/\lambda_\theta=2$ pairs at $20000$ and $87000$. 
Hence, we interpret the energy peaks in the streamwise spectra as the 
signature of cross-flow structures.  
Moreover, the spanwise size of these structures is comparable with $\delta$, 
thus we can ascribe to them also the peaks in the spanwise spectra 
at $\lambda_z/\delta\approx2$. 
 
%
\begin{figure}
\centering
(a)\includegraphics[width=.30\textwidth]{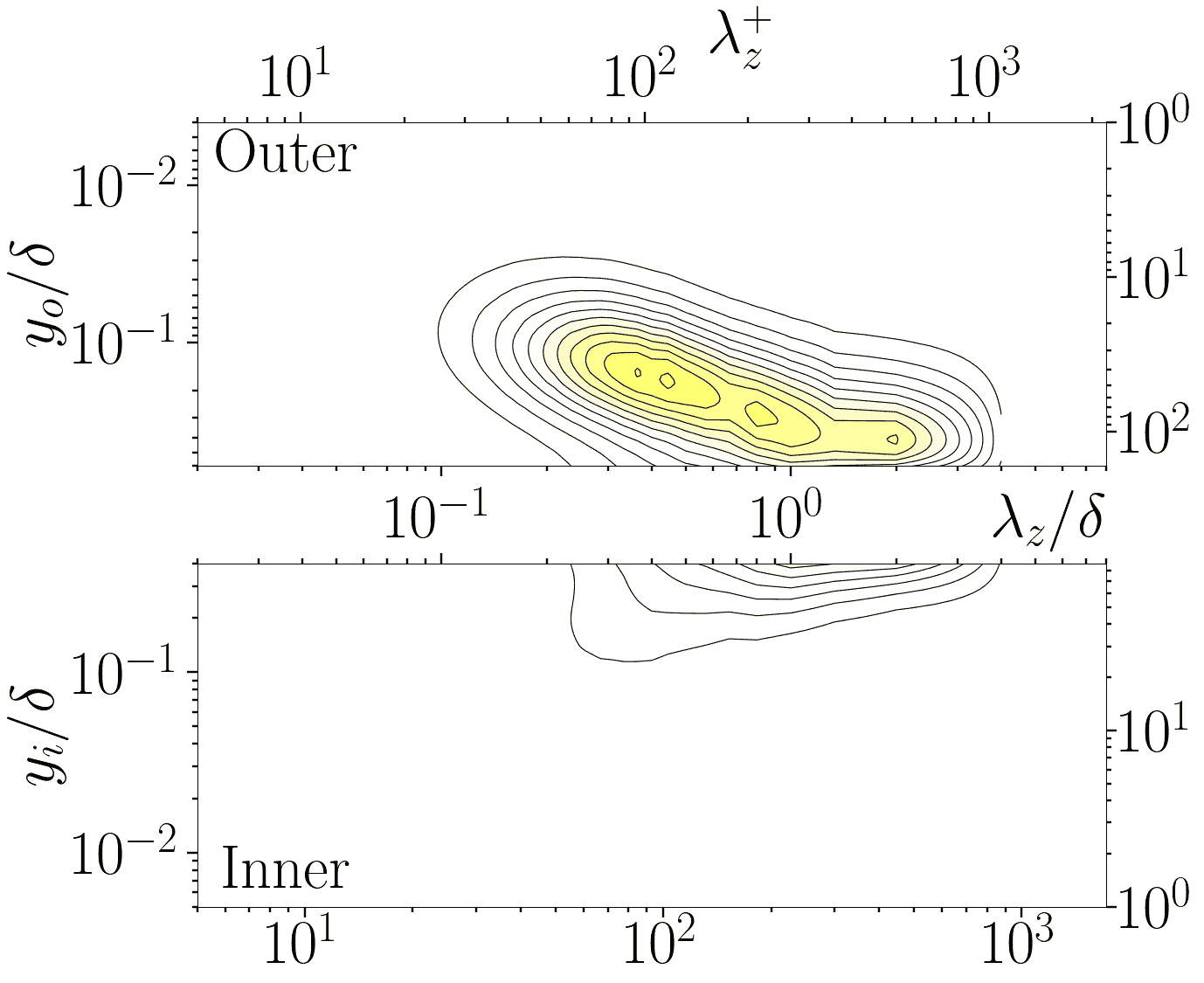}\label{vsp_re4_r40}
(b)\includegraphics[width=.29\textwidth]{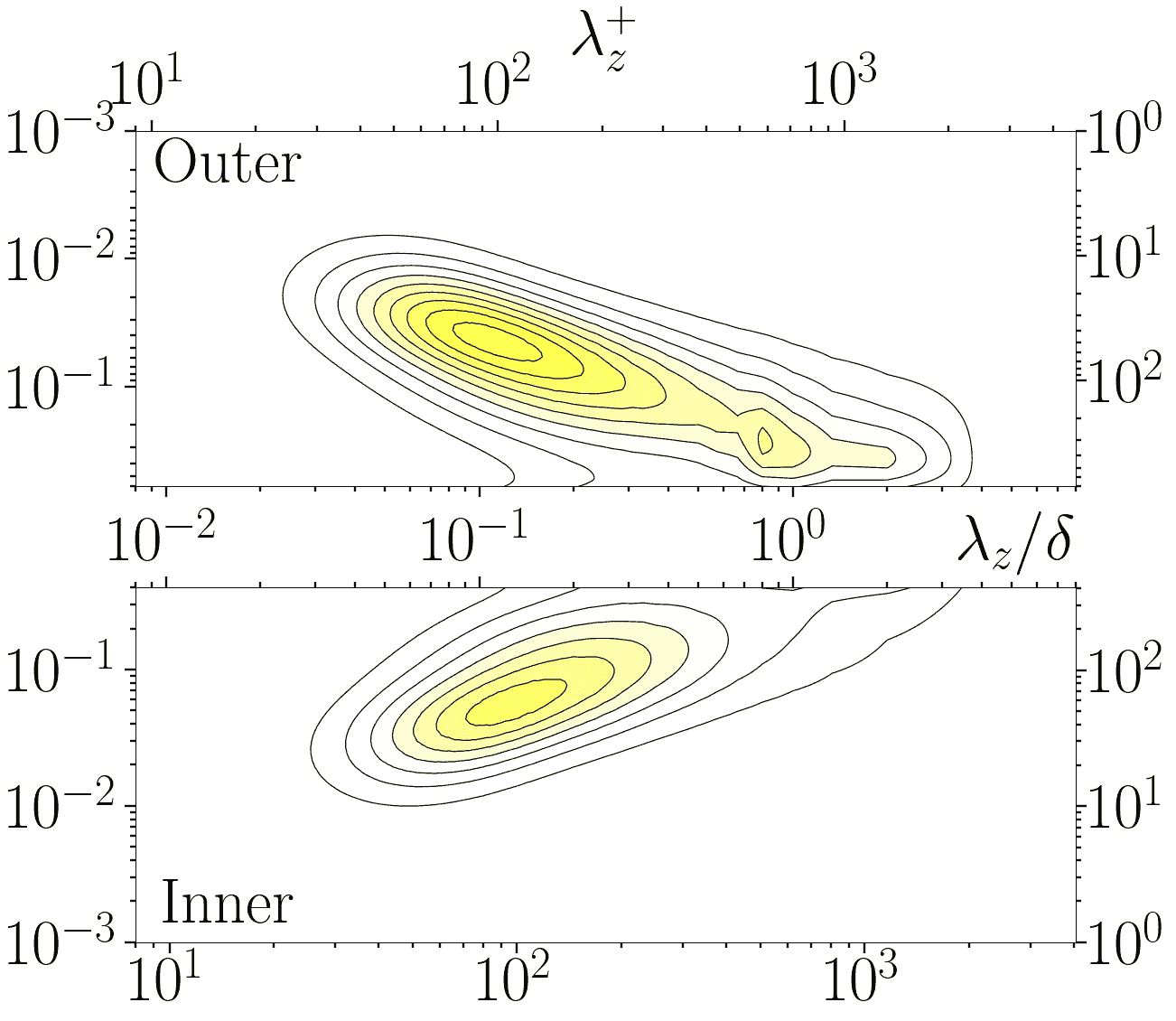}\label{vsp_re20_r40}
(c)\includegraphics[width=.30\textwidth]{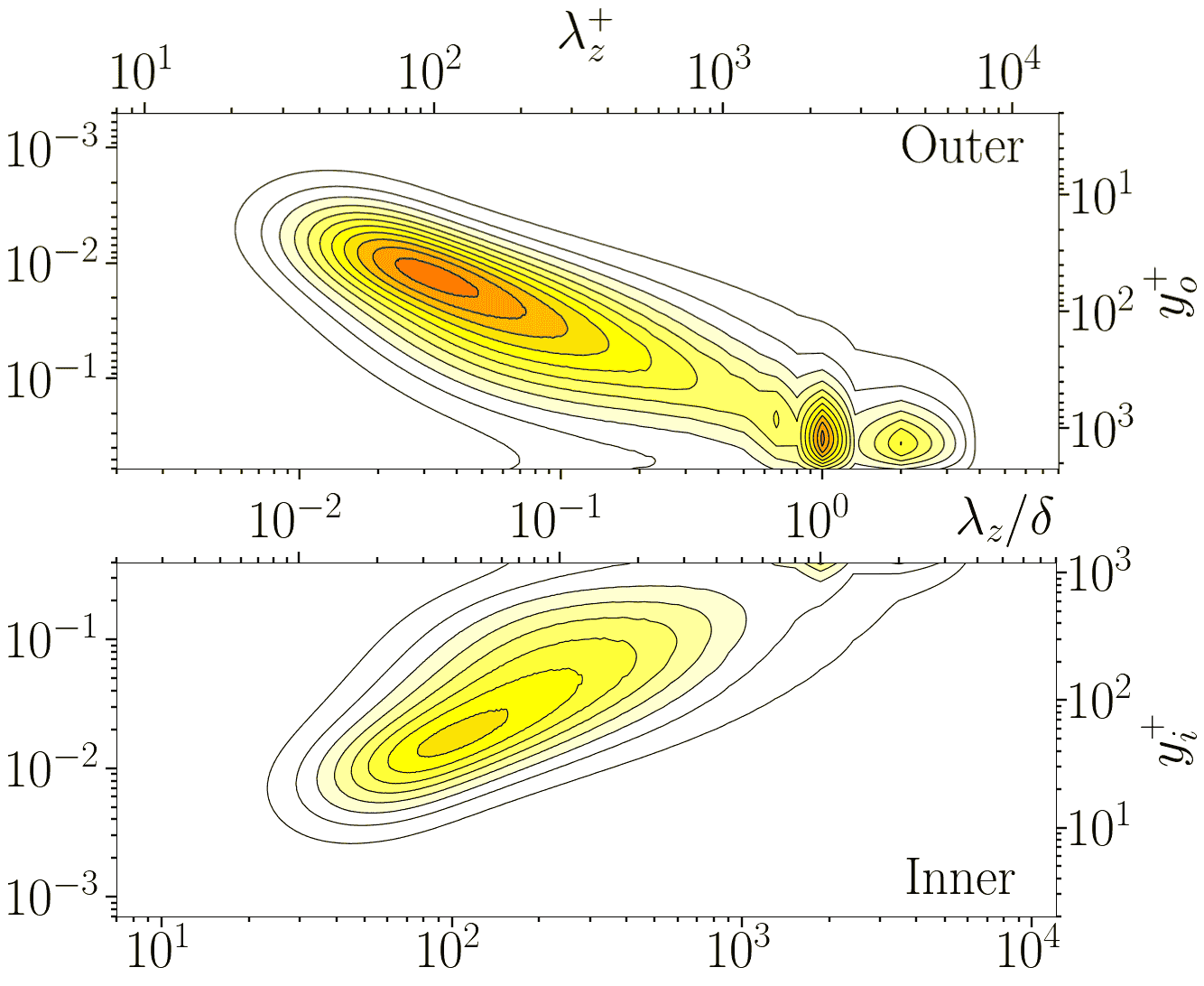}\label{vsp_re87_r40}\\
(d)\includegraphics[width=.30\textwidth]{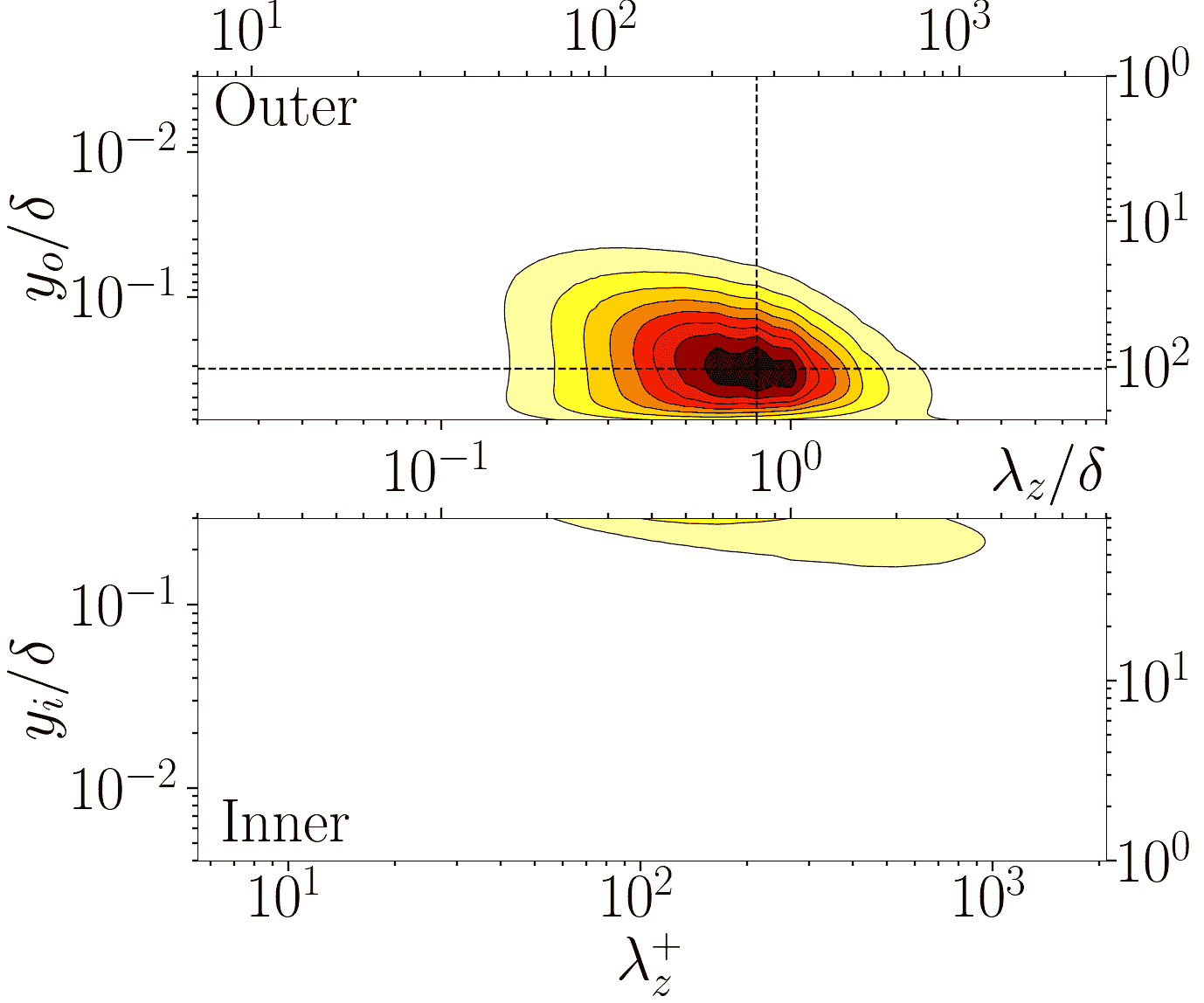}\label{vsp_re4_r1}
(e)\includegraphics[width=.29\textwidth]{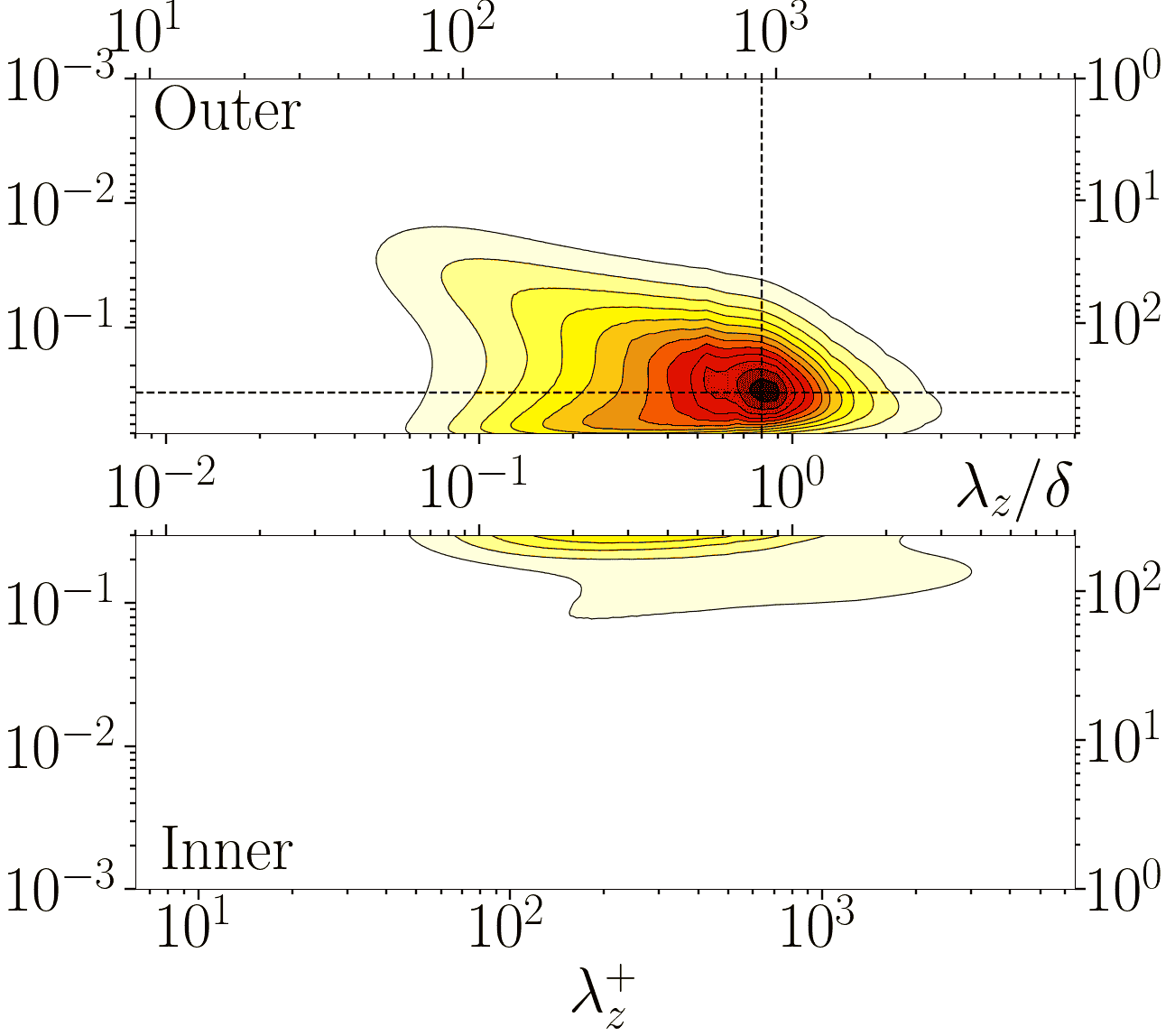}\label{vsp_re20_r1}
(f)\includegraphics[width=.30\textwidth]{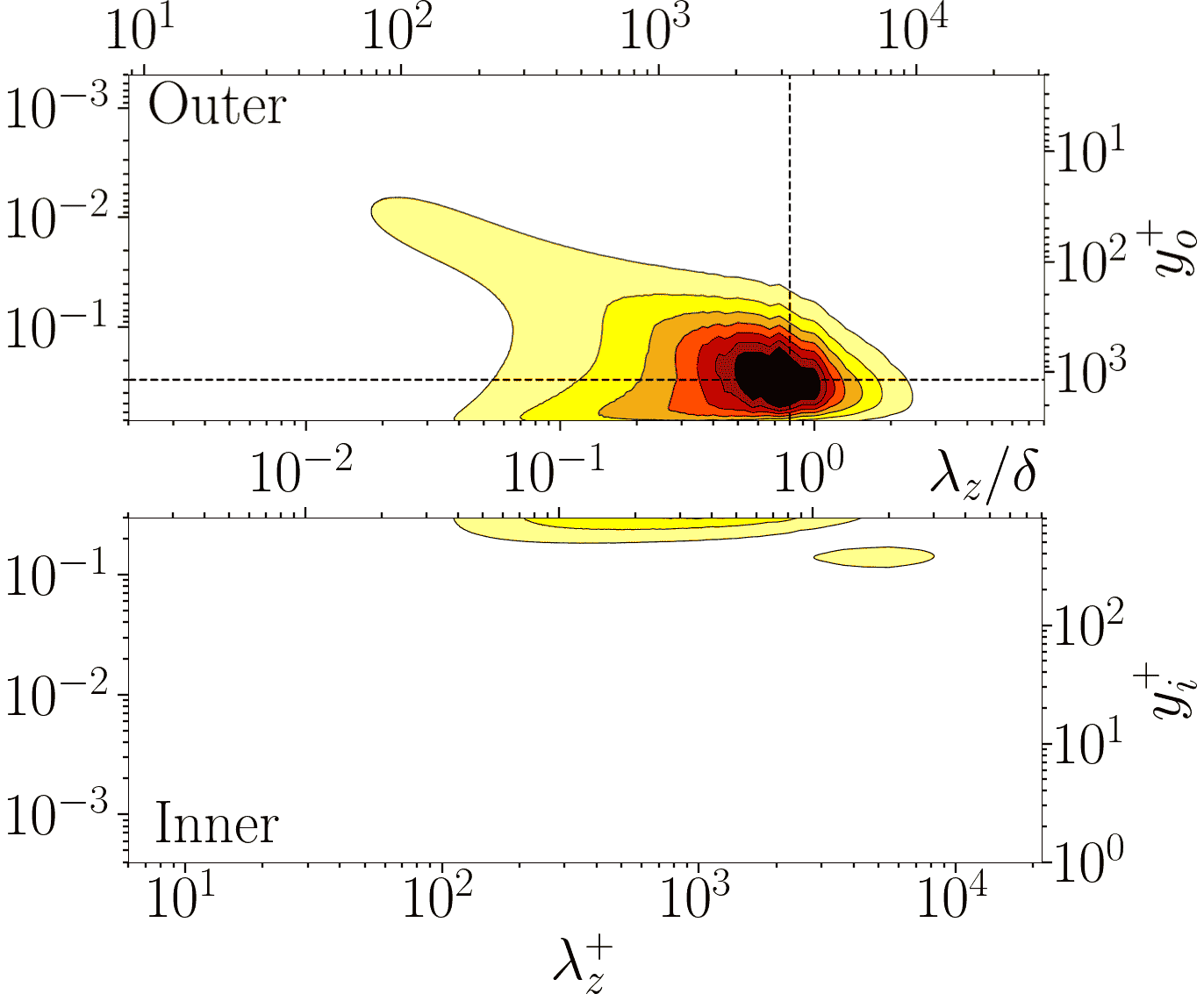}\label{vsp_re87_r1}
\caption{Pre-multiplied spectra of fluctuating wall-normal velocity ($k_z^* E^*_{vv}$) 
as a function of the spanwise wavelength and wall distance 
for the R40 flow cases (upper panels) and R1 flow cases (lower panels). 
From left to right, the panels correspond 
to $\Rey_b=4000$~(a, d), $20000$~(b, e), $87000$~(c, f). 
Wall distance from the inner and outer wall and spanwise wavelength are reported 
in outer units ($y_i/\delta$, $y_o/\delta$, $\lambda_z/\delta$) 
and local wall units ($y_i^+$, $y_o^+$, $\lambda_z^+$), respectively.
Dashed lines mark the wavelength and wall-distance of the energy 
peak related to longitudinal large-scale structures.} 
\label{fig:vsp}
\end{figure}
 
In figure~\ref{fig:vsp} we show the pre-multiplied spectral 
densities of wall-normal velocity fluctuations in the spanwise 
direction ($k_z^* E^*_{vv}$). 
In the R40 flow cases (upper panels) energy peaks related to 
near-wall turbulent structures form at $y_i^+\approx y_o^+\approx50$, 
$\lambda_z^+\approx100$, 
except for the inner wall at $\Rey_b=4000$, 
in agreement with the flow visualisations. 
Two secondary peaks in the channel core, more prominent at $\Rey_b=87000$, 
highlight the presence of energetic modes with $\lambda_z/\delta\approx1$ 
and $\lambda_z/\delta\approx2$ at $y_o/\delta\approx0.3$. 
Following the same reasoning about the streamwise velocity spectra,
we trace these energetic modes to longitudinal roll cells, 
which generate large-scale radial sweeps and ejections. 
%
The effect of strong curvature on the energy distribution 
of the wall-normal velocity fluctuations (lower panels) 
is rather striking. Contrary to streamwise velocity spectra,
the energy content is now much higher in the case of strong curvature,
which points to substantial structural changes.
No distinct peak is observed in the inner part of the channel,
which conveys that the transverse large-scale structures that we noted 
above have no impact (on average) on the wall-normal velocity fluctuations. 
A prominent energy peak is located at $y_o/\delta\approx0.3$,
$\lambda_z/\delta\approx0.8$, marked by the intersection of the dashed lines. 
This energetic mode is not present in plane channel flow  
\citep[see e.g.][]{cho2018scale} and cannot be related to cross-flow 
structures, which are more intense near the inner wall. Hence,
we interpret them as the signatures of longitudinal roll cells.
This would mean that the typical configuration of longitudinal large-scale 
structures consists of $n=L_z/\lambda_z=10$ pairs of roll cells 
(with $L_z=8$, $\lambda_z/\delta=0.8$), which we will ascertain 
in the following. 

\subsection{Longitudinal large-scale structures}\label{sec:long}

Flow visualisations and energy spectra reveal the presence of 
longitudinal large-scale structures, which are akin to the 
Dean vortices observed in laminar flow~\citep{dean1928fluid}.
To understand how these structures depend on curvature 
and Reynolds number and how they affect the flow field, 
we separate the coherent contribution from the underlying 
turbulence, where coherence is intended in the sense given 
by~\citet{hussain1986coherent}. For that purpose we use
the triple decomposition of \citet{hussain1970mechanics}, whereby
a generic field variable, $\varphi(\theta,r,z,t)$, is decomposed as
\begin{equation}
\varphi(\theta,r,z,t)={\Phi}(r)+\tilde{\varphi}(r,z,t)+\varphi''(\theta,r,z,t),
\label{triple}
\end{equation}
where 
${\Phi}$ is the space average over the homogeneous directions
($\theta$, $z$) and time ($t$), 
\begin{equation}
\tilde{\varphi}(r,z,t)=\langle\varphi(\theta,r,z,t)\rangle_\theta-{\Phi}(r),
\label{eq:coh}
\end{equation}
is the coherent contribution from the longitudinal vortices,
$\langle\varphi(\theta,r,z,t)\rangle_\theta$ is the average over the 
streamwise direction, 
and 
\begin{equation}
\varphi''(\theta,r,z,t)=\varphi(\theta,r,z,t)-\langle\varphi(\theta,r,z,t)\rangle_\theta 
\end{equation}
is the instantaneous turbulent fluctuation.
The total fluctuation is then $\varphi'=\tilde{\varphi}+\varphi''$. 
The instantaneous coherent fields ($\tilde{\varphi}$) 
are averaged over time to obtain the mean coherent contribution
($\overline{\tilde{\varphi}}$). 
Since the time average is taken over a finite window, 
it is impossible to determine whether the longitudinal vortices are moving on 
a timescale longer than the averaging window. If they are slowly drifting, 
the time averages would underestimate their strength~\citep{moser1987effects}. 
%
\begin{figure}
\centering
(a)\includegraphics[width=.3\textwidth]{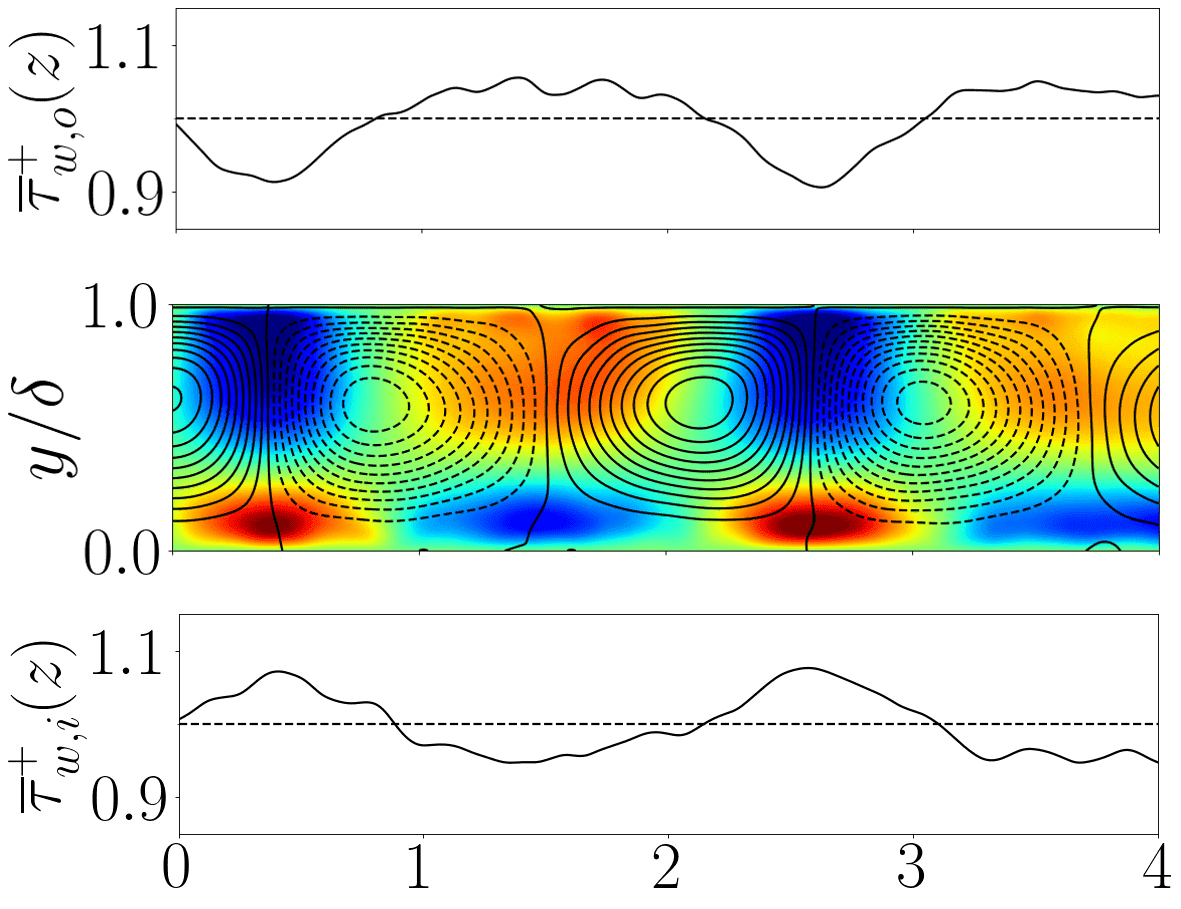}\label{rz_re4_r40}
(b)\includegraphics[width=.3\textwidth]{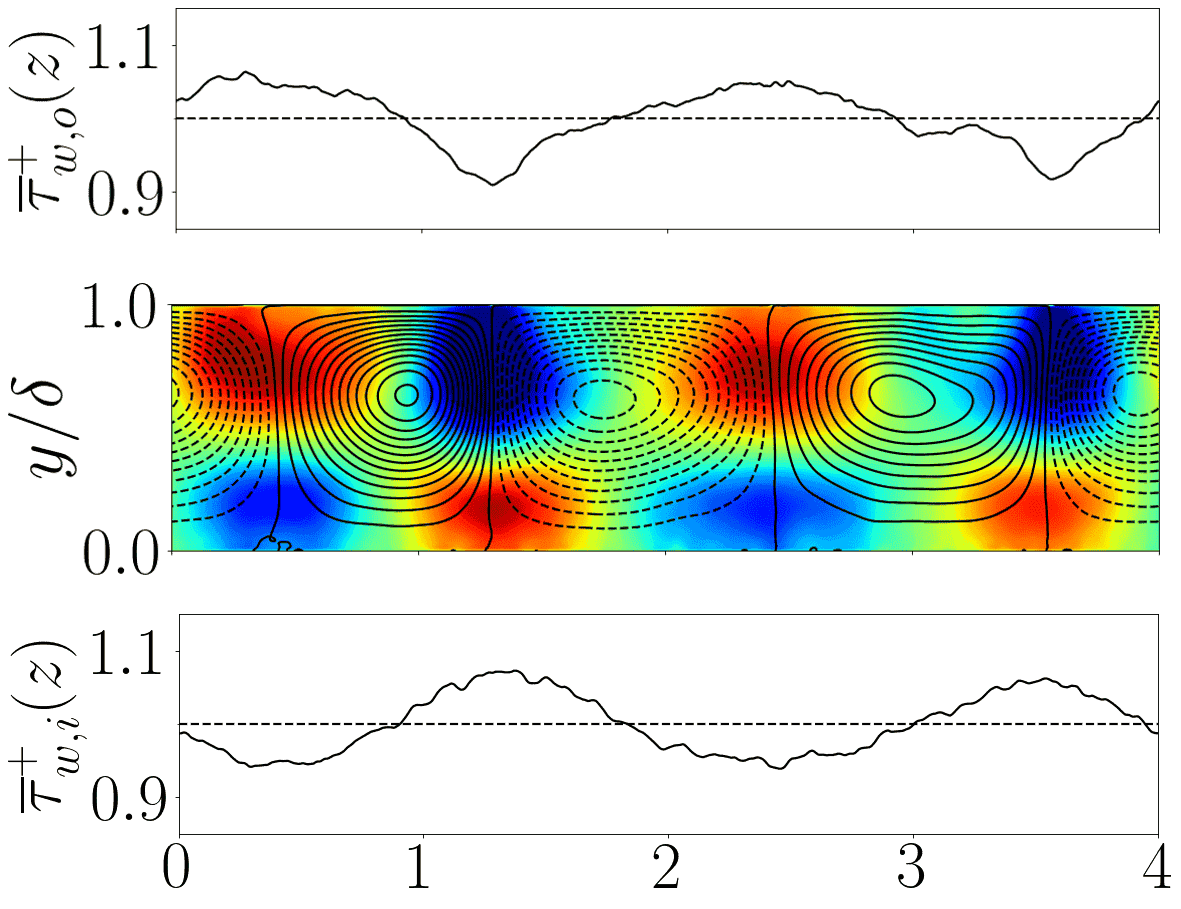}\label{rz_re20_r40}
(c)\includegraphics[width=.3\textwidth]{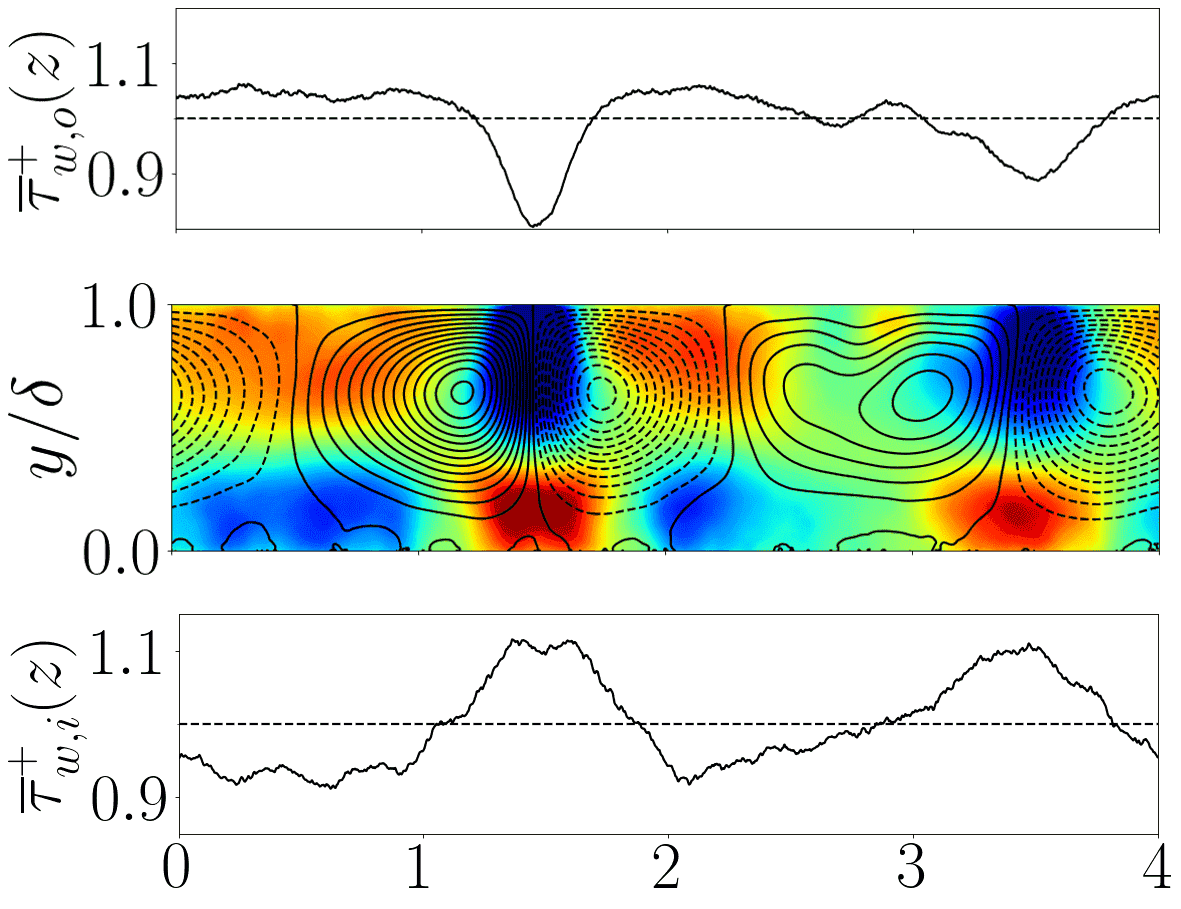}\label{rz_re87_r40}
(d)\includegraphics[width=.3\textwidth]{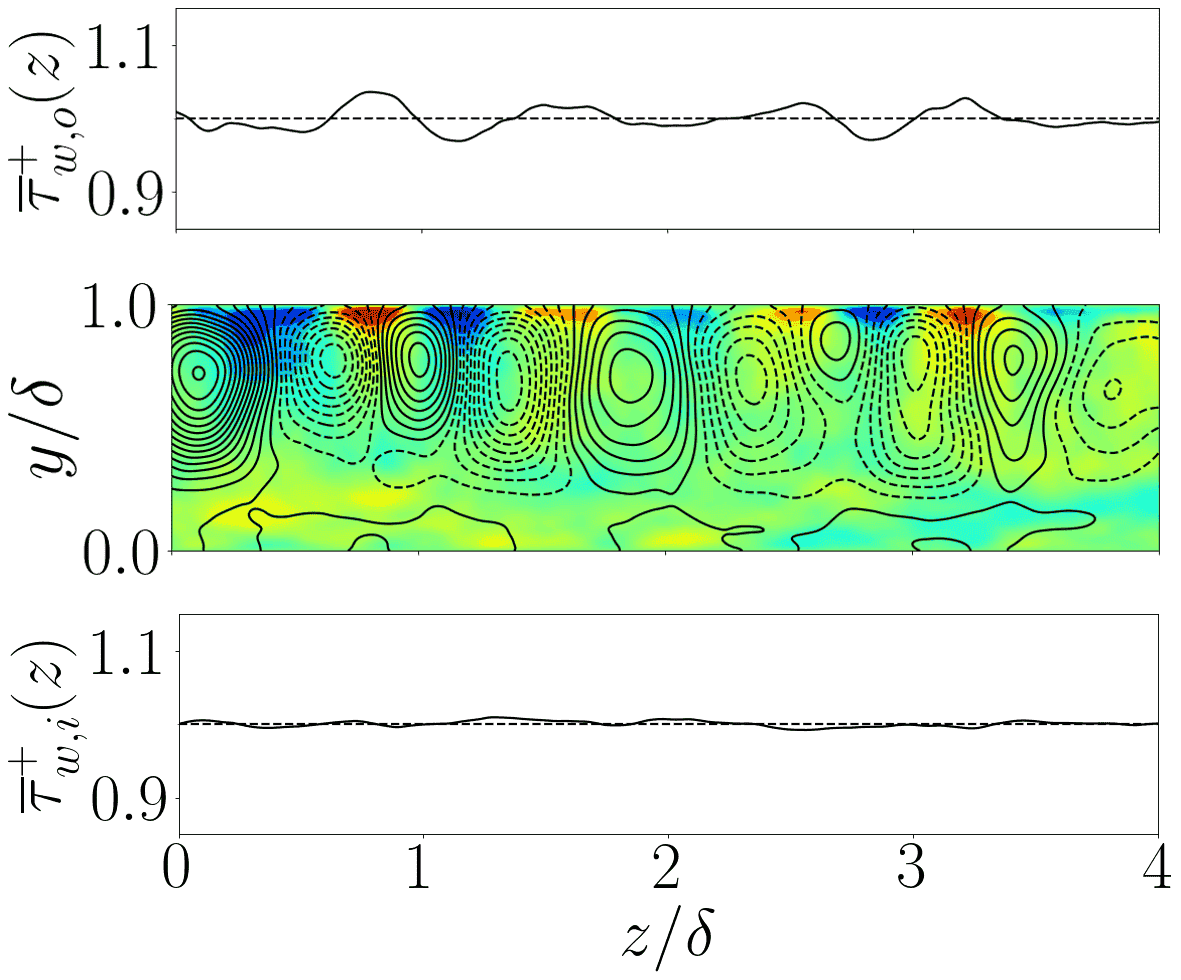}\label{rz_re4_r1}
(e)\includegraphics[width=.3\textwidth]{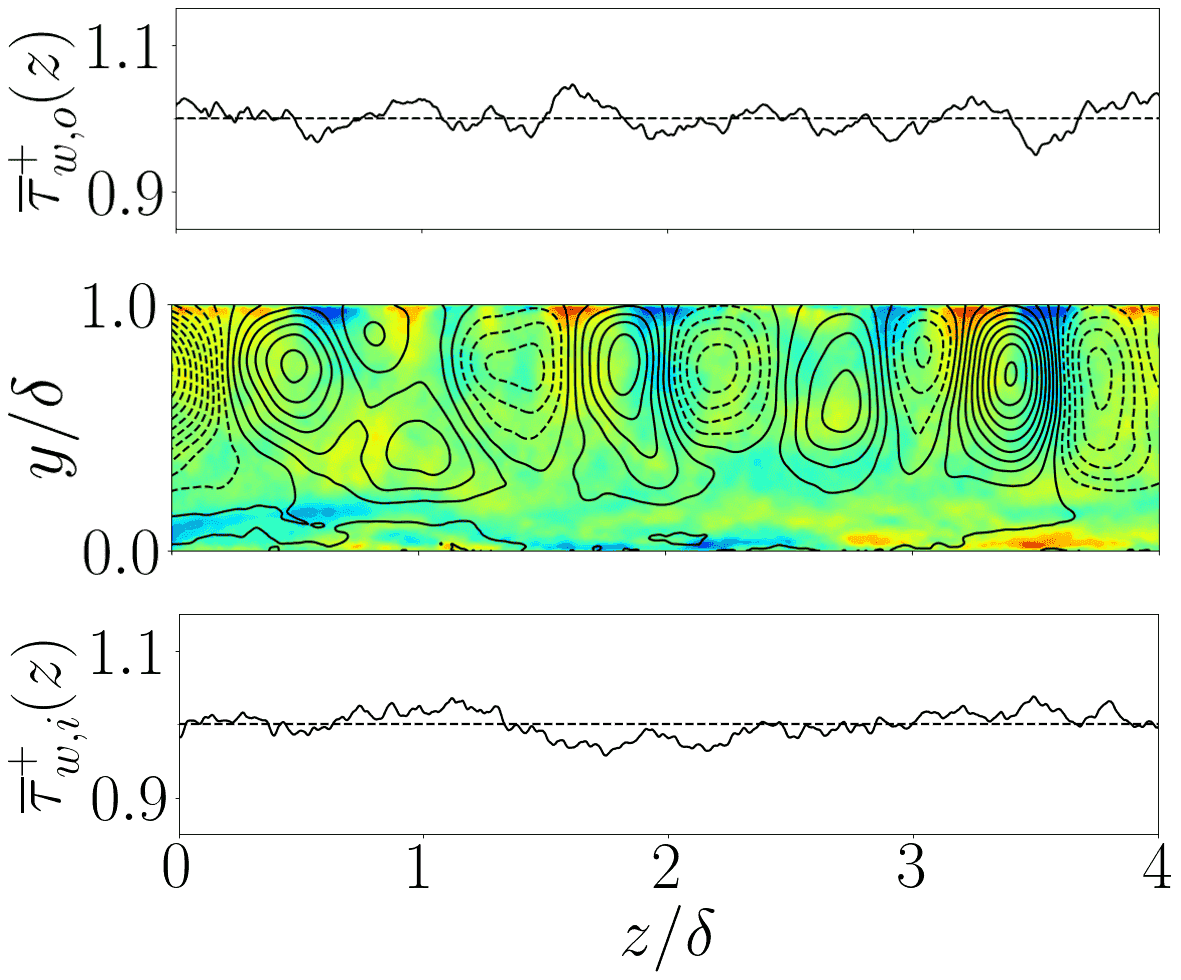}\label{rz_re20_r1}
(f)\includegraphics[width=.3\textwidth]{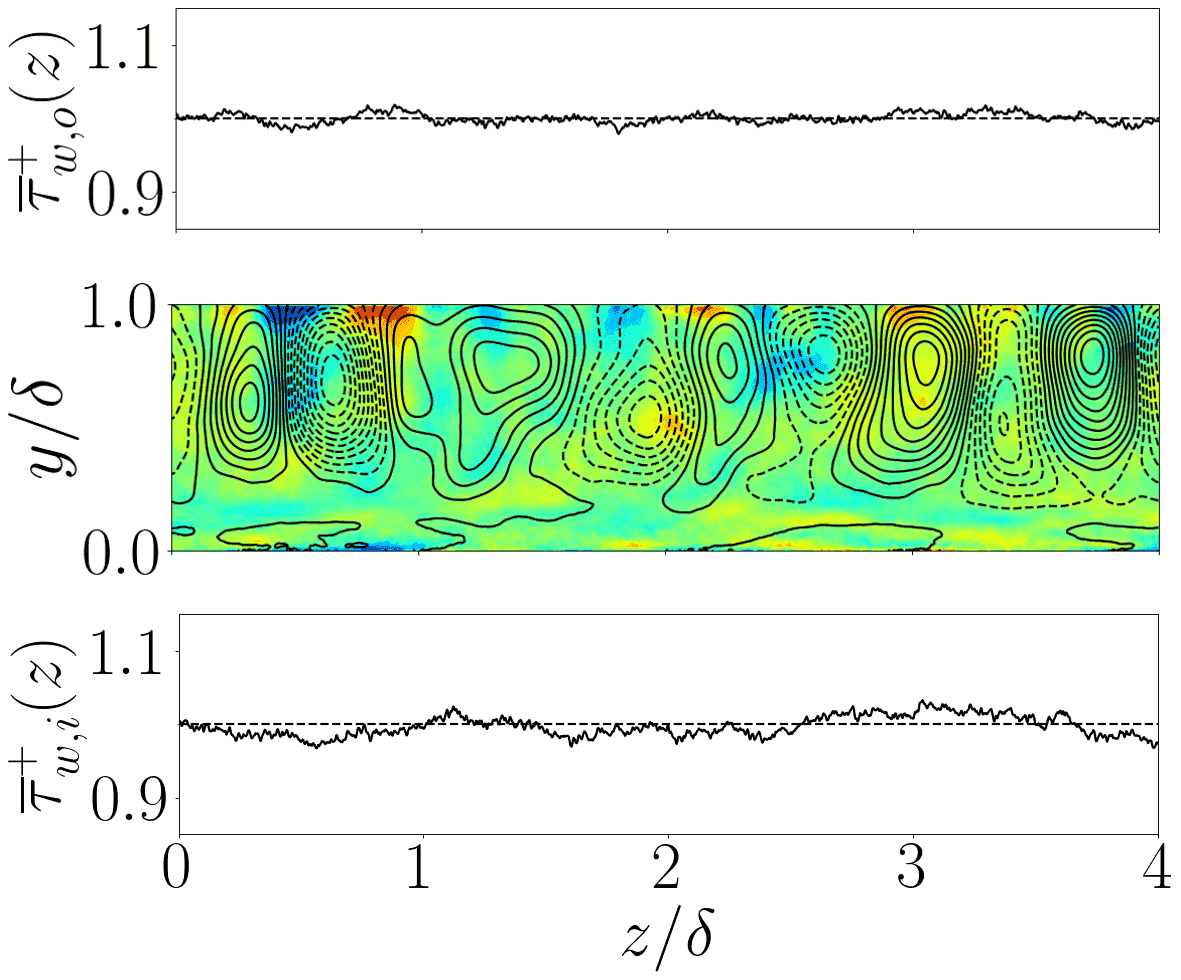}\label{rz_re87_r1}
\caption{Coherent Stokes streamfunction ($\overline{\tilde{\psi}}$) 
in an $(r,z)$-plane overlaid to flooded contours of coherent 
streamwise velocity ($\overline{\tilde{u}}^*$),
for the R40 flow cases (upper panels) and for the R1 flow cases (lower panels). 
From left to right, the panels correspond 
to $\Rey_b=4000$ (a, d), $20000$ (b, e), $87000$ (c, f). 
Positive values of ${\tilde{\psi}}$ (solid lines) indicate a 
clockwise-rotating roll cell, and vice-versa for negative values (dashed lines).
The flooded contours range from -1.5 (blue) to 1.5 (red). 
Each panel shows the spanwise distribution of the mean shear stress 
at the two walls, $\overline{\tau}_{w,i}^+(z)$ and $\overline{\tau}_{w,o}^+(z)$,
defined in~\eqref{eq:tauw}. 
Only half of the domain is shown for the R1 flow cases.} 
\label{fig:rz}
\end{figure}
The results of the eduction procedure are presented in figure~\ref{fig:rz}, 
where we show the mean coherent Stokes streamfunction 
($\overline{\tilde{\psi}}$), 
defined such that 
$\tilde{w}= {\partial \tilde\psi}/{\partial r}$, 
$\tilde{v}=-{\partial \tilde\psi}/{\partial z}$. 
The overlaid flooded contours represent the mean coherent streamwise velocity 
($\overline{\tilde{u}}^*$).
In addition, we report the spanwise distribution of the mean shear at the inner 
and outer wall, 
\begin{equation}
\overline{\tau}_{w,i}^+(z)=\frac{\nu}{u_{\tau,i}^2}\de{\langle u\rangle_{\theta,t}}{r}\bigg|_{r_i},\qquad
\overline{\tau}_{w,o}^+(z)=\frac{\nu}{u_{\tau,o}^2}\de{\langle u\rangle_{\theta,t}}{r}\bigg|_{r_o}.
\label{eq:tauw}
\end{equation}
In the R40 flow cases (upper panels) two pairs of counter-rotating vortices 
appear which nearly span the whole channel thickness and with a spanwise 
wavelength $\lambda_z/\delta\approx2$, which is in agreement with the
energy spectra in figure~\ref{fig:usp40}. 
Longitudinal roll cells with the same size were detected by~\cite{moser1987effects}. 
The spanwise inhomogeneity due to these secondary eddies
has a strong impact on the wall shear. 
Indeed, large-scale ejections are generated
between any pair of counter-rotating vortices 
(blue contours), 
where low-speed fluid is diverted away from the wall, and
local friction attains a minimum.
Correspondingly, large-scale sweeps generate 
at the opposite wall whereby high-speed fluids is pushed towards the wall 
(red contours), causing local increase of wall friction.
This tendency is clearer as the Reynolds number increases, 
as one can infer from the spanwise distribution of the local wall shear,
which shows excursions of about $10\%$ at $\Rey_b=4000$ (a) 
and $\Rey_b=20000$ (b), and reaching up to $20\%$ at $\Rey_b=87000$ (c). 
 
As for the R1 flow cases (lower panels), 
the number of pairs of counter-rotating vortices increases to about ten  
(only five are visible as half of the domain is shown).  
Hence, the spanwise wavelength of the roll cells is $\lambda_z/\delta\approx0.8$, 
which corresponds to what found from the spectral analysis 
in figure~\ref{fig:vsp}. 
Spanwise shortening of the longitudinal roll cells is also observed in  
rotating channel flows, in which similar vortices develop on account of Coriolis 
forces~\citep{matsson1990curvature}. 
Through DNS of rotating channel flow,~\cite{brethouwer2017statistics} found that 
the size of the longitudinal roll cells is smaller at higher rotation numbers 
(the rotation number in rotating channel flow is the counterpart of the 
curvature ratio in curved channels), in agreement with previous numerical 
studies~\citep{kristoffersen1993direct,yang2012channel}.
In addition,~\cite{brethouwer2017statistics} found that the size of the roll 
cells is independent of the Reynolds number, which is also the case here. 
In fact, due to strong channel curvature, the roll cells are pushed 
towards the outer wall  and cannot fill the entire channel height. 
Hence, the flow region near the inner wall (say, $y/\delta<0.2$) 
is not affected, as seen in the spanwise distribution of the wall shear,
which is nearly flat. As for the shear at the outer wall, spanwise excursions 
of $\tau_{w,o}^+(z)$ have a maximum amplitude of about $4\%$ at $\Rey_b=4000$~(d) 
and $\Rey_b=20000$~(e), whereas they are very small at $\Rey_b=87000$~(f), 
at which the effect of the longitudinal vortices is outweighed by turbulence.   
%
The reason why the outer-wall shear is barely affected by longitudinal 
vortices is their unsteadiness. 
The coherent streamfunction shows indeed that roll cells 
are less organised in the R1 flow cases than in R40, 
and on average they do not affect the streamwise velocity. 
Nonetheless, the impact of those eddies on the instantaneous flow is certainly 
not negligible in the case of strong curvature.  
The strength of longitudinal large-scale structures can be 
quantified in terms of the maximum amplitude of the coherent wall-normal 
velocity~\citep{canton2016large}, which results in 
$\text{max}|\tilde{v}|/u_b\approx6\%$ for the R40 flow cases, 
nearly independent of the Reynolds number. 
Longitudinal vortices in the R1 flow cases are much stronger, 
their strength being
$\text{max}|\tilde{v}|/u_b\approx20\%$ at $\Rey_b=4000$, 
$14\%$ at $\Rey_b=20000$, $12\%$ at $\Rey_b=87000$. 

\subsection{Splitting and merging events}\label{sec:split}
%
\begin{figure}
\centering
\includegraphics[width=\textwidth]{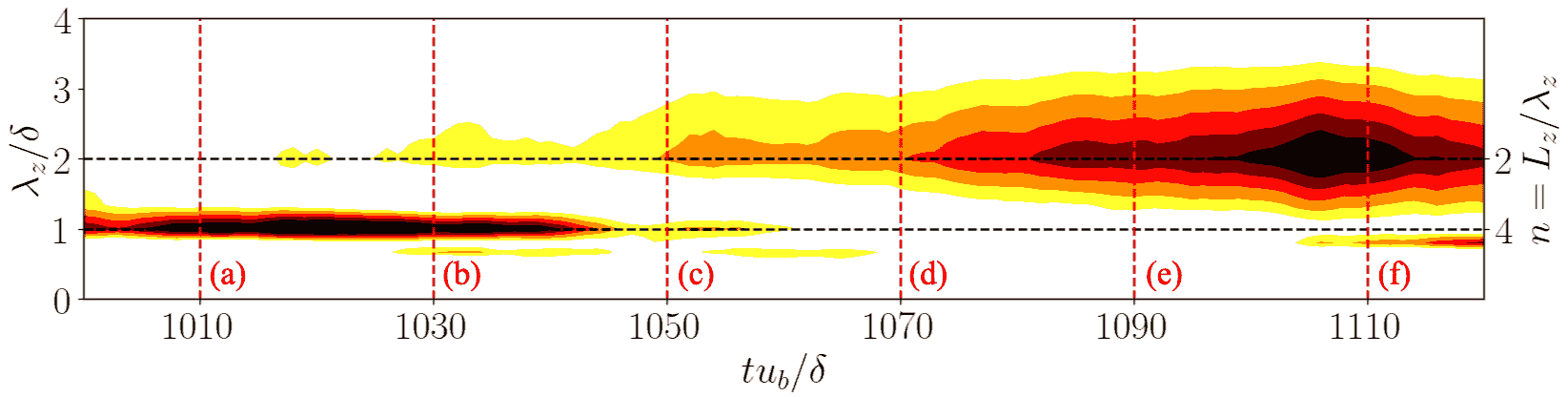}\\
\caption{
Time history of the pre-multiplied spanwise energy spectra of fluctuating 
streamwise velocity ($k_z^*E^*_{uu}$) for the R40 flow case at $\Rey_b=87000$,  
fixed the wall distance at $y/\delta=0.8$. 
The black dashed lines correspond to $\lambda_z/\delta=2$ and 
$\lambda_z/\delta=1$, 
at which the expected number of longitudinal vortices 
are indicated on the right vertical axis.} 
\label{fig:sptime40}
\end{figure}
\begin{figure}
\centering
(a)\includegraphics[width=.30\textwidth]{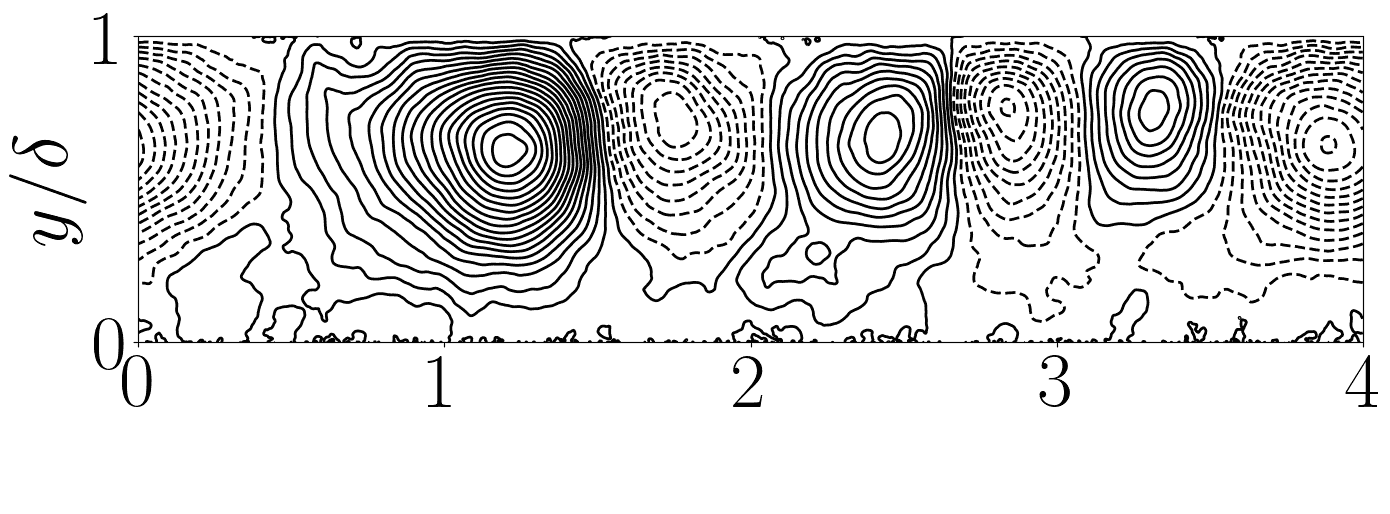}
(b)\includegraphics[width=.30\textwidth]{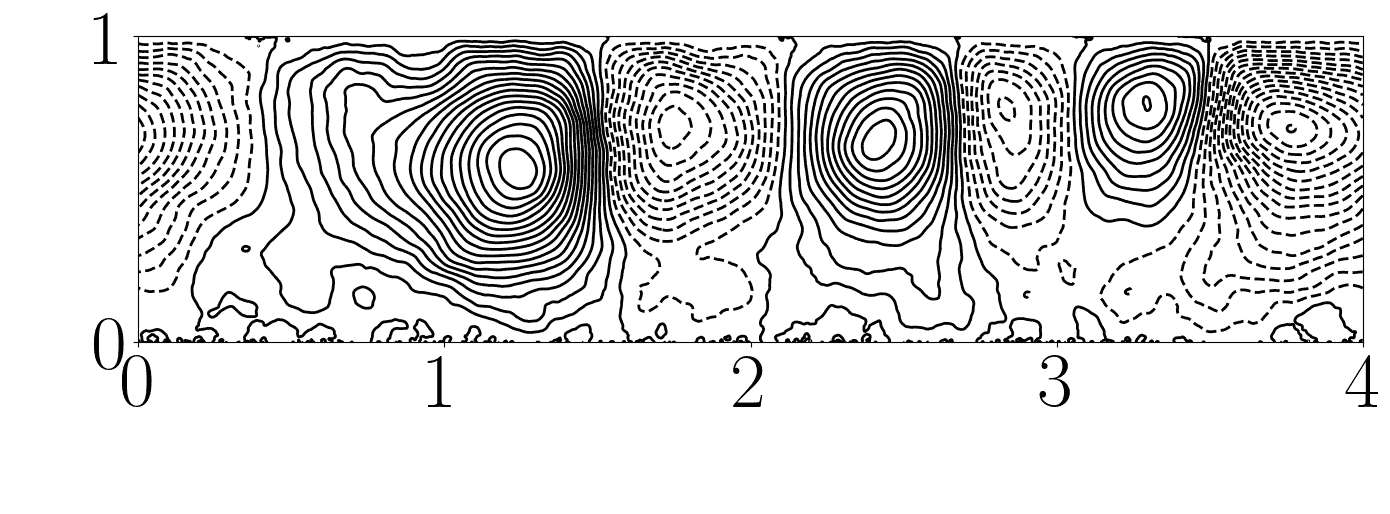}
(c)\includegraphics[width=.30\textwidth]{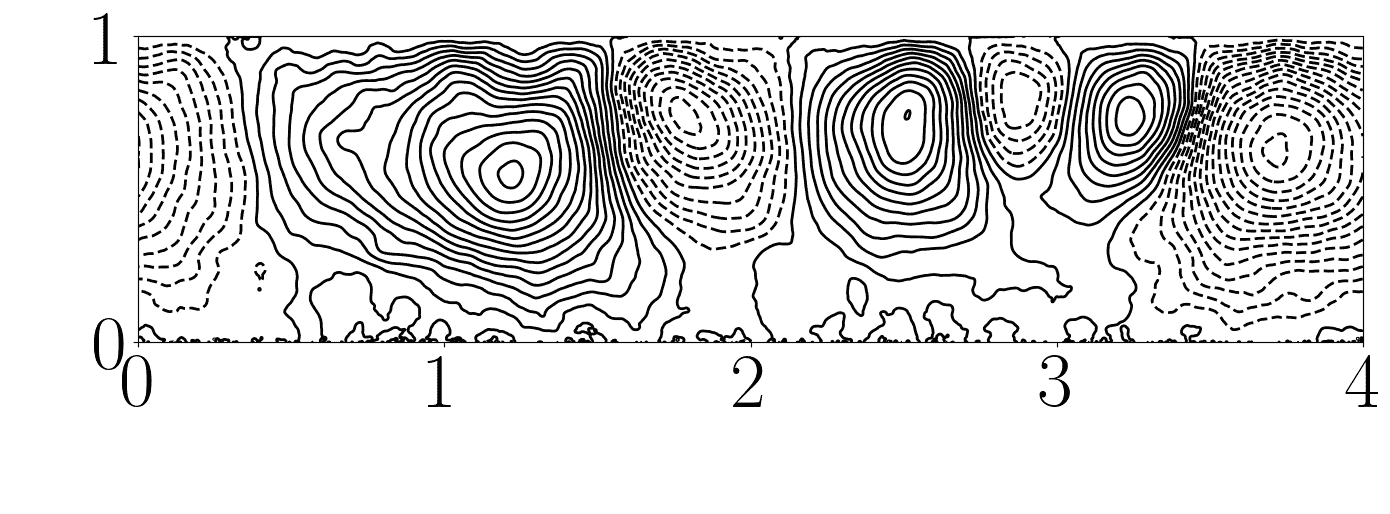}\\
(d)\includegraphics[width=.30\textwidth]{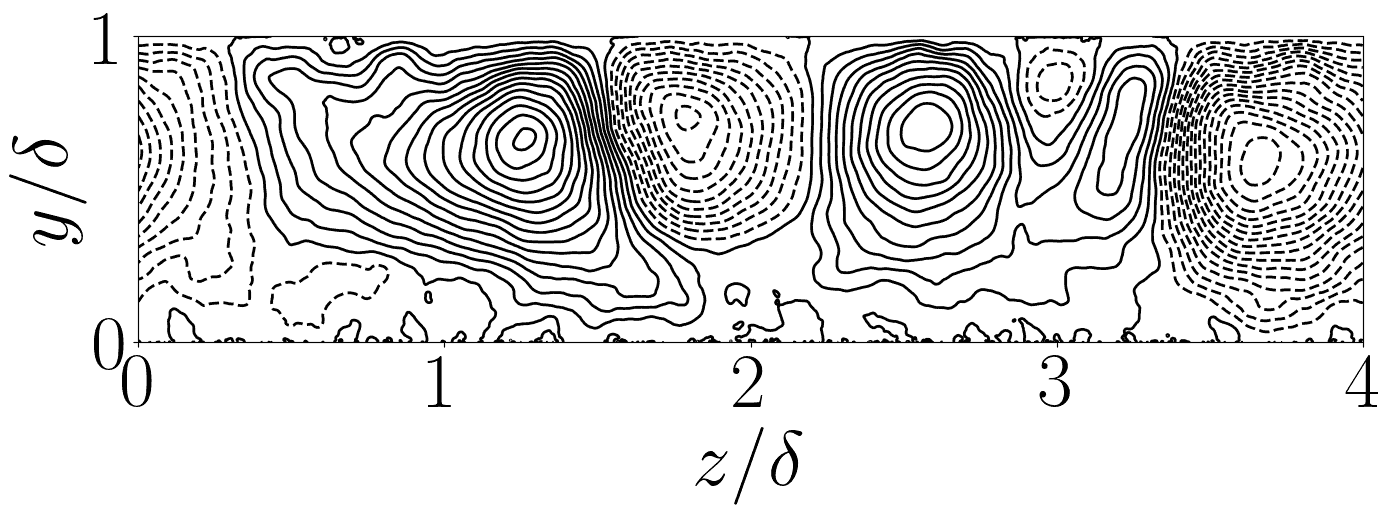}
(e)\includegraphics[width=.30\textwidth]{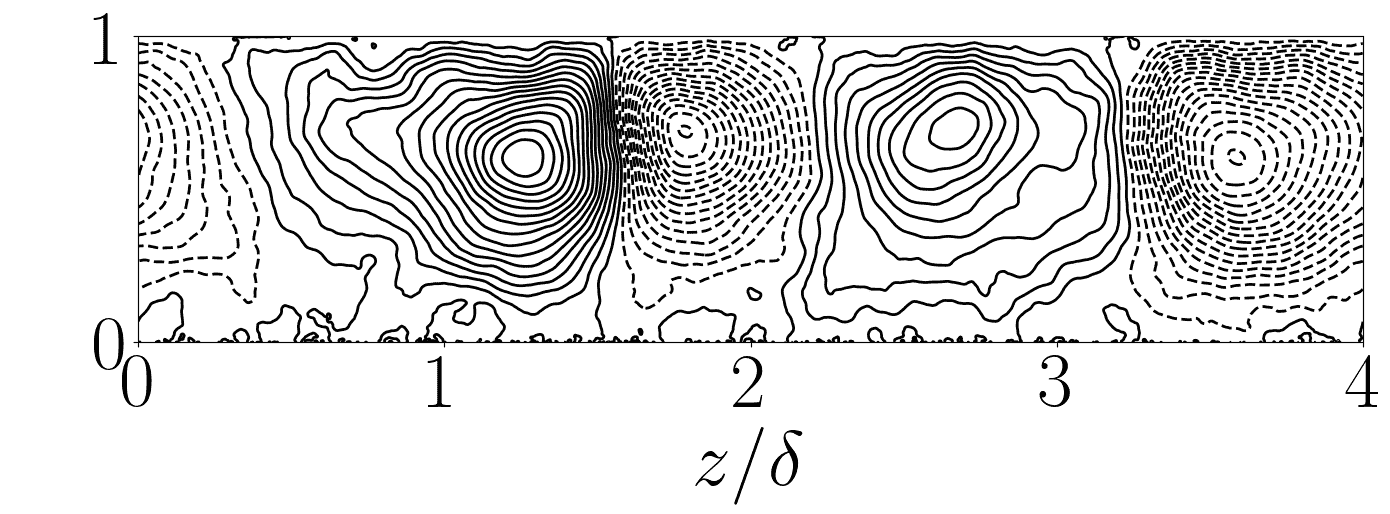}
(f)\includegraphics[width=.30\textwidth]{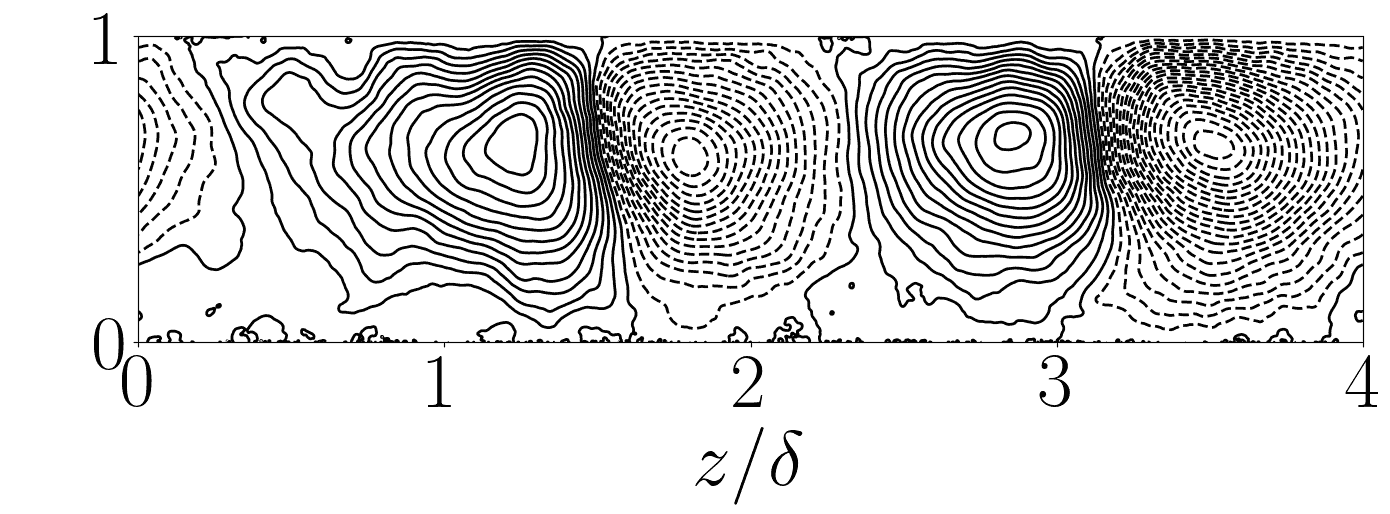}
\caption{
Coherent streamfunction (${\tilde{\psi}}$) at the time instants marked by 
the red dashed lines in figure~\ref{fig:sptime40},
for the R40 flow case at $\Rey_b=87000$.}
\label{fig:stokes40}
\end{figure}
The eduction procedure based on triple decomposition 
allowed to quantify the mean configuration of longitudinal 
large-scale structures. However, the energy spectra highlighted 
the presence of multiple energy peaks at large wavelengths, 
which are particularly evident for the R40 flow case at $\Rey_b=87000$ 
(figure~\ref{fig:usp40}). Multiple peaks suggest either 
the coexistence of large-scale structures of different sizes, 
or the occurrence of splitting and merging events. 
To clarify this point, we inspect the time history of the spanwise 
energy spectra of streamwise velocity fluctuations, shown in 
figure~\ref{fig:sptime40} for the R40 flow case at $\Rey_b=87000$,  
%
at the wall distance where the peak of the energy associated with the 
longitudinal vortices occurs.
A time window from $tu_b/\delta=1000$ to $1120$ was selected for the analysis 
and we took six subsequent snapshots of ${\tilde{\psi}}$, shown in 
figure~\ref{fig:stokes40}, at intervals of $20\delta/u_b$ 
marked by the red dashed lines in figure~\ref{fig:sptime40}. 
%
From $tu_b/\delta\approx1000$ to $1040$ most energy is clustered 
around $\lambda_z/\delta=1$, hence $n=4$ pairs of vortices would be expected 
(the number of vortices pairs is indicated on the right vertical axis). 
However, the snapshots of ${\tilde{\psi}}$ at 
$tu_b/\delta\approx1010$ (a) and $1030$ (b) reveal the occurrence 
of $n=3$ pairs. 
In this respect, we note that the roll cells are not uniform in size, 
as the pair centred at $z/\delta\approx0.5$ has a wavelength 
$\lambda_z/\delta\approx2$, whereas the two remaining pairs 
have a wavelength $\lambda_z/\delta\approx1$. A possible explanation 
is that the two pairs of small-size vortices are stronger than the large-size one, 
hence more energy is concentrated at $\lambda_z/\delta=1$. 
A vortex-merging event is observed between $tu_b/\delta\approx1050$ and $1090$. 
Panels~(c) and (d) 
depict that the pair of smallest vortices located at $z/\delta\approx3$ 
decrease in size and strength, as the iso-lines of the streamfunction get sparser.
This process continues until 
the smallest pair is embedded within the adjacent clockwise rotating vortex~(e), 
which appears more regular and strong at $tu_b/\delta\approx1110$~(f). 
From $tu_b/\delta\approx1080$ on, the energy peak is clustered around 
$\lambda_z/\delta=2$ and, as expected, $n=2$ pairs of roll cells are found.  

\begin{figure}
\centering
\includegraphics[width=\textwidth]{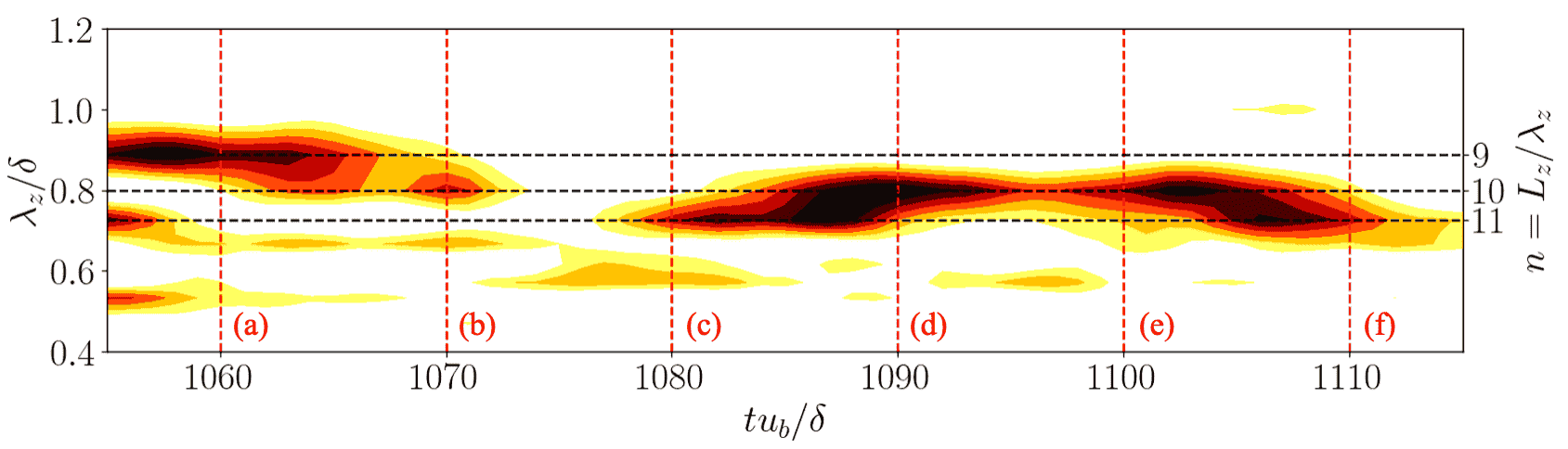}\\
\caption{
Time history of the pre-multiplied spanwise energy spectra of fluctuating 
wall-normal velocity ($k_z^*E^*_{vv}$) for the R1 flow case at $\Rey_b=87000$, 
fixed the wall distance at $y/\delta=0.7$. 
The black dashed lines correspond to 
$\lambda_z/\delta=0.73$, 
$\lambda_z/\delta=0.8$ and 
$\lambda_z/\delta=0.89$, 
at which the expected number of longitudinal vortices 
are indicated on the right vertical axis.} 
\label{fig:sptime1}
\end{figure}
\begin{figure}
\centering
(a)\includegraphics[width=.30\textwidth]{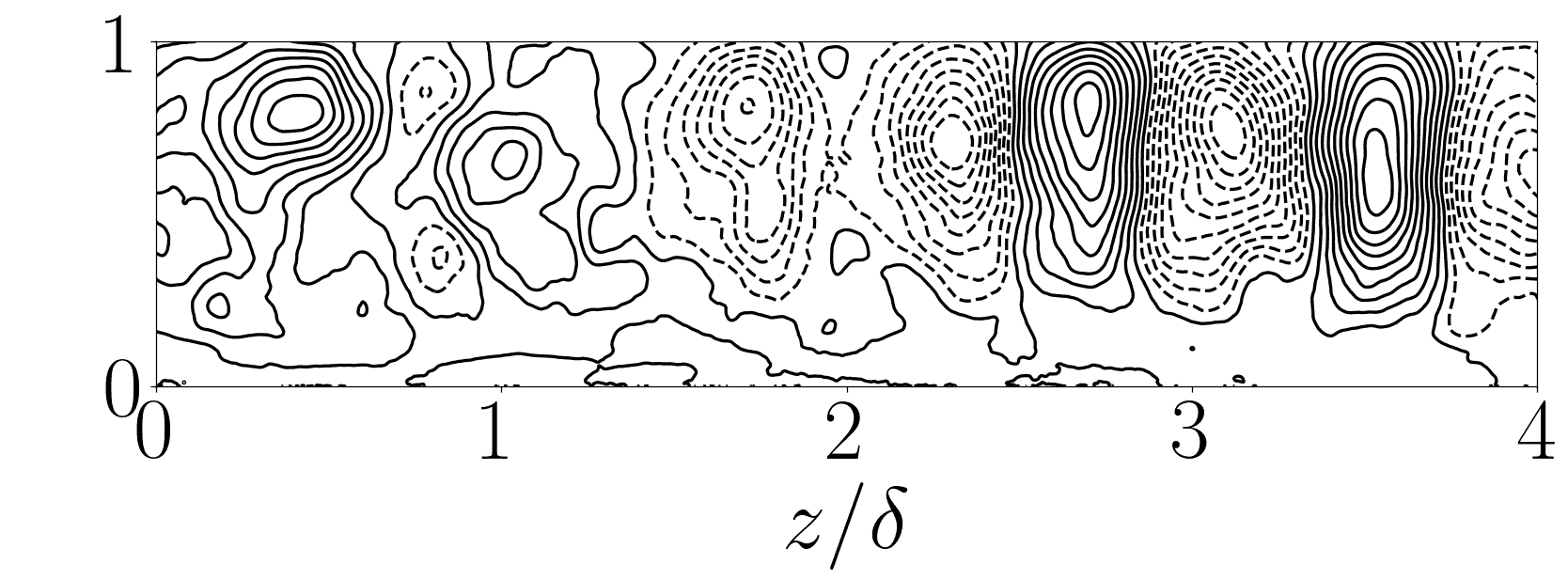}
(b)\includegraphics[width=.30\textwidth]{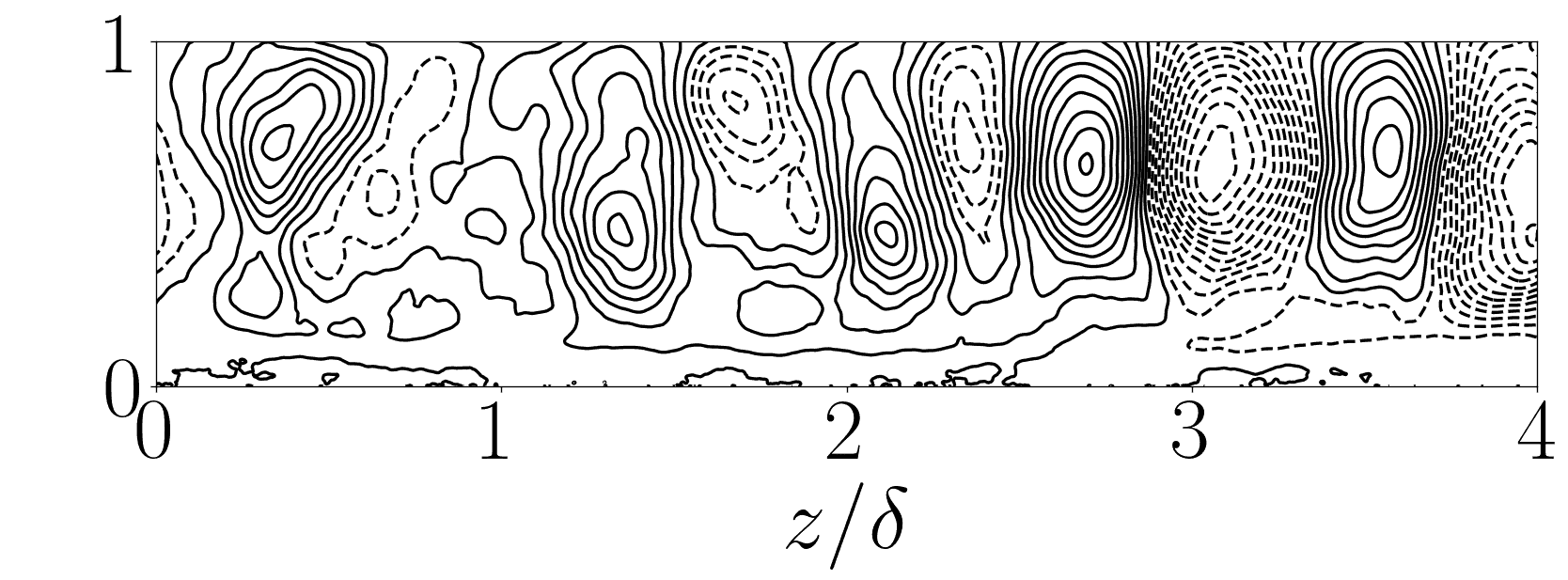}
(c)\includegraphics[width=.30\textwidth]{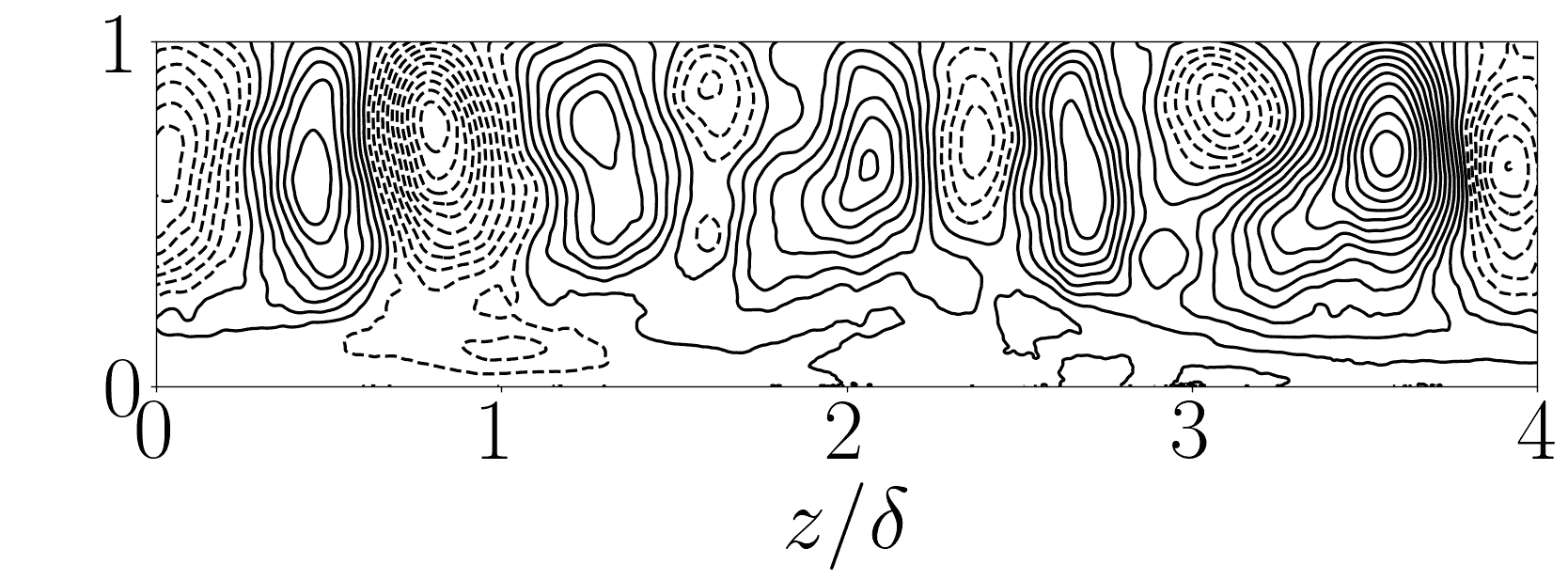}\\
(d)\includegraphics[width=.30\textwidth]{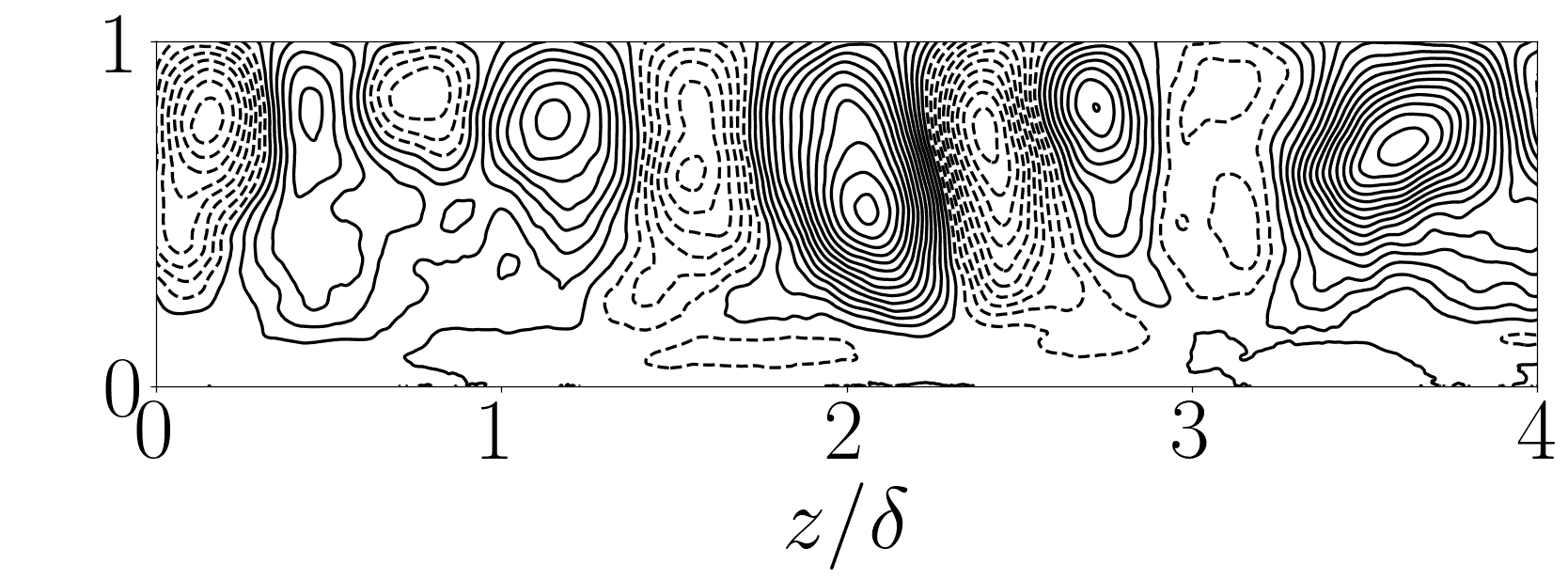}
(e)\includegraphics[width=.30\textwidth]{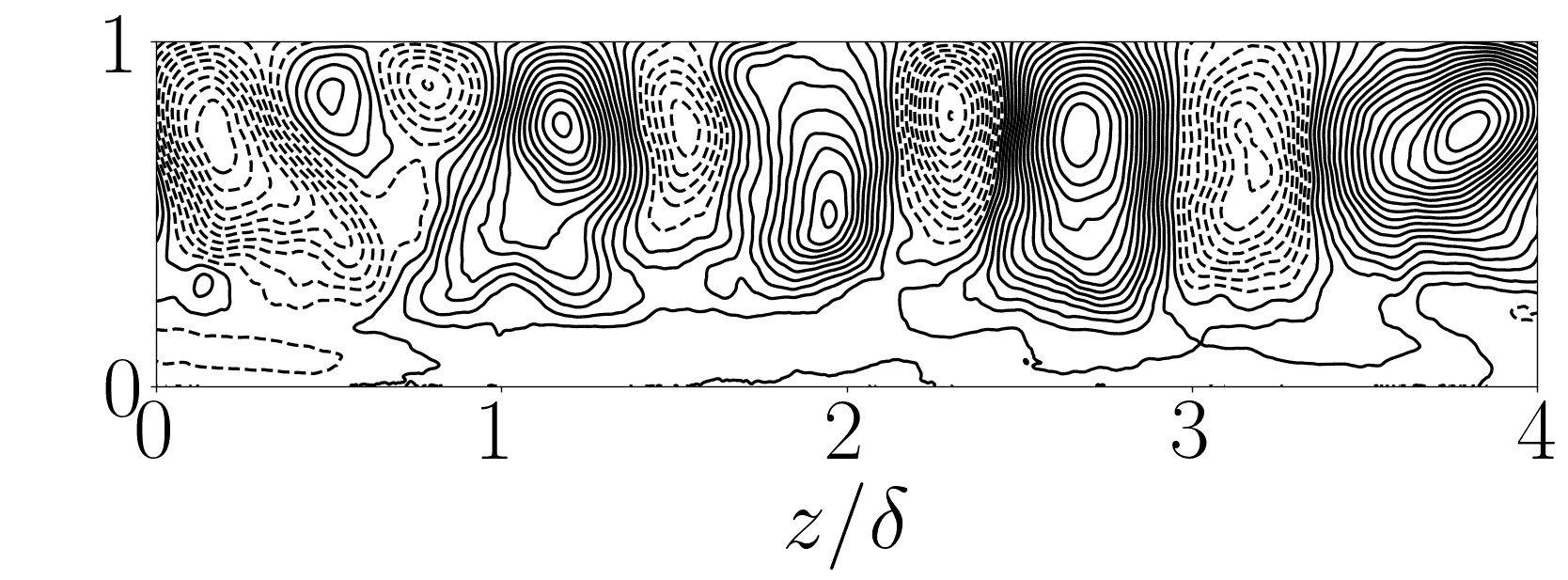}
(f)\includegraphics[width=.30\textwidth]{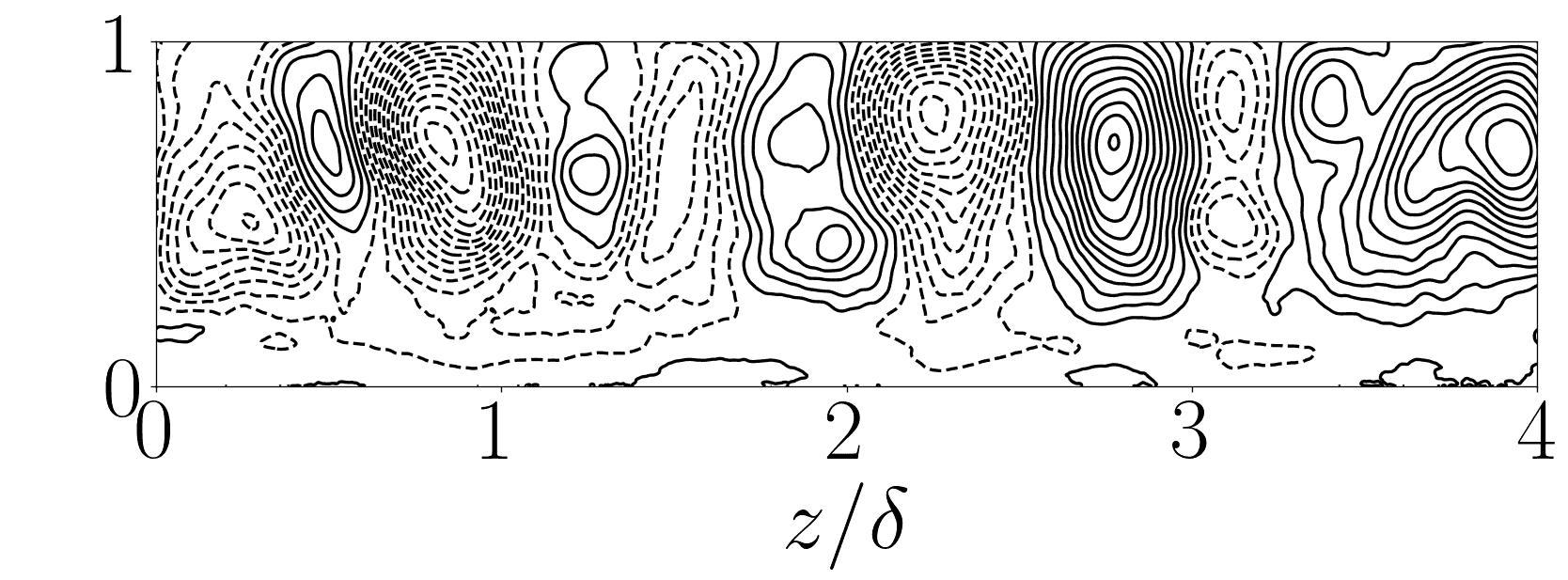}
\caption{
Coherent streamfunction (${\tilde{\psi}}$) at the time instants marked by 
the red dashed lines in figure~\ref{fig:sptime1} 
for the R1 flow case at $\Rey_b=87000$. Only half of the domain is shown.}
\label{fig:stokes1}
\end{figure}
The results obtained for the R1 flow case at $\Rey_b=87000$ are presented 
in figure~\ref{fig:sptime1}, which reports the time evolution 
of the spanwise spectra of the wall-normal velocity 
at the peak position of energy associated with
the longitudinal vortices. 
The time window under scrutiny is half as for the R40 flow case, 
and the six subsequent snapshots of ${\tilde{\psi}}$, shown in 
figure~\ref{fig:stokes1}, are sampled at intervals of $10\delta/u_b$ 
instead of $20$ on account of stronger unsteadiness in cases with
strong curvature.
The time evolution of the spectra shows that the energy peak is clustered  
around $\lambda_z/\delta\approx0.89$ from $tu_b/\delta=1055$ to $1065$.
Consistently, figure~\ref{fig:stokes1}(a) displays nine vortices
in the region under scrutiny 
($n=9$ pairs are present in the whole domain). 
This configuration seems to be unstable because of the ill-defined structure 
of the vortices pair located at $z/\delta\approx1$, and because two 
counterclockwise vortices are adjacent straddling $z/\delta\approx2$. 
At $tu_b/\delta=1070$ the energy peak shifts at $\lambda_z/\delta\approx0.8$, 
indeed the related figure~\ref{fig:stokes1}(b) shows that a new clockwise rotating 
vortex emerged at $z/\delta\approx2.2$ fitting in between the two counterclockwise 
rotating vortices and increasing the number of vortices to $10$ (i.e. $n=10$). 
The number of vortex pairs increase further to $n=11$ at $tu_b/\delta\approx1080$, 
as visible in figure~\ref{fig:stokes1}(c), and the energy peak 
decreases accordingly to $\lambda_z/\delta\approx0.73$. 
From $tu_b/\delta\approx1085$ to $1105$ the energy peak settles at 
$\lambda_z/\delta\approx0.8$, and panels (d) and (e) highlight that the 
vortex configuration consists of $n=10$ pairs. 
The energy peak shifts again to $\lambda_z/\delta\approx0.73$ at $tu_b/\delta=1110$. 
Figure~\ref{fig:stokes1}(f) shows that the main vortexes are still ten, 
however small secondary vortices tend to split off from the primary ones, 
bringing more energy to smaller scales. 
In the case of strong curvature, the transitions from one vortex configuration 
to the other can be attributed to the unsteady dynamics of the vortices, 
which can be inferred from their spanwise motions, distorted shapes 
and different sizes, rather than to splitting and merging phenomena, 
which could not be clearly identified.

\subsection{Role of longitudinal vortices on velocity fluctuations}\label{sec:fluc}

\begin{figure}
\centering
(a)\includegraphics[width=.45\textwidth]{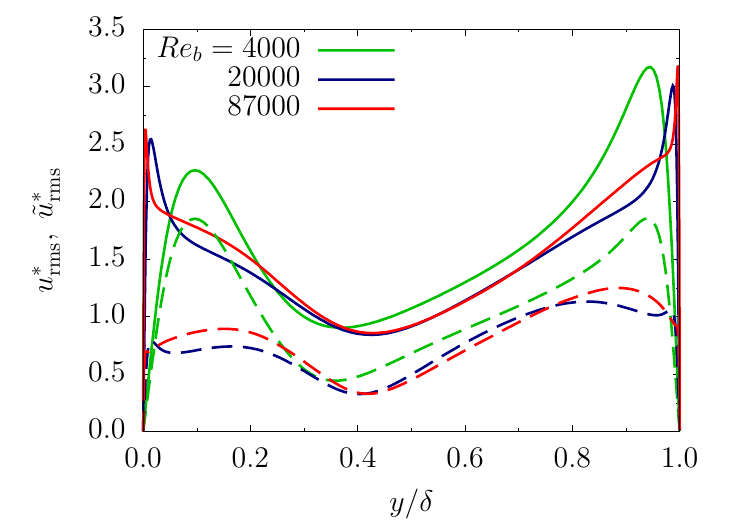}\label{urms40}
(b)\includegraphics[width=.45\textwidth]{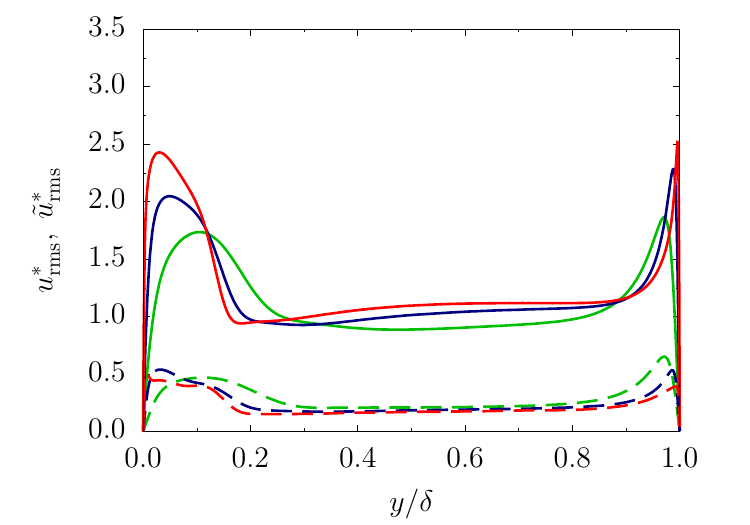}\label{urms1}\\
(c)\includegraphics[width=.45\textwidth]{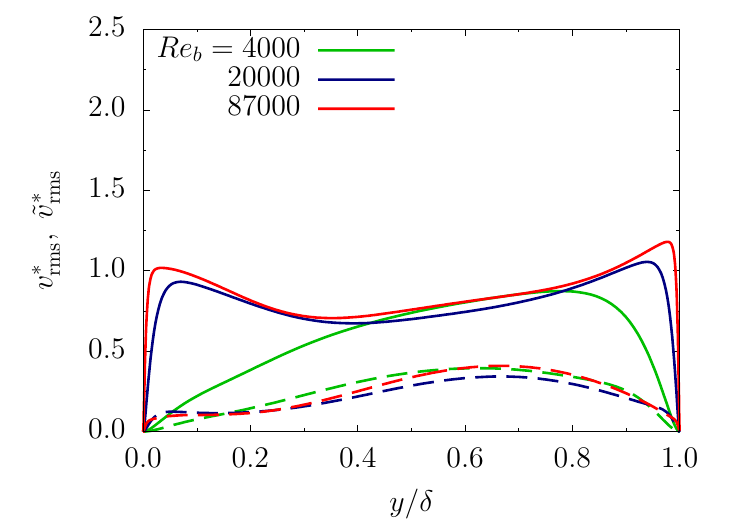}\label{vrms40}
(d)\includegraphics[width=.45\textwidth]{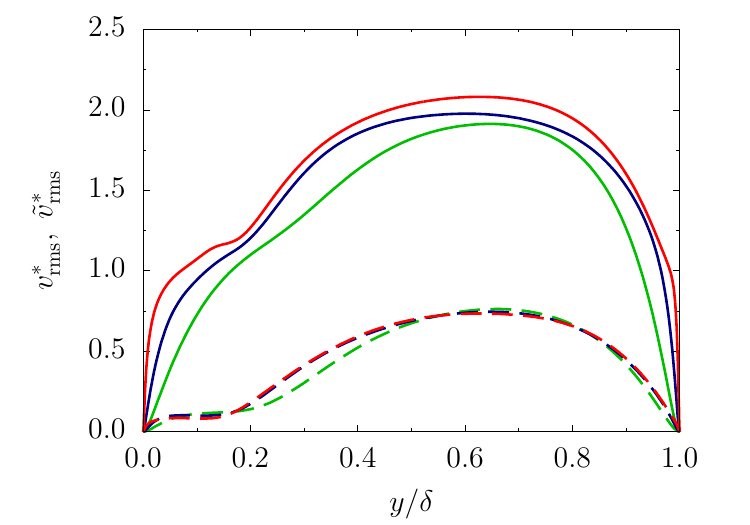}\label{vrms1}\\
(e)\includegraphics[width=.45\textwidth]{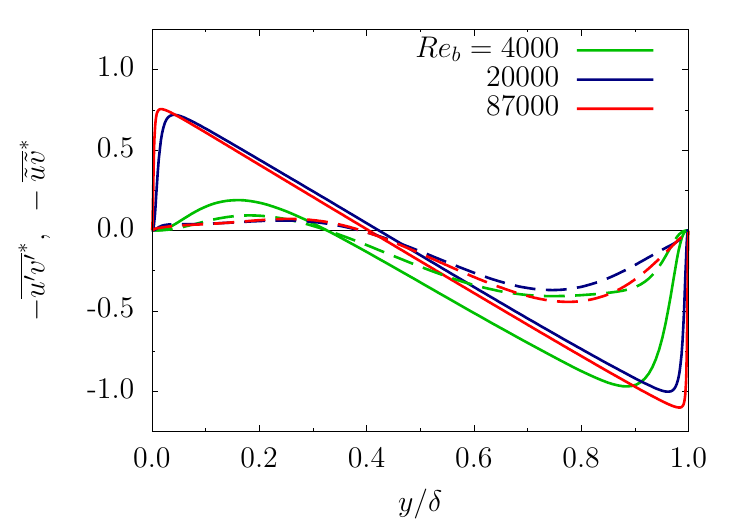}\label{uv40}
(f)\includegraphics[width=.45\textwidth]{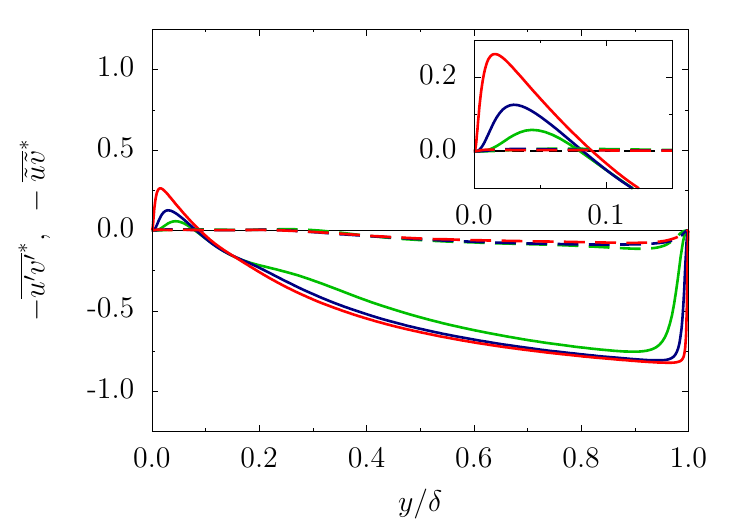}\label{uv1}\\
\caption{Profiles of root-mean-square streamwise~(a, b) 
and wall-normal~(c, d) velocity fluctuations and mean turbulent shear stress~(e, f) 
at various Reynolds numbers for the R40 flow cases (left panels) and R1 flow 
cases (right panels). Solid lines refer to total fluctuations and
	dashed lines refer to coherent fluctuations due to longitudinal vortices.}
\label{fig:rms}
\end{figure}

In figure~\ref{fig:rms} we show the root-mean-square (RMS) 
of the streamwise (a, b) and wall-normal (c, d) velocity 
fluctuations, as well as the turbulent shear stress (e, f). 
Total fluctuations are reported along with
the contributions due to the longitudinal vortices,
which we have determined by taking the root-mean-square 
of the coherent contribution~\eqref{eq:coh} 
along the spanwise direction and in time.
%
For the R40 flow cases, the streamwise velocity fluctuations~(a) 
are higher at the outer than at the inner wall. 
As made clear from the increased intensity of the coherent fluctuations, 
this asymmetry is mainly due to the 
longitudinal vortices, which are stronger near the outer wall. 
An exception is the case at $\Rey_b=4000$ (green line), 
in which the coherent contribution at the outer wall  
is comparable to that on the inner wall. 
In general, the coherent contribution is about half of the total. 
The peak of the coherent contribution occurs between 
$y/\delta\approx0.8$ and $y/\delta\approx0.9$, corresponding to a 
bump in the profile of the total fluctuations. 
As for the R1 flow cases, the profiles of streamwise velocity fluctuations~(b) 
show a near-wall peak on the concave side comparable to that 
of the corresponding R40 flow cases, 
and little contribution from coherent fluctuations. 
The profile of the velocity fluctuations is almost flat in the channel core, 
and the near-wall peak on the convex side is very different 
from the case with mild curvature, extending much farther from the wall.
%
Wall-normal velocity fluctuations are only relevant in the channel core,
with a peak at $y/\delta\approx0.7$. 
For the R1 flow cases~(d) fluctuations of wall-normal velocity
are up to two limes larger than the streamwise velocity. 
Substantial contribution is found to be provided by the coherent fluctuations, 
which also attain a peak at $y/\delta\approx 0.7$, as observed in the
velocity spectra.

As for the turbulent shear stress, it is nearly symmetrical 
in fully-turbulent R40 flow cases~(e), 
resembling the case of a plane channel~\citep[see e.g.][]{kim1987turbulence}. 
However, two differences should be noted: 1) the point of zero crossing
is shifted towards the inner wall, at $y/\delta\approx0.4$, and 
2) the peak value increases near the outer wall 
and decreases near the inner wall.  
The effect of convex curvature is particularly evident 
at $\Rey_b=4000$, at which the zero crossing is even closer 
to the inner wall. 
The coherent turbulent stress is about half of the total at 
$y/\delta\approx0.8$, pointing to significant contribution of 
longitudinal vortices to momentum transport. 
Similar conclusions were also reached by~\cite{moser1987effects} 
and~\cite{brethouwer2022turbulent}. 
The effects of strong curvature are substantial, as shown in panel (f).  
In the channel core, where viscous effects are negligible, 
the turbulent stress profile is no longer linear but rather quadratic, 
as after the analytical distribution of the total shear 
stress~\eqref{eq:tost}. As seen in the inset of the panel~(f), 
the peak value is greatly reduced near the inner wall, 
which is in accordance with the findings reported by \cite{so1973experiment} 
from experiments on turbulent boundary layers over a convex surface. 
The reason of the Reynolds stress reduction near the inner wall will be 
investigated in the following. The coherent shear stress 
becomes very small near the inner wall, since the longitudinal vortices 
are pushed towards the outer wall, as explained in \S\ref{sec:long}. 
  
\subsection{Transverse large-scale structures}\label{sec:transv}
%
Visualisations of the flow near the inner wall of strongly 
curved channels (figure~\ref{fig:ufluc1}) revealed the presence of 
alternating regions of positive/negative velocity fluctuations 
elongated along the spanwise direction. Those wavy patters are the 
footprint of transverse large-scale structures, which are originated 
from streamwise instabilities~\citep{finlay1988instability} 
and which are convected at the mean flow speed~\citep{matsson1992experiments}. 
As we did for the longitudinal vortices, we exploit triple 
decomposition to separate the effects of the cross-stream structures 
from those of turbulence. 
A field variable, $\varphi(\theta,r,z,t)$, is decomposed as
\begin{equation}
\varphi(\theta,r,z,t)={\Phi}(r)+\hat{\varphi}(\theta,r,t)+\varphi'''(\theta,r,z,t)
\end{equation}
where 
\begin{equation}
\hat{\varphi}(\theta-tu_c/r,r,t)=\langle\varphi(\theta,r,z,t)\rangle_z-{\Phi}(r) 
\label{eq:trcoh}
\end{equation}
is the contribution of the transverse structures, 
$\langle\varphi(\theta,r,z,t)\rangle_z$ is the average along the 
spanwise direction, and $\varphi'''(\theta,r,z,t)$ is the instantaneous 
turbulent fluctuation. Since the transverse structures are advected with the 
flow, phase alignment is required to educe the associated coherent 
contribution.
The time averages are evaluated by shifting each subsequent $z$-averaged field 
by an angle $\Delta\theta=\Delta tu_c/r_c$ 
where $\Delta t=\delta/u_b$ 
is the inverse of the sampling rate, $u_c$ is the convection velocity 
and $r_c$ is the curvature radius. 
The convection velocity was preliminarily estimated as the speed needed to retain maximum
coherence in time, which resulted in $u_c\approx0.67u_b$. 
This result was corroborated from the analysis of the 
wavenumber-frequency spectra of the streamwise velocity fluctuations 
for the R1 flow case at $\Rey_b=4000$ 
(further details are given in the appendix~\ref{app:conv}). 
When scaled by the friction velocity at the inner wall, 
the convection velocity at $\Rey_b=4000$ is $u_c^+\approx11$, which is 
comparable with the convection velocity of the near-wall energy-containing 
eddies~\citep{kim1993propagation, jimenez2001turbulent}.

\begin{figure}
\centering
(a)\includegraphics[width=.3\textwidth]{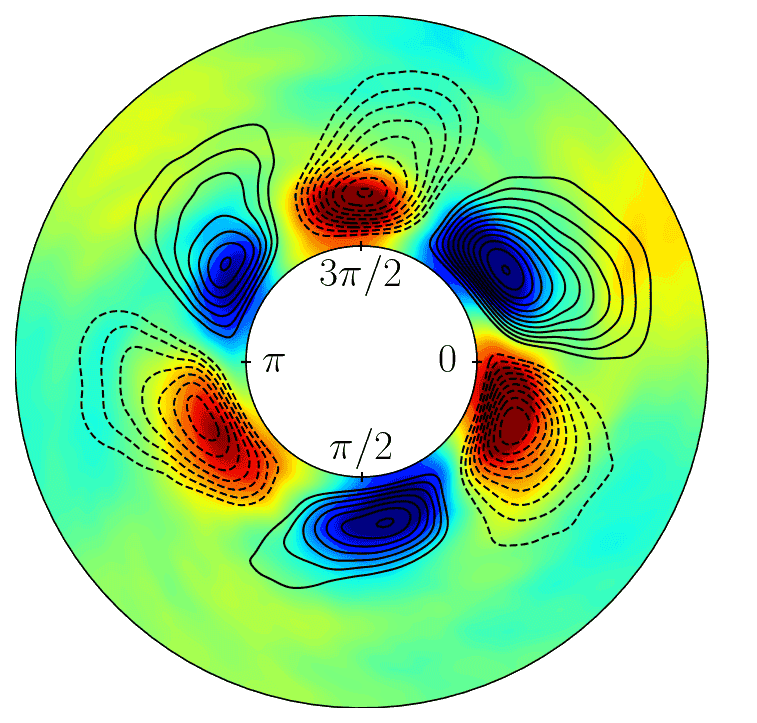}
(b)\includegraphics[width=.3\textwidth]{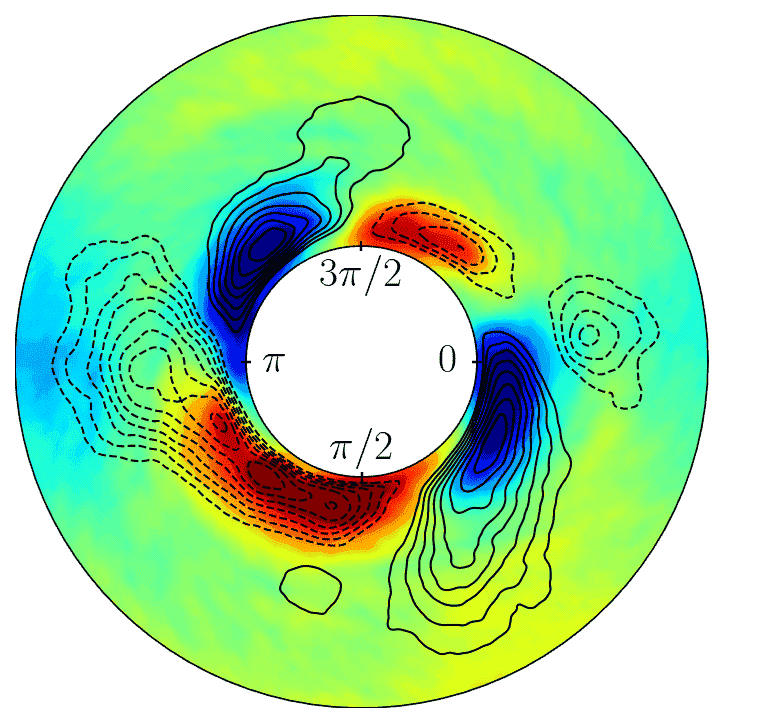}
(c)\includegraphics[width=.3\textwidth]{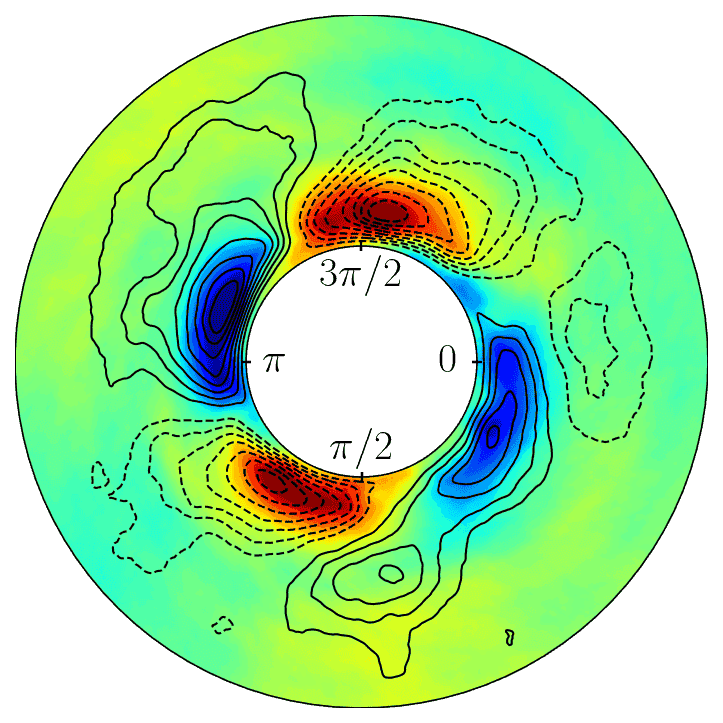}
\caption{Mean coherent Stokes streamfunction ($\overline{\hat{\psi}}$)  
overlaid to flooded contours of the mean coherent pressure ($\overline{\hat{p}^*}$)
in a $(\theta,r)$-plane for the R1 flow cases at $\Rey_b=4000$~(a), $20000$~(b), 
$87000$~(c). 
Positive values of ${\hat{\psi}}$ (solid lines), indicate a clockwise-rotating
roll cell, associated with negative coherent pressure (blue contours), 
whereas negative (dashed lines) correspond to counter-clockwise rolls and
positive coherent pressure (red contours).
The mean flow is clockwise.} 
\label{fig:tr}
\end{figure}
In figure~\ref{fig:tr} we show the mean coherent Stokes streamfunction 
($\overline{\hat{\psi}}$),  defined such that 
$\hat{v}= ({\partial \hat\psi}/{\partial \theta})/r$, 
$\hat{u}=-{\partial \hat\psi}/{\partial r}$, 
overlaid to flooded contours of the mean coherent pressure 
($\overline{\hat{p}^*}$) in a $(\theta,r)$-plane. 
Alternating high- and low-pressure regions are observed, marking the presence of 
the transverse large-scale structures. Those are organised into three pairs 
of roll cells at $\Rey_b=4000$~(a), whereas only two pairs are found at 
$\Rey_b=20000$~(b) and $87000$~(c). This result is in agreement with 
our interpretation of the streamwise energy spectra (figure~\ref{fig:usp1}). 
Although the radial extension of the transverse structures is comparable with
$\delta$, they are most intense near the inner wall and weaker towards 
the outer wall. The streamfunction shows that the cross-stream structures 
have an irregular shape and tend to split at high Reynolds number. 
This more chaotic organisation yields reduced strength of the 
transverse structures, which can be measured by the maximum amplitude 
of the coherent streamwise velocity, $\mathrm{max}|\hat{u}|$.  
Indeed, we found $\mathrm{max}|\hat{u}|/u_b\approx9\%$ at $\Rey_b=4000$ and 
$\mathrm{max}|\hat{u}|/u_b\approx7\%$ at $20000$ and $87000$,  
hence transverse large-scale structures are about half as strong as 
longitudinal (see~\S\ref{sec:long}). The centres of the roll 
cells are approximately at the same location as the local minima of the 
mean shear rate (figure~\ref{fig:shear}), 
supporting the idea that the transverse structures originate from a 
shear-layer instability~\citep{finlay1988instability, yu1991secondary}. 

\begin{figure}
\centering
(a)\includegraphics[width=.47\textwidth]{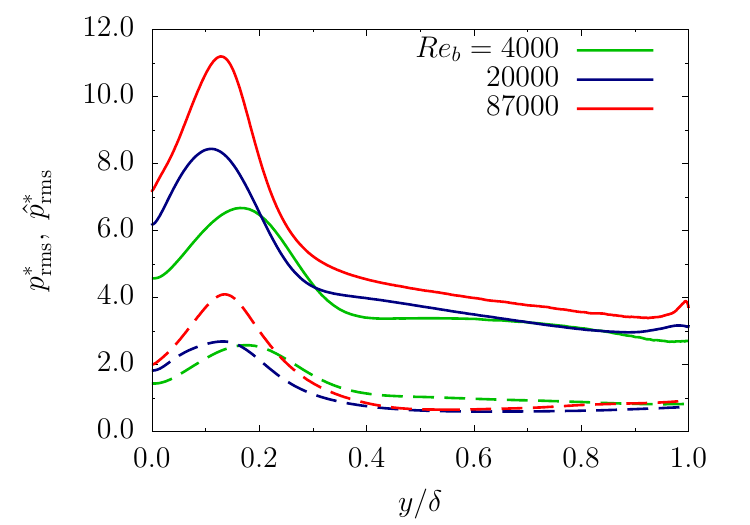}
(b)\includegraphics[width=.47\textwidth]{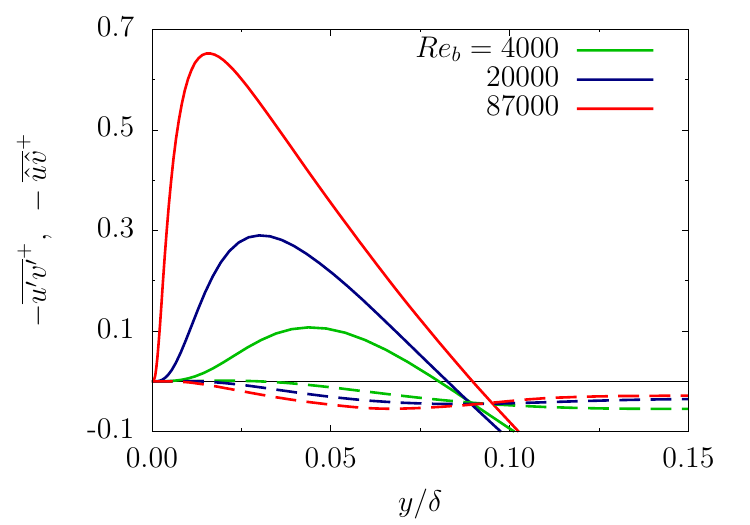}\\
\caption{Profiles of RMS pressure fluctuations (a) 
and mean turbulent stress near the inner wall (b) at various 
Reynolds numbers, for the R1 flow cases. Solid lines refer to total fluctuations 
and dashed lines refer to coherent fluctuations due to transverse large-scale 
structures (the latter are denoted with `hat').} 
\label{fig:prms}
\end{figure}
Figure~\ref{fig:prms}(a) displays the distributions of the RMS pressure fluctuations 
for the R1 flow cases, and includes the contribution of coherent 
fluctuations due to the transverse structures,
which are obtained from the spanwise-coherent contribution as from equation~\eqref{eq:trcoh}.
Near the inner wall, pressure fluctuations attain a peak which is twice 
as high as the outer-wall at $\Rey_b=4000$, and almost four times 
as high at $\Rey_b=87000$.  
The impact of the transverse coherent structures on pressure fluctuations is 
substantial, as the peak value of the coherent contribution 
near the inner wall is about one third of the total. 
Another effect of spanwise structures, which is related to the increase of 
pressure fluctuations, is the suppression of the turbulent shear stress. 
This can be ascertained in figure~\ref{fig:prms}(b), where we show the 
total turbulent stress (solid lines) and the coherent turbulent stress 
due to transverse structures (dashed lines). The results is that the coherent 
turbulent stress yields a negative contribution, meaning that the transverse 
structures tend to suppress ejections and sweeps near the inner wall. 
A similar result was reported by~\cite{kuwata2022dissimilar} simulating a 
turbulent flow over high-aspect-ratio longitudinal ribs, in which case 
the spanwise large-scale structures (originated by a Kelvin-Helmholtz instability) 
were found to increase locally pressure fluctuations and suppress the turbulent 
shear stress. 

\subsection{Role of transverse structures on wall shear and pressure}\label{sec:wall}
%
Suppression of the turbulent shear stress associated with spanwise-coherent
structures was found to yield frictional drag reduction by many 
authors~\citep{koumoutsakos1999vorticity, fukagata2005feedback, mamori2011drag}. 
In addition, experiments on turbulent boundary layers revealed that 
characteristic and identifiable variation of the wall pressure accompanies 
the advection of large organised structures, which are
responsible for variation of the wall shear stress~\citep{thomas1983role}. 
To understand whether transverse large-scale structures play a role 
in drag reduction at the inner wall, 
we then investigate if strong pressure fluctuations due to the spanwise-coherent
structures have an impact on friction at the inner wall.
\begin{figure}
\centering
(a)\includegraphics[width=.3\textwidth]{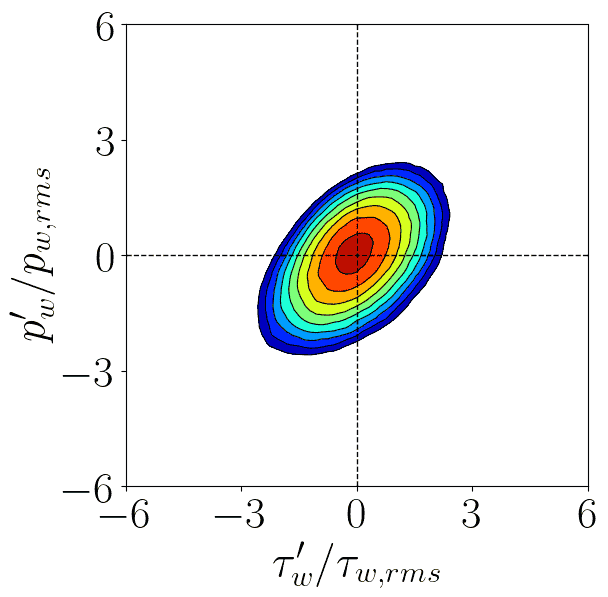} 
(b)\includegraphics[width=.3\textwidth]{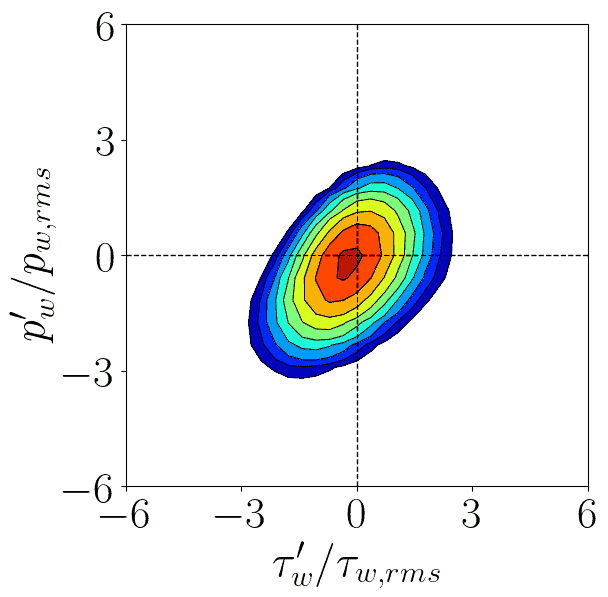}
(c)\includegraphics[width=.3\textwidth]{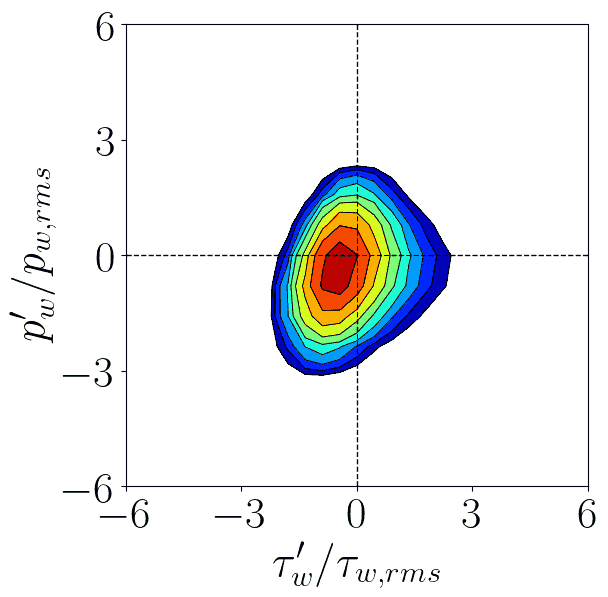}
\caption{Joint PDF of wall shear and pressure fluctuations at the inner wall, 
$P(\tau_w', p_w')$, for the R1 flow cases at $\Rey_b=4000$~(a), $20000$~(b) 
and $87000$~(c).} 
\label{fig:tpjpdf}
\end{figure}
For that purpose, we preliminarily verify whether those quantities are correlated 
by inspecting the joint PDF of the wall shear stress and of the fluctuating pressure 
at the inner wall, $P(\tau_w', p_w')$, which we report in figure~\ref{fig:tpjpdf}. 
Strong positive correlation emerges
at $\Rey_b=4000$ (a) and $20000$ (b), at which flow near the inner wall is dominated
by spanwise-coherent structures. This correlation becomes less distinct at $87000$ (c), 
at which turbulent fluctuations start reach down to the near-wall region. 
Hence, the effect of spanwise-coherent structures at the inner wall is to enhance 
pressure fluctuations, which are strongly correlated 
with shear stress fluctuations.

\begin{figure}
\includegraphics[width=.47\textwidth]{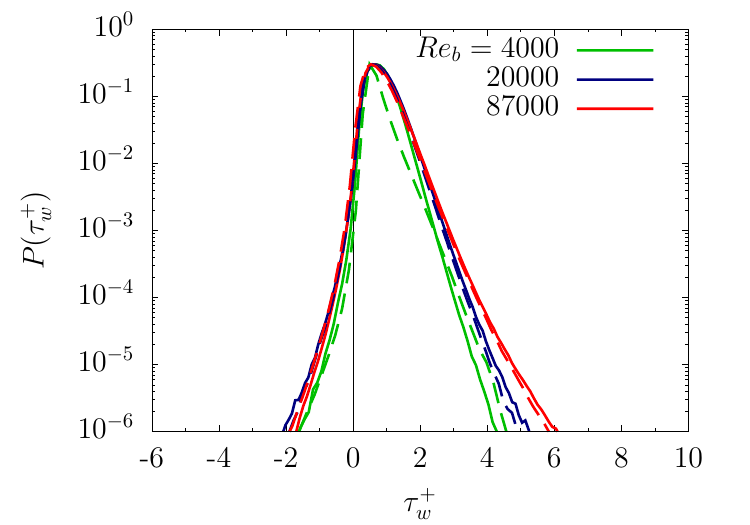}
\includegraphics[width=.47\textwidth]{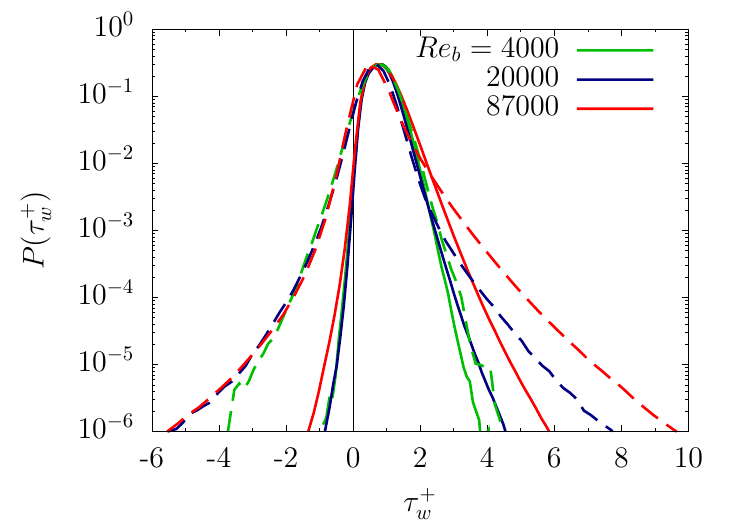}
\caption{PDF of the wall shear stress, $P(\tau_w^+)$, at various Reynolds numbers 
for flow cases R40 (a) and R1 flow cases (b).   
Dashed lines refer to the inner wall and solid lines refer to the outer wall.}
\label{fig:pdf}
\end{figure}
\begin{table}
\begin{center}
\begin{tabular}{cccc}
	$\quad \Rey_b\quad$ 
	& $\quad r_c/\delta\quad$ 
	& $P(\tau_{w,i}^+<0)$
	& $P(\tau_{w,o}^+<0)$ \\
\midrule
	4000  & 40.5 & 0.05 & 0.08 \\  
	20000 & 40.5 & 0.23 & 0.22 \\
	87000 & 40.5 & 0.36 & 0.22 \\
\midrule
	4000  & 1.0  & 4.38 & 0.10 \\  
	20000 & 1.0  & 3.92 & 0.09 \\
	87000 & 1.0  & 4.50 & 0.16 \\
\midrule
\end{tabular}
\caption{Probability of backflow events at the two walls at various 
Reynolds numbers, for the R40 and R1 flow cases.}
\label{tab:4}
\end{center}
\end{table}
Strong fluctuations of the wall shear can contribute to friction 
reduction at the inner wall by increasing the number of backflow 
events. This insight is supported by the PDF 
of the wall shear stress reported in figure~\ref{fig:pdf}. 
The mildly-curved cases, 
in which spanwise-coherent structures are not detected, can serve as a comparison. 
In addition, in table~\ref{tab:4} we list the probability of backflow 
events.
In the R40 flow cases (a) the PDFs of the inner-wall shear 
and of the outer-wall shear are nearly identical, 
with increasing probability of large values of positive shear 
as the Reynolds number increases. Backflow events are very rare, 
their probability not exceeding $0.3\%$. 
The picture is quite different for the R1 flow cases (b). 
As for the outer wall (solid lines), the probability of negative shear 
is even smaller than for the R40 flow cases. This results points to a 
possible analogy between the effect of favourable pressure gradient 
and concave curvature, since backflow probability decreases with 
flow acceleration~\citep{zaripov2023backflow}. 
In contrast, the PDF tails widen at the inner wall (dashed lines),
showcasing the enhancement of wall-shear fluctuations due to spanwise-coherent
structures. The widening of the negative tail is associated with 
increase of the backflow events at the inner wall, 
whose probability exceeds $4\%$. 

\begin{figure}
\centering
(a)\includegraphics[width=.3\textwidth]{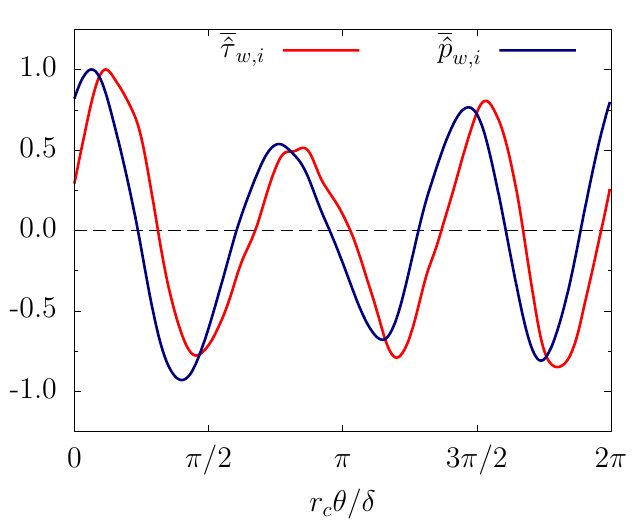}
(b)\includegraphics[width=.3\textwidth]{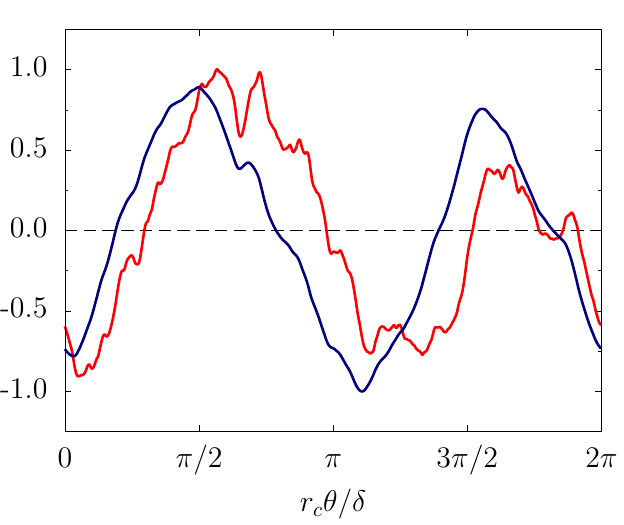}
(c)\includegraphics[width=.3\textwidth]{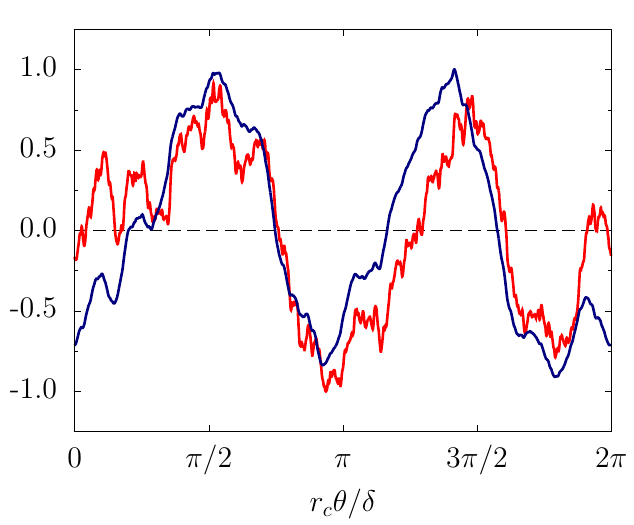}
\caption{Streamwise distribution of mean coherent 
shear stress ($\overline{\hat{\tau}}_{w,i}$, red lines),  
and mean coherent pressure ($\overline{\hat{p}}_{w,i}$, blue lines) 
at the inner wall, for the R1 flow cases at $\Rey_b=4000$~(a), $20000$~(b), $87000$~(c).
Both quantities are normalised by their maximum value.} 
\label{fig:tauwt}
\end{figure}

Further insights into the relationship between wall pressure and wall shear 
are provided in figure~\ref{fig:tauwt}, where we show the streamwise 
distribution of the coherent pressure and 
of the coherent shear stress at the inner wall,
both normalised by their maximum value.
Streamwise inhomogeneity induced by spanwise-coherent structures 
imposes clear imprint on the wall pressure 
(blue lines), which features peaks and troughs corresponding 
to the high- and low-pressure regions observed in figure~\ref{fig:tr}. 
The coherent wall shear stress (red lines) 
also features peaks and troughs, which we explain as follows. 
The coherent streamwise velocity of a clockwise-rotating roll cell 
(positive values of the streamfunction in figure~\ref{fig:tr}) 
opposes the mean flow near the inner wall, reducing locally 
the streamwise velocity and hence the wall shear. Counter-clockwise 
rotating roll cells act in the opposite way. 
The small-scale oscillations of the wall-shear trend overlapping 
with the large-scale ones at $\Rey_b=87000$~(c) are due to the 
presence of turbulent activity, which explains lower correlation 
between wall shear and wall pressure (see figure~\ref{fig:tpjpdf}). 
The effects of spanwise-coherent structures on the flow field 
can be further characterised by analysing the phase shift 
between the wall shear stress and the wall pressure. The wavy pressure 
distribution generates local pressure gradients in the streamwise direction, 
which tend to accelerate and decelerate the fluid and result in the 
alternating regions of high and low wall shear. 
Looking back at figure~\ref{fig:tr}, one can see that high-pressure regions 
correspond to counter-clockwise rotating eddies, vice-versa for 
the low-pressure regions. Hence, between any pair of counter-rotating 
eddies where the flow is locally subjected to an adverse pressure 
gradient ($u'<0$) high-speed fluid is pushed toward the inner wall ($v'<0$). 
Between any neighbouring pair, there is a favourable pressure 
gradient ($u'>0$) and simultaneously low-speed fluid is diverted away from 
the inner wall ($v'>0$). In both cases, the combination of these motions yields 
to $-u'v'<0$, explaining why spanwise structures make a negative contribution 
to the production of Reynolds shear stress, which we have highlighted 
in figure~\ref{fig:prms}(b). 

\subsection{Quadrant analysis}\label{sec:rey}
To provide a quantitative basis for the qualitative analysis above, 
we consider the joint probability density function (JPDF) of 
streamwise and wall-normal fluctuations, $P(u',v')$, such that
\begin{equation}
-\overline{u'v'}=\int_{-\infty}^{+\infty} u'v' P(u',v') \mathrm{d}u'\mathrm{d}v',
\end{equation}
where the covariance integrand, $u'v' P(u',v')$, is a measure of the 
contribution of each pair of $u'$ and $v'$ to the turbulent
shear stress~\citep{wallace1977reynolds}. 
Each quadrant of the plane of the $(u',v')$ plane
corresponds to a class of motion, specifically
Q2 and Q4 quadrant events correspond to `ejections' ($u'<0$ and $v'>0$) 
and `sweeps' ($u'>0$ and $v'<0$), yielding positive contribution to the
turbulent shear stress,
whereas Q1 and Q3 quadrants correspond to `outward interactions' ($u'>0$ 
and $v'>0$), and `inward interactions' ($u'<0$ and $v'<0$), which 
yield negative contribution to it.
\begin{figure}
\centering
(a)\includegraphics[width=.3\textwidth]{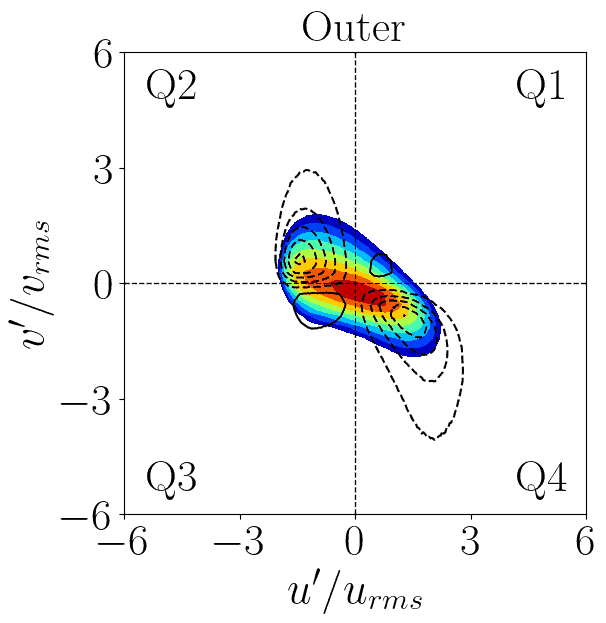}\label{uvjpdf_re4_r40_out}
(b)\includegraphics[width=.3\textwidth]{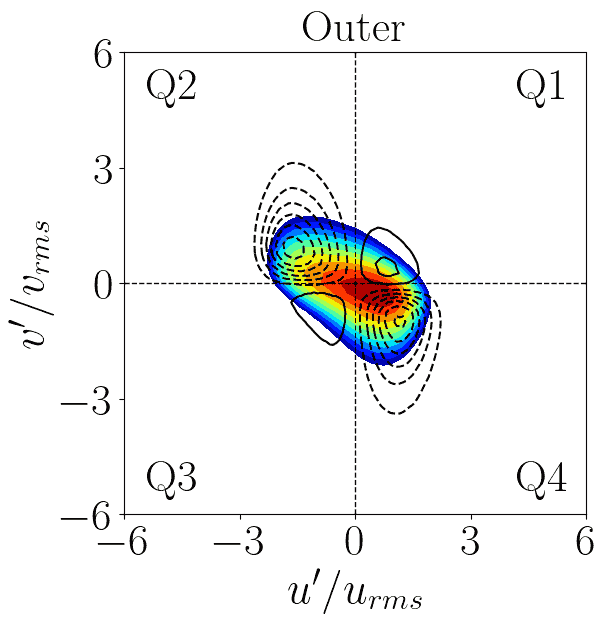}\label{uvjpdf_re20_r40_out}
(c)\includegraphics[width=.3\textwidth]{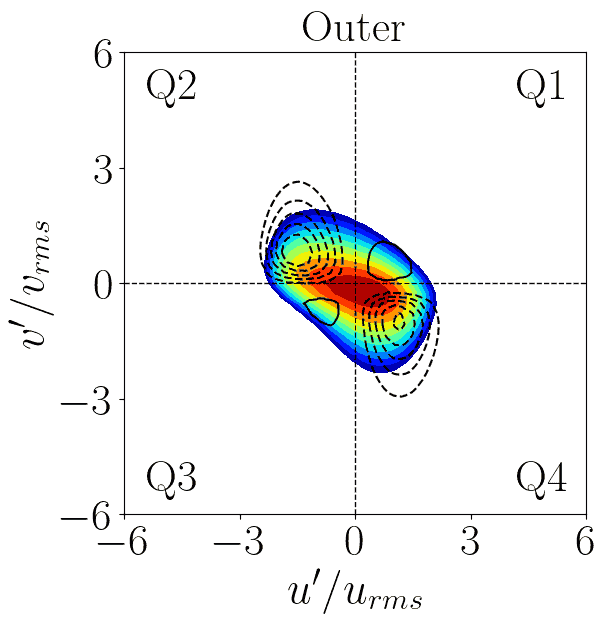}\label{uvjpdf_re87_r40_out}\\
(d)\includegraphics[width=.3\textwidth]{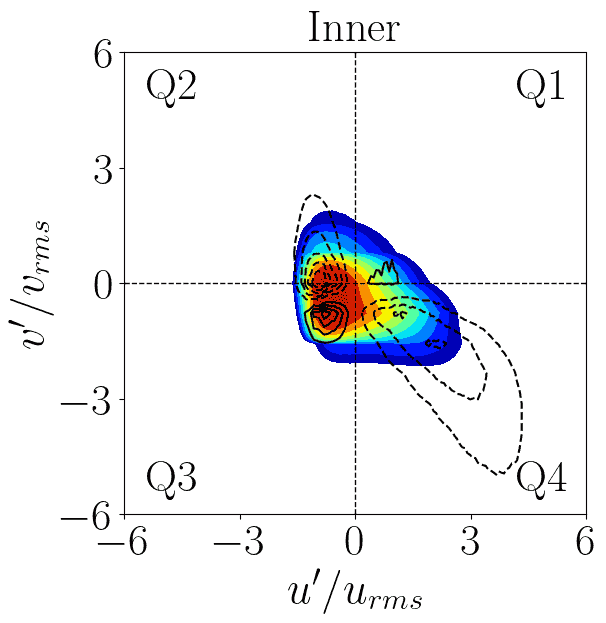}\label{uvjpdf_re4_r40_inn}
(e)\includegraphics[width=.3\textwidth]{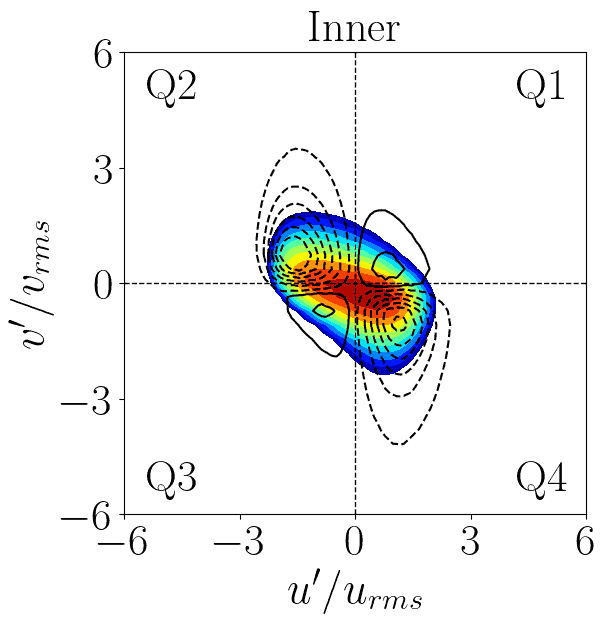}\label{uvjpdf_re20_r40_inn}
(f)\includegraphics[width=.3\textwidth]{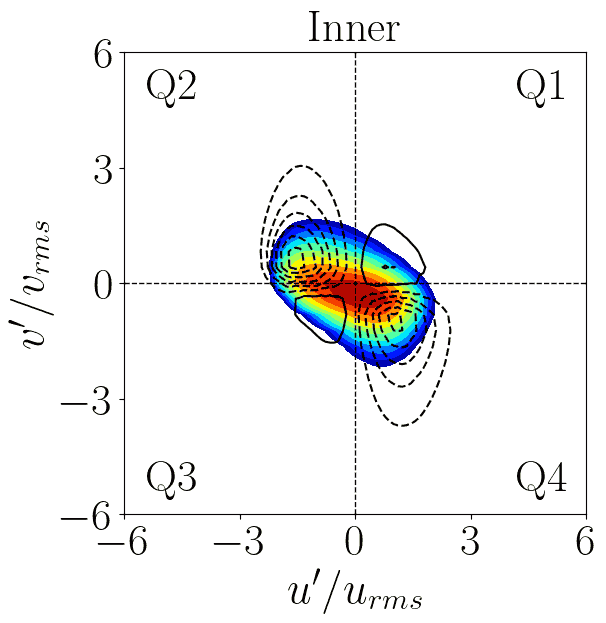}\label{uvjpdf_re87_r40_inn}
\caption{
Joint PDF of streamwise and wall-normal velocity fluctuations, 
superimposed to flooded contours of the covariance integrand, 
near the outer wall (upper panels) and the inner wall 
(lower panels) at $y^+\approx 12$, for the R40 flow cases. 
From left to right, the panels correspond to $\Rey_b=4000$~(a, d), 
$20000$~(b, e), $87000$~(c, f).} 
\label{fig:jpdf40}
\end{figure}
\begin{figure}
\centering
(a)\includegraphics[width=.3\textwidth]{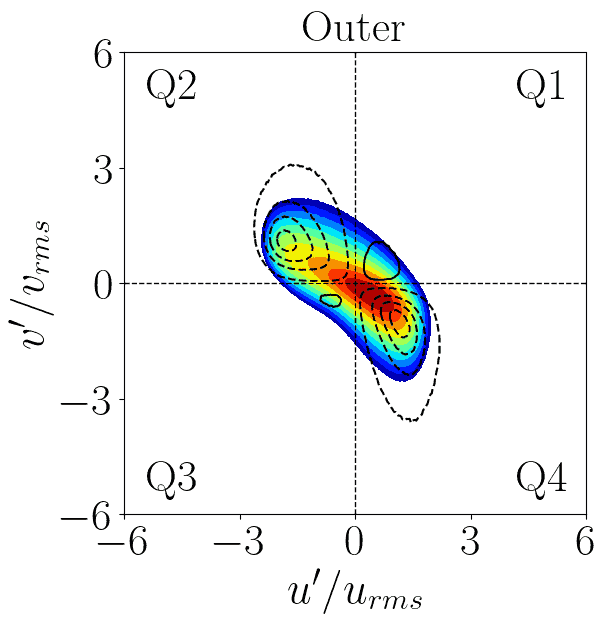}\label{uvjpdf_re4_r1_out}
(b)\includegraphics[width=.3\textwidth]{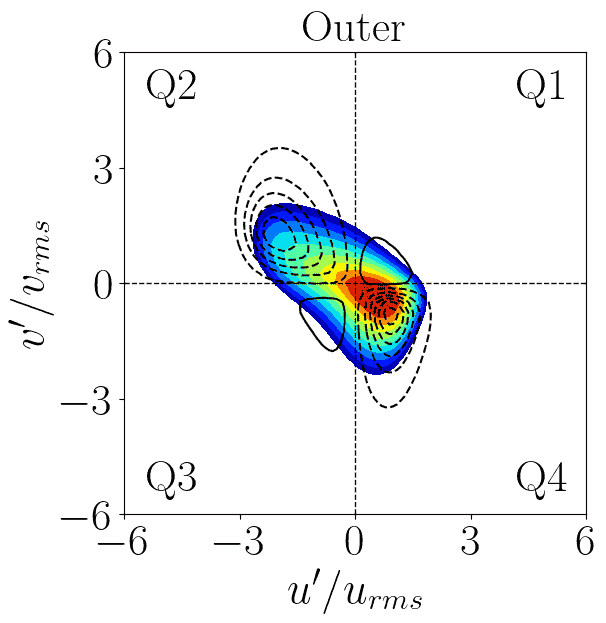}\label{uvjpdf_re20_r1_out}
(c)\includegraphics[width=.3\textwidth]{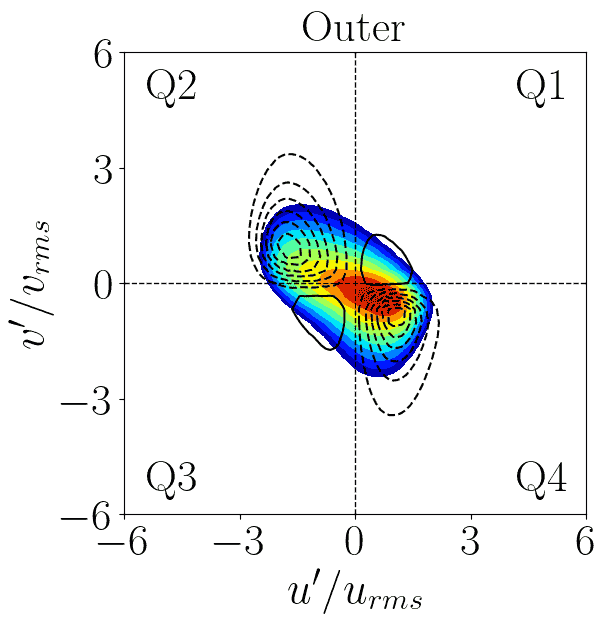}\label{uvjpdf_re87_r1_out}\\
(d)\includegraphics[width=.3\textwidth]{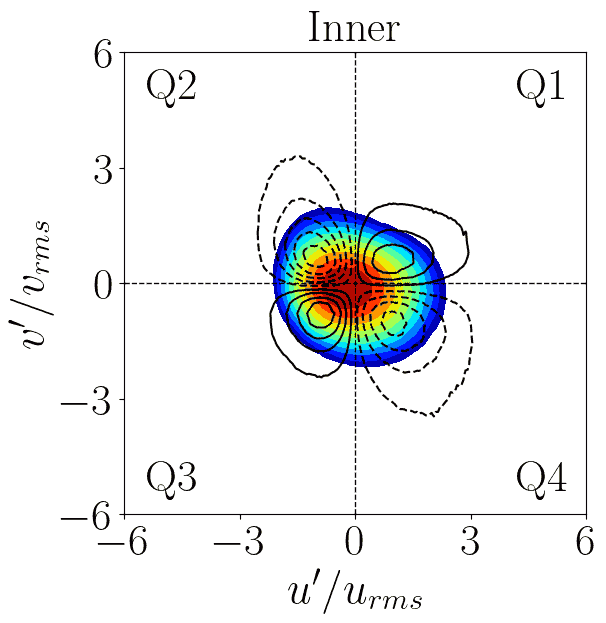}\label{uvjpdf_re4_r1_inn}
(e)\includegraphics[width=.3\textwidth]{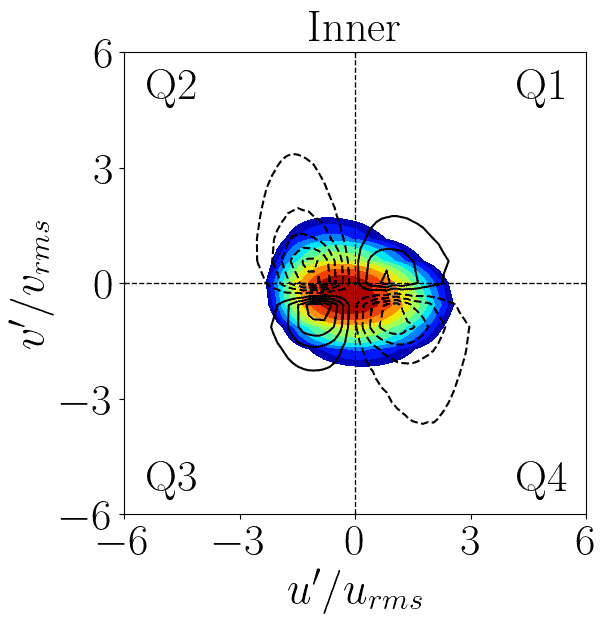}\label{uvjpdf_re20_r1_inn}
(f)\includegraphics[width=.3\textwidth]{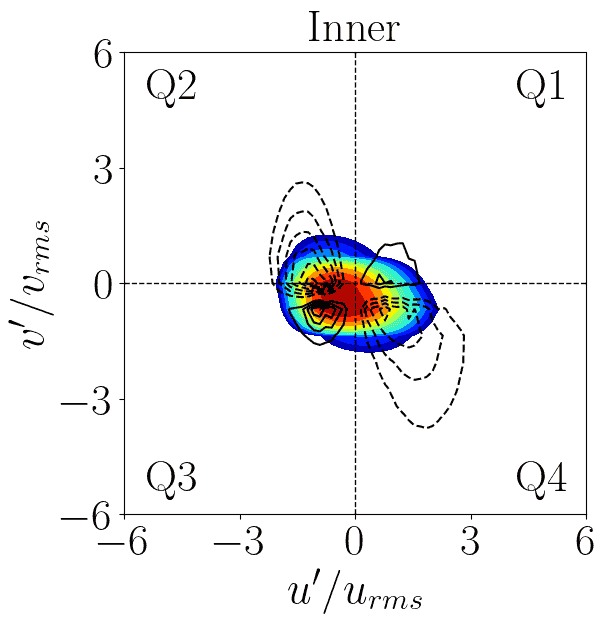}\label{uvjpdf_re87_r1_inn}
\caption{
Joint PDF of streamwise and wall-normal velocity fluctuations, 
superimposed to flooded contours of the covariance integrand, 
near the outer wall (upper panels) and the inner wall 
(lower panels) at $y^+\approx 12$, for the R1 flow cases. 
From left to right, the panels correspond to $\Rey_b=4000$~(a, d), 
$20000$~(b, e), $87000$~(c, f).} 
\label{fig:jpdf1}
\end{figure}
 
In figures~\ref{fig:jpdf40} and~\ref{fig:jpdf1} we show flooded contours
of $P(u',v')$ superimposed to iso-lines of $u'v'P(u',v')$ ,
for the R40 and R1 flow cases, respectively. 
The joint PDF is evaluated in wall-parallel planes near the outer and the 
inner wall at $y^+\approx 12$, which in plane channels is the 
`balance point' where the contributions of ejections and sweeps 
are equal~\citep{kim1987turbulence}. In the R40 flow cases, the JPDF 
at the outer wall (upper panels) has a roughly elliptical shape 
with major axis inclined along the Q2 and Q4 quadrants, 
pointing to high probability of sweeps and ejections. 
Similar observations apply to the fully turbulent R40 
flow cases near the inner wall~(panels (e) and (f) of figure~\ref{fig:jpdf40}), 
and for the R1 flow cases near the outer wall
(upper panels of figure~\ref{fig:jpdf1}).
However, strong curvature 
seems to increase the probability of sweeps more than ejections, which 
can be inferred from the shift of the JPDF peak towards the Q4 quadrant. 
The scenario is different at the inner wall. 
In the R40 flow case at $\Rey_b=4000$, depicted in figure~\ref{fig:jpdf40}(d), 
the peak probability is mainly concentrated in the Q2 and Q3 quadrants, 
and to a lesser extent in the Q4 quadrant,
on account of higher and less probable values of $u'$ and $v'$. 
The roughly circular shape of the JPDF at the inner wall in the 
R1 flow cases~(figure \ref{fig:jpdf1}, lower panels) shows that the negative 
correlation between $u'$ and $v'$, typical of the near-wall region of 
turbulent flows, vanishes. In addition, the covariance integrand 
highlights that the contribution to the turbulent stress is not dominated 
by the Q2 and Q4 motions, but strong contributions also come from 
the Q1 and Q3 motions. 
%
\begin{table}
\begin{center}
\begin{tabular}{ccccc}
$\quad \Rey_b\quad$ & $\quad$Q1$\quad$ & $\quad$Q2$\quad$ & $\quad$Q3$\quad$ & $\quad$Q4$\quad$\\
\toprule
	\multicolumn{5}{c}{R40 outer wall} \\ 
\midrule
	4000  &  -8.90 & +47.56 &  -7.90 & +69.24 \\  
	20000 & -12.67 & +66.64 & -10.10 & +56.13 \\
	87000 & -16.60 & +66.59 & -11.63 & +61.64 \\
\midrule
	\multicolumn{5}{c}{R40 inner wall} \\ 
\midrule
	4000  & -30.76 & +31.38 & -11.82 & +111.19\\  
	20000 & -14.10 & +63.82 &  -9.57 & +59.86 \\
	87000 & -17.26 & +63.75 & -10.81 & +64.33 \\
\midrule
	\multicolumn{5}{c}{R1 outer wall} \\ 
\midrule
	4000  & -5.09 & +56.27 & -2.43 & +51.29 \\  
	20000 & -7.01 & +72.47 & -7.09 & +41.64 \\
	87000 & -9.58 & +68.95 & -9.51 & +50.14 \\
\midrule
	\multicolumn{5}{c}{R1 inner wall} \\ 
\midrule
	4000  & -75.25 &+115.39 & -65.18 & +125.04 \\  
	20000 & -60.01 & +95.88 & -49.38 & +113.52 \\
	87000 & -39.40 & +59.72 & -21.18 & +100.86 \\
\bottomrule
\end{tabular}
\caption{Percentage contribution of quadrants to turbulent
shear stress ($\overline{u'v'}_{Qi}/\overline{u'v'}$),
for the R40 and R1 flow cases near the inner and outer wall, at $y^+\approx 12$.}
\label{tab:3}
\end{center}
\end{table}
 
More quantitative results are presented in table~\ref{tab:3}, 
where we list the integrated contributions to the turbulent shear stress 
from each quadrant ($\overline{u'v'}_{Qi}/\overline{u'v'}$, $i=1,2,3,4$) 
at $y^+\approx 12$. For the R40 fully turbulent flow cases the results are 
comparable with experimental results for plane channel flow~\citep{wallace1972wall}, 
namely the contribution of both
ejections and sweeps is about $70\%$,
whereas inward and outward interactions each contribute negatively 
by about $20\%$. At $\Rey_b=4000$, instead, the Q2 contribution near the 
inner wall reduces to $30\%$, which is comparable with the Q1 contribution but 
opposite in sign, whereas the Q4 contribution exceeds $110\%$ of 
the shear stress. Comparing this value with the shape of the JPDF 
in figure~\ref{fig:jpdf40}(d), one can infer that larger, energetic but 
infrequent motions are the main contributors to the turbulent shear stress. 
These strong sweeps can be attributed to the
longitudinal vortices 
pushing the high-speed mean flow from the channel core towards the inner wall. 
As for the R1 flow cases, the quadrant contributions at the outer wall 
are similar to those of the R40 flow cases, except for greater contribution 
of ejections as compared to sweeps. Hence, the most probable motions 
(i.e. sweeps, as visible in figure~\ref{fig:jpdf1}) do not contribute as much
to the turbulent shear stress as the ejections, see table~\ref{tab:3}.
At the inner wall, the fractional contributions of the outward (Q1) 
and inward (Q3) interactions exceeds half the contributions of ejections 
(Q2) and sweeps (Q4). This result confirms that the transverse large-scale 
structures, which were found to generate Q1 and Q3 motions in \S\ref{sec:wall}, 
play a key role in the strong attenuation of the turbulent shear stress 
near the inner wall of strongly curved channels.  

\subsection{Energy production reversal}\label{sec:reversal}

\begin{figure}
\centering
(a)\includegraphics[width=.45\textwidth]{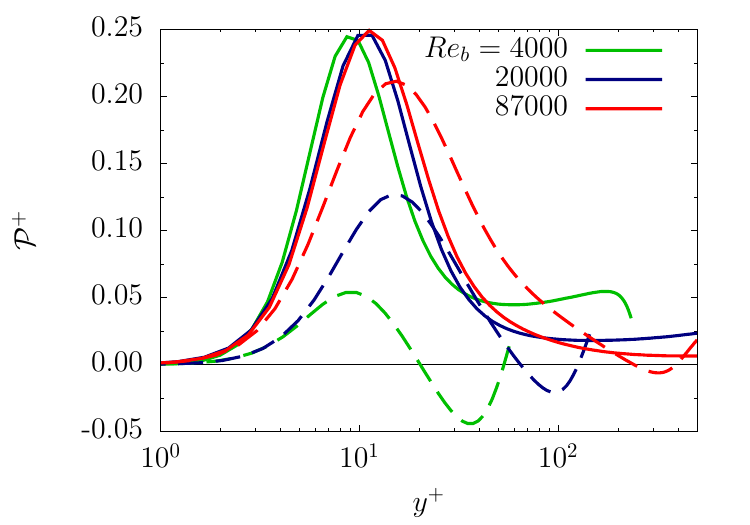}
(b)\includegraphics[width=.45\textwidth]{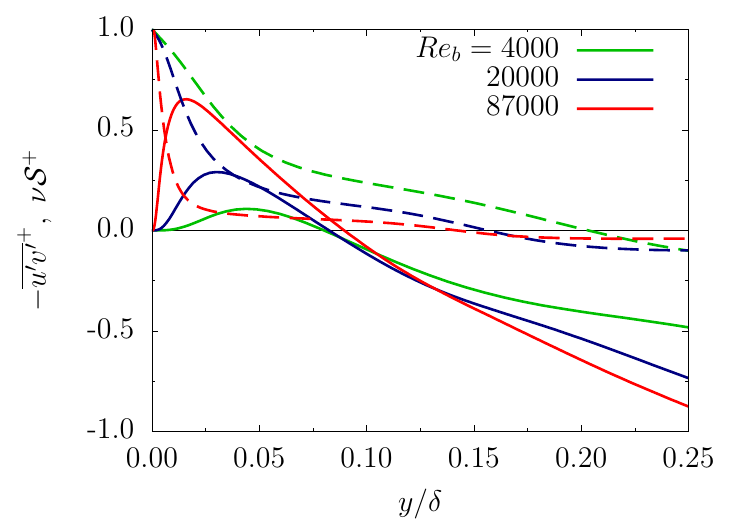}\\
\caption{(a) TKE production ($\mathcal{P}^+$) 
near the outer wall (solid lines) and near the inner wall (dashed lines). 
(b) Turbulent shear stress ($-\overline{u'v'}^+$, solid lines) and viscous 
shear stress ($\nu\mathcal{S}^+$, dashed lines) near the inner wall. 
All quantities are reported in local wall units for the R1 flow cases.} 
\label{fig:prod}
\end{figure}

Outward and inward interactions contribute negatively to turbulent
shear stress, hence they yield negative contribution to the production 
of turbulence kinetic energy (TKE), $\mathcal{P}=-\overline{u'v'}\mathcal{S}$, 
where $\mathcal{S}=\mathrm{d}U/\mathrm{d}r-U/r$ is the mean shear rate. 
Based on the analysis in \S\ref{sec:rey}, we expect that TKE production 
be reduced near the inner wall in flow cases with strong curvature. 
In figure~\ref{fig:prod}(a) we then show the TKE production 
for the R1 flow cases. Near the outer wall 
(solid lines) all curves tend to collapse, especially at high Reynolds 
number, revealing that the flow similarity is preserved near a 
concave surface. 
The peak production is located at $y^+\approx12$ (except at $\Rey_b=4000$, 
for which the peak occurs at $y^+\approx9$) and the peak value is 
$\mathcal{P}^+\approx0.25$, similar to
plane channel flow~\citep{kim1987turbulence, laadhari2002evolution}. 
In contrast, TKE production near the inner wall (dashed lines) depends 
heavily on the Reynolds number, implying that classical wall scaling 
no longer holds near highly convex surfaces. Furthermore, a region where 
production is negative appears at each Reynolds number. 
For the production to be positive everywhere, the turbulent shear stress 
and the viscous shear stress must have the same sign and vanish 
at the same location. 
This is not the case for the strongly curved channel, as clearly illustrated in 
figure~\ref{fig:prod}(b), which shows profiles of the turbulent shear 
stress, $-\overline{u'v'}^+$ (solid lines), and of the viscous shear stress, 
$\nu\mathcal{S}^+$ (dashed lines), near the inner wall of the R1 flow cases. 
%
The displacement between the zero-crossings of the turbulent shear stress 
and of the viscous shear, the amplitude of which decreases slightly 
as the Reynolds number increases, is related to asymmetry of the 
mean velocity profile~\citep{beguier2005negative} and  
leads to a region of opposing shear where $-\overline{u'v'}^+<0$ and 
$\nu\mathcal{S}^+>0$, hence $\mathcal{P}^+<0$. 
As pointed out in \S\ref{sec:fric}, 
the point 
of vanishing shear marks the interface between the two 
distinct flow structures developing at each wall. Straddling this 
interface there is a diffusive transfer of shear stress from the 
outer- to the inner-wall region that has a sign opposite to the 
locally produced shear stress, leading to negative net 
TKE production~\citep{hanjalic1972fully}. 
In the region with negative production a local energy reversal 
mechanism takes place, whereby energy is transferred from turbulent 
fluctuations to the mean flow~\citep{eskinazi1969energy}. Besides its 
physical significance, this result provides useful caveat for 
use of RANS models for simulations of turbulent flows over convex walls. 
In fact, standard eddy-viscosity models would clearly fail if
turbulent shear stress and mean shear do not go to zero at the same location.

\section{Conclusions}\label{sec:conclusions}
%
We have investigated fully developed flow in a curved channel 
to get insight into turbulence bounded by curved surfaces. 
This setup showcases rich physics due to the interplay of turbulence 
with large-scale coherent structures driven by centrifugal instabilities, 
which break the symmetry of the flow resulting in different behaviour 
near the convex and the concave walls. We have focused on the effects of curvature 
by examining two extreme cases, a mildly curved channel with radius of 
curvature $r_c/\delta=40.5$ and a strongly curved channel with $r_c/\delta=1$, 
where $\delta$ is the channel height. 
For each geometry, we have studied the effect of Reynolds number ($\Rey_b$) through 
an extensive series of DNS, covering flow regimes from laminar up to the 
moderately high value of $\Rey_b=u_b\delta/\nu=87000$, 
where $u_b$ is the bulk velocity and $\nu$ the kinematic viscosity. 
Our analysis has shown that the friction coefficient is somewhat higher 
as compared to the case of a plane channel. In addition, we have found 
that flow transition is anticipated by concave wall curvature and 
delayed by convex curvature, thus preventing turbulence to fully develop 
near the inner wall of strongly curved channels. 
Visualisations of the flow field have shown the presence 
of fine-scale turbulent structures as well as large-scale coherent structures 
elongated either in the streamwise and in the cross-stream directions. 
Through the spectral analysis of velocity fluctuations we detected 
the clear imprinting 
of the near-wall turbulence cycle at both walls in the case of mildly 
curved channels, whereas wall turbulence is virtually absent at the 
inner (convex) wall of strongly curved channels. Clear footprints of 
longitudinal and transverse large-scale structures were also found. 

Longitudinal large-scale structures, originating from centrifugal 
instabilities and resembling the Dean vortices found in laminar flow, 
were identified through a vortex eduction based on triple decomposition, 
which allowed to quantify their effects on turbulence. Specifically, 
longitudinal coherent structures were found to depend weakly on Reynolds 
number and highly on the channel geometry. As channel curvature increases, 
vortices are displaced toward the outer wall, their size reduces 
and they become more unsteady. 
As a consequence, the mean spanwise distribution of the wall shear 
is modified by longitudinal vortices only in mildly curved cases. 
The effects of longitudinal coherent vortices 
on velocity fluctuations and turbulent shear stress were quantified, 
revealing that in mildly curved channels streamwise velocity fluctuations 
and momentum transport are greatly affected, whereas in strongly curved 
channels the wall-normal velocity component is the most influenced. 
Through the combined use of spectral analysis and triple decomposition, 
we have identified unsteady splitting and merging of the longitudinal 
coherent eddies.
Transverse large-scale structures were found near the convex wall
of strongly curved channels, whose footprint consists of pressure waves 
advected downstream at the mean flow speed. 
These structures were also characterised by using triple decomposition 
and phase averaging, showing that 
their effect is to enhance pressure fluctuations near the inner wall. 
This enhancement is correlated with increased fluctuations 
of the wall shear stress, which is responsible for friction 
reduction at the inner wall. The relationship between increased pressure 
fluctuations and decreased wall friction has been attributed 
to the generation 
of alternating favourable and adverse pressure gradients, 
which combine with upward and downward radial motions, respectively, 
suppressing the turbulent shear stress near the convex wall. 
This insight has been further explored and quantified 
through quadrant analysis of streamwise and wall-normal velocity 
fluctuations. 
An enhancement of inward and outward interactions 
has been found, leading to the appearance of a region with negative production 
of turbulence kinetic energy in the strongly curved cases. 
 
These results offer a valuable dataset for fundamental research on the interaction between turbulence and curved walls, 
as well as for turbulence modelling, e.g. subgrid-scale models for LES and RANS closures. 
We also believe that insights gained from the time-evolving curved channel flow 
may be relevant to boundary layers over both convex and concave surfaces. 
In fact, our analysis has shown that the interaction between the inner and outer wall regions 
is concentrated in the channel core, 
whereas the inner (outer) wall region is dominated by convex (concave) curvature effects. 
In this work, we focused on two values of curvature. 
Further investigation of additional curvatures is necessary 
to determine the critical value at which the coherence of near-wall streaks 
at the convex wall breaks down, and whether significant changes in flow physics 
occur with increasing curvature. 

\backsection[Acknowledgements]{The results reported in this paper have been achieved 
using the EuroHPC Research Infrastructure resource LEONARDO based at CINECA, 
Casalecchio di Reno, Italy, under project EuroHPC $02044$ and IscraB CONCORDE.}

\backsection[Funding]{This research received financial support from ICSC-Centro Nazionale di Ricerca in 
`High Performance Computing, Big Data and Quantum Computing', funded by European Union-NextGenerationEU.}

\backsection[Declaration of interests]{The authors report no conflict of interest.}

\backsection[Data availability statement]{All data that support the findings of this 
study are available from the corresponding author upon request.}

\backsection[Author ORCIDs]{\\
G. Soldati, https://orcid.org/0009-0007-4597-756X;\\
P. Orlandi, https://orcid.org/0000-0002-0305-5723;\\
S. Pirozzoli, https://orcid.org/0000-0002-7160-3023.
}

\appendix

\section{Effects of domain size}\label{app:domain}
\begin{figure}
\centering
(a)\includegraphics[width=.47\textwidth]{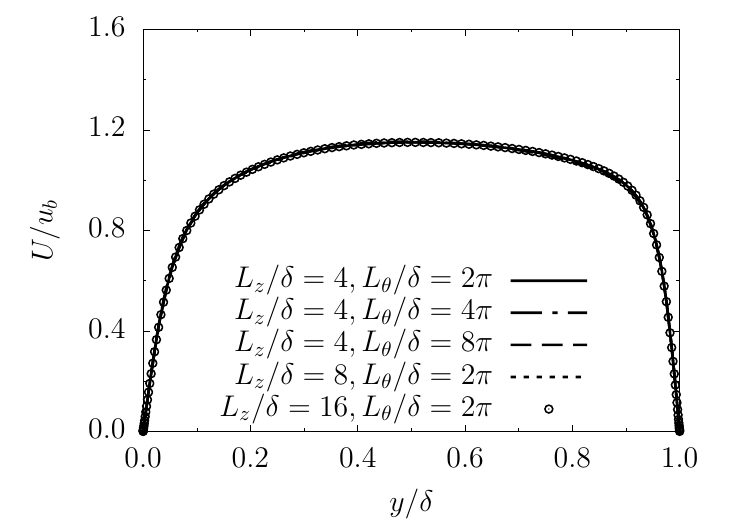}
(b)\includegraphics[width=.47\textwidth]{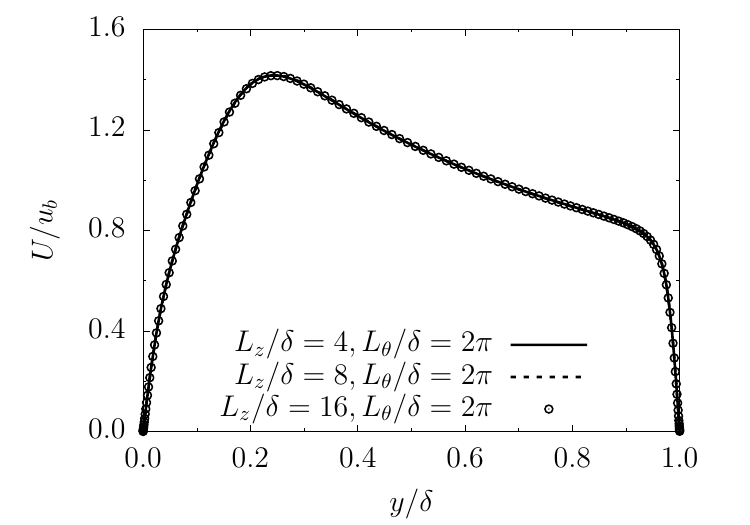}
(c)\includegraphics[width=.47\textwidth]{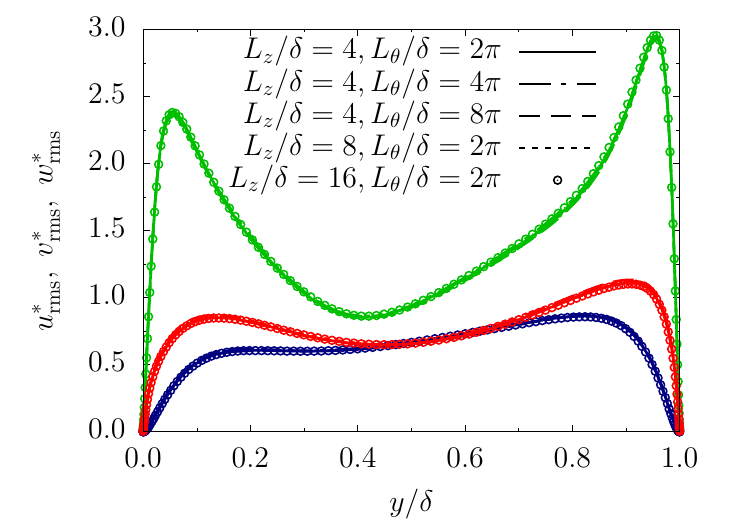}
(d)\includegraphics[width=.47\textwidth]{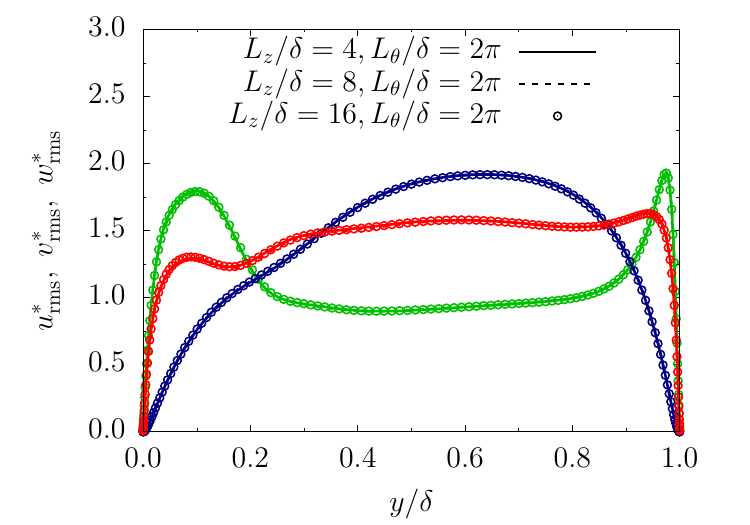}
\caption{Mean streamwise velocity ($U/u_b$) 
for the R40 flow cases~(a) and R1 flow cases~(b);
RMS profiles of streamwise (green), 
wall-normal (blue) and spanwise (red)
velocity fluctuations 
for the R40 flow cases~(c) and R1 flow cases~(d).
The Reynolds number is fixed at $\Rey_b=5000$ and the domain size is 
varied as marked, case by case, by the different line types.}
\label{fig:U_sens}
\end{figure}
Six additional simulations have been carried out to verify a-posteriori 
the adequacy of the domain sizes used in the DNS campaign.  
Here, we consider the Reynolds number $\Rey_b=5000$, which is the least
to heave fully turbulent flow. As for the R40 flow cases, from the baseline 
configuration ($L_\theta/\delta=2\pi$, $L_z/\delta=4$) we have doubled 
and quadrupled both $L_\theta$ (for fixed $L_z$) and $L_z$ 
(for fixed $L_\theta$), by retaining the same grid spacing.
As for the R1 flow cases, the streamwise extent of the domain resolves 
a full circumference, hence it was not increased. 
From $L_z/\delta=8$ used in the baseline DNS, the spanwise extent 
was decreased by a factor two ($L_z/\delta=4$) and 
increased by a factor two ($L_z/\delta=16$). 
In figure~\ref{fig:U_sens} we show the mean streamwise velocity (a, b) 
and the RMS profiles of streamwise (green), wall-normal (blue) and 
spanwise (red) velocity fluctuations (c, d) for both the R40 flow cases 
(a, c) and R1 flow cases (b, d). One will see that all the profiles 
obtained with the different combinations of domain size  
(marked by the different line types) nearly collapse to a single line, 
showing that the domain size used in the DNS campaign is adequate.  

\section{Global friction velocity}\label{app:utaug}
\begin{figure}
\centering
\includegraphics[width=.6\textwidth]{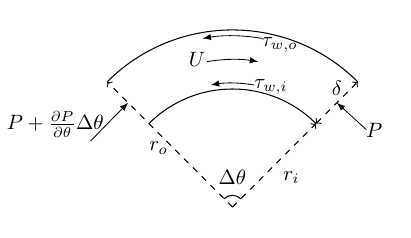}
\caption{Sketch of the curved channel flow.}
\label{fig:sketch}
\end{figure}
The derivation of the global friction velocity has been shown 
by~\cite{brethouwer2022turbulent} starting from the mean momentum balance, 
yet we believe that it may be instructive to illustrate briefly 
how it relates to the mean-pressure gradient.
The analytical expression of the streamwise pressure gradient 
as a function of the shear stresses at the two walls 
can be derived in the simplest way from the balance of moments of forces 
about the centre of curvature, 
which, referring to the sketch of figure~\ref{fig:sketch}, reads
\begin{equation}
	\int_{r_i}^{r_o}\left(-P-\frac{\partial P}{\partial \theta}\Delta\theta+P\right)r\mathrm{d}r
	+ \left(\tau_{w,i}r_i^2+\tau_{w,o}r_o^2\right)\Delta\theta = 0.
\end{equation}
Hence, recalling that $r_c=(r_o+r_i)/2$ and $\delta=r_o-r_i$, we obtain 
\begin{equation}
	-\dt{P}{\theta}=\frac{\tau_{w,i} r_i^2+\tau_{w,o} r_o^2}{r_c\delta}.
	\label{dpdt}
\end{equation}
To define a global friction velocity, $u_{\tau,g}$, 
an equivalent length of the curved channel must be defined. 
Considering the length of the channel at the centreline,
$L_\theta=r_c\Delta\theta$, one obtains
\begin{equation}
	-\frac{1}{\rho r_c}\dt{P}{\theta} = 2\frac{u_{\tau,g}^2}{\delta}\quad\Rightarrow\quad
	u_{\tau,g}^2=\frac{u_{\tau,i}^2 r_i^2+u_{\tau,o}^2 r_o^2}{2r_c^2}
\label{eq:utau_g}
\end{equation}
where $u_{\tau,i}^2 = \tau_{w,i}/\rho$ and $u_{\tau,o}^2 = \tau_{w,o}/\rho$.

\section{Convergence of flow statistics}\label{app:stats}
Whether the statistics are well-converged can be verified by comparing 
the analytical profile of the total shear stress with the DNS results. 
As shown by~\cite{brethouwer2022turbulent}, 
starting from the mean momentum equation in streamwise direction 
for the curved channel flow, 
which reads 
\begin{equation}
	0=-\frac{1}{\rho r}\de{P}{\theta}+\nu\left(\frac{1}{r}\de{U}{r}
	-\frac{U}{r^2}+\deq{U}{r}\right)-
	\left(\de{\overline{uv}}{r}+\frac{2}{r}\overline{uv}\right),
\label{mombal}
\end{equation}
one can derive the following equation for the total shear 
stress distribution
\begin{equation}
	\tau(r) = \frac{r_i^2\tau_{w,i}(r_o^2-r^2)
	+r_o^2\tau_{w,o}(r_i^2-r^2)}{2r^2r_c\delta}.
	\label{eq:tost}
\end{equation}
\begin{figure}
\centering
	(a)\includegraphics[width=.47\textwidth]{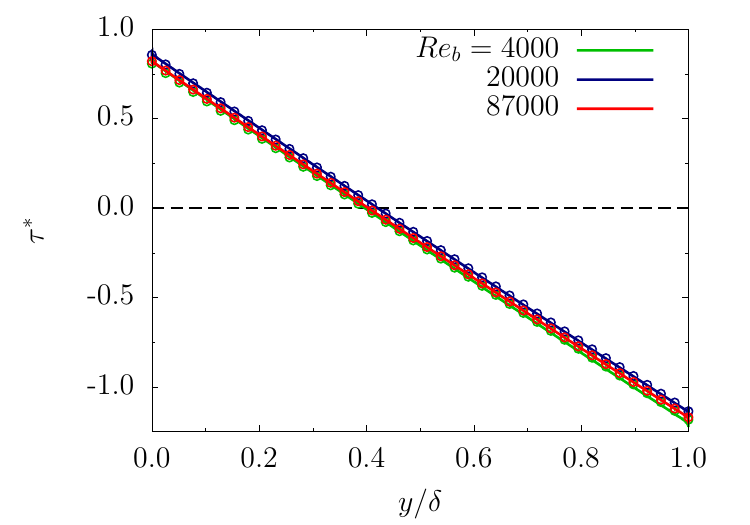}
	(b)\includegraphics[width=.47\textwidth]{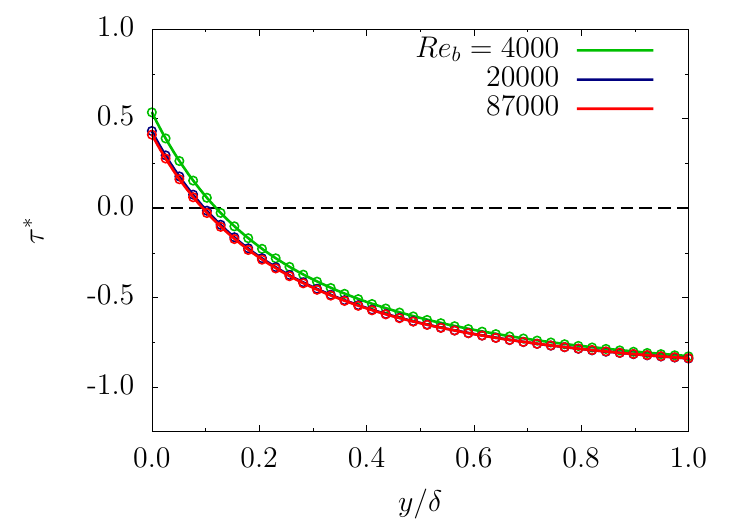}
\caption{Mean profile of the total shear stress ($\tau^*$) 
at various Reynolds numbers for the R40 flow cases~(a) and 
R1 flow cases~(b). Circles denote the analytical profile~\eqref{eq:tost}.}
\label{fig:tost}
\end{figure}
In figure~\ref{fig:tost} we show the analytical profile of the total shear stress 
(circles) and the DNS results at the corresponding Reynolds numbers 
(solid lines). The DNS results effectively collapse to the analytical profile, 
which corroborates that satisfactory statistical convergence is acheived.
%
%
\begin{figure}
	(a)\includegraphics[width=.46\textwidth]{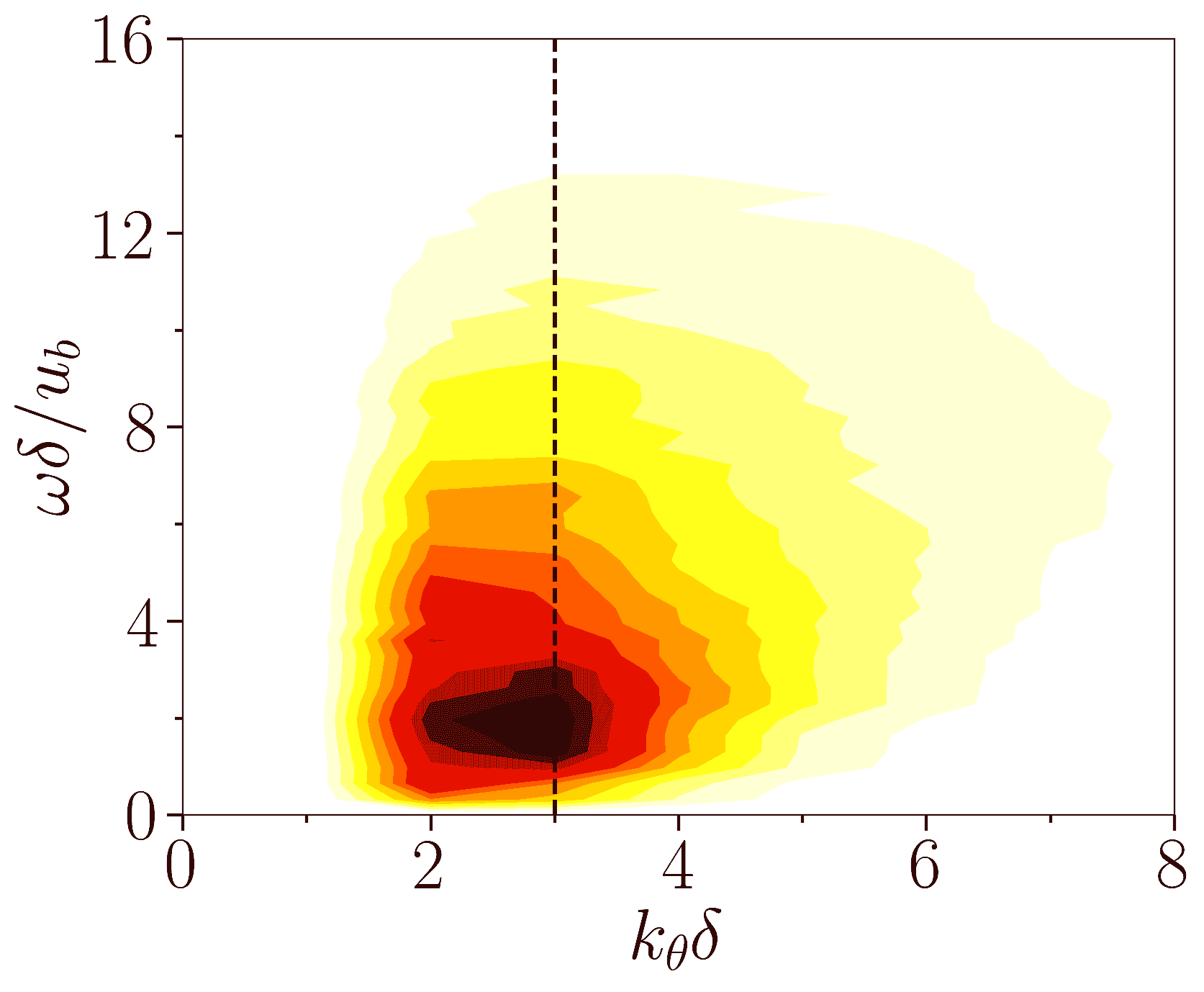}
	(b)\includegraphics[width=.48\textwidth]{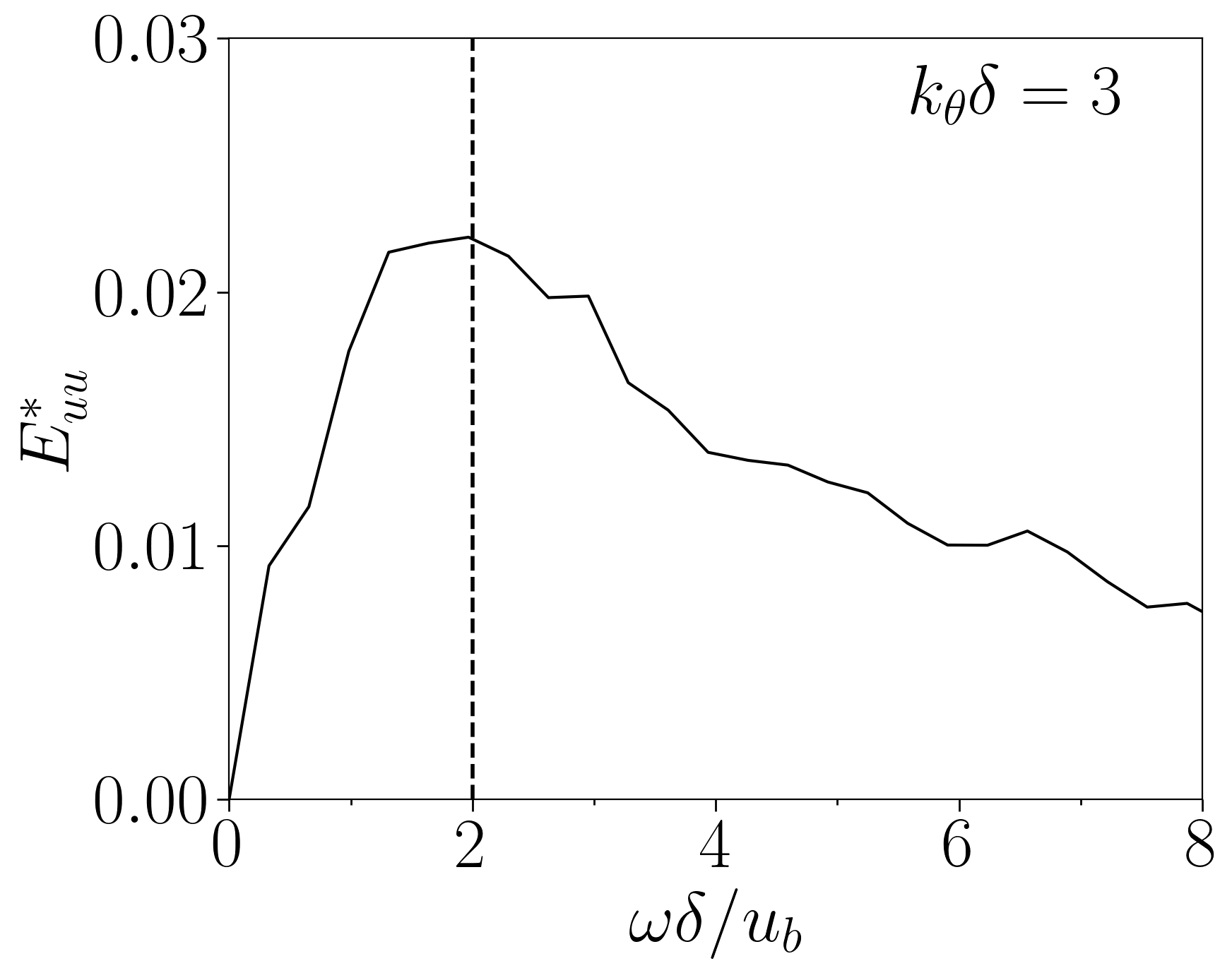}
	\caption{(a) Wavenumber-frequency spectra of the fluctuating 
	streamwise velocity,
	$E^*_{uu}(k_\theta,\omega)$, for the R1 flow case at $\Rey_b=4000$; 
	(b) $E^*_{uu}(\omega)$ at $k_\theta\delta=3$.
	The wall distance is fixed at $y/\delta=0.09$.}
\label{fig:freqwave}
\end{figure}
\section{Estimation of the convection velocity}\label{app:conv}
In figure~\ref{fig:freqwave}(a) we show the wavenumber-frequency spectra 
of the streamwise velocity fluctuations, $E^*_{uu}(k_\theta,\omega)$, 
for the R1 flow case at $\Rey_b=4000$. The wall distance is kept fixed at 
$y/\delta=0.09$, which is the peak location of $k_\theta^*E_{uu}^*$ 
related to transverse structures (see figure~\ref{fig:usp1}). 
An energy peak appears in the spectral density at the wavenumber 
$k_\theta\delta\approx3$ (marked by the vertical dashed line) 
corresponding to $\lambda_\theta/\delta\approx2\pi/3$, which is the 
wavelength of transverse structures at $\Rey_b=4000$ 
(see figure~\ref{fig:usp1}). 
As seen in figure~\ref{fig:freqwave}(b), 
showing $E^*_{uu}(\omega)$ fixed the wavenumber at $k_\theta\delta=3$, 
the energy peak is at the angular frequency $\omega\approx2u_b/\delta$. 
Hence, the convection velocity of transverse 
structures can be estimated as $u_c=\omega/k_\theta\approx0.67u_b$. 

\vspace{1cm}
%
%
%

\bibliographystyle{jfm}
\bibliography{jfm}
\end{document}